\documentclass[11pt,twoside]{article}
\usepackage{latexsym}

\makeatletter
									       
\def\icite{\@ifnextchar [{\@tempswatrue\@citey}{\@tempswafalse\@citey[]}}
\def\@citex[#1]#2{%
\if@filesw \immediate \write \@auxout {\string \citation {#2}}\fi 
\@tempcntb\m@ne \let\@h@ld\relax \def\@citea{}%
\@cite{%
  \@for \@citeb:=#2\do {%
    \@ifundefined {b@\@citeb}%
      {\@h@ld\@citea\@tempcntb\m@ne{\bf ?}%
      \@warning {Citation `\@citeb ' on page \thepage \space undefined}}%
      {\@tempcnta\@tempcntb \advance\@tempcnta\@ne%
      \@tempcntb\number\csname b@\@citeb \endcsname \relax%
      \ifnum\@tempcnta=\@tempcntb 
	\ifx\@h@ld\relax%
	  \edef \@h@ld{\@citea\csname b@\@citeb\endcsname}%
	\else%
	  \edef\@h@ld{\ifmmode{-}\else--\fi\csname b@\@citeb\endcsname}%
	\fi%
      \else
	\@h@ld\@citea\csname b@\@citeb \endcsname%
	\let\@h@ld\relax%
      \fi}%
    \def\@citea{,\penalty\@highpenalty\,}%
  }\@h@ld
}{#1}}
\def\@citey[#1]#2{%
\if@filesw \immediate \write \@auxout {\string \citation {#2}}\fi 
\@tempcntb\m@ne \let\@h@ld\relax \def\@citea{}%
\@icite{%
  \@for \@citeb:=#2\do {%
    \@ifundefined {b@\@citeb}%
      {\@h@ld\@citea\@tempcntb\m@ne{\bf ?}%
      \@warning {Citation `\@citeb ' on page \thepage \space undefined}}%
      {\@tempcnta\@tempcntb \advance\@tempcnta\@ne%
      \@tempcntb\number\csname b@\@citeb \endcsname \relax%
      \ifnum\@tempcnta=\@tempcntb 
	\ifx\@h@ld\relax%
	  \edef \@h@ld{\@citea\csname b@\@citeb\endcsname}%
	\else%
	  \edef\@h@ld{\ifmmode{-}\else--\fi\csname b@\@citeb\endcsname}%
	\fi%
      \else
	\@h@ld\@citea\csname b@\@citeb \endcsname%
	\let\@h@ld\relax%
      \fi}%
    \def\@citea{,\penalty\@highpenalty\,}%
  }\@h@ld
}{#1}}
\def\@cite#1#2{{$^{#1}$\if@tempswa , #2\fi }}
\def\@icite#1#2{{$#1$\if@tempswa , #2\fi }}

\gdef\@publabel{\hfil}
\gdef\@pubdate{\null}
\gdef\@pubnumber{\null}
\gdef\@author{\null}
\gdef\@title{\null}
\gdef\@abstract{\null}
\long\def\pubdate#1{\gdef\@pubdate{#1}}
\long\def\pubnumber#1{\gdef\@pubnumber{#1}}
\long\def\publabel#1{\gdef\@publabel{#1}}
\long\def\author#1{\gdef\@author{#1}}
\long\def\title#1{\gdef\@title{#1}}
\long\def\abstract#1{\gdef\@abstract{#1}}
\def\titlerelax{
\let\maketitle\relax
\let\settitleparameters\relax
\let\consolidatetitle\relax
\let\inittitlepage\relax
\let\finishtitlepage\relax
\let\titlepagecontents\relax
\let\multithanks\relax
\let\titlebaselines\relax
\let\@makepub\relax
\let\@maketitle\relax
\let\@makeauthor\relax
\let\@makeabstract\relax
\let\@maketitlenote\relax
\let\thanks\relax
\let\titlerelax\relax}

\def\titleclean
{\gdef\@titlenote{}
\gdef\@abstract{}
\gdef\@author{}
\gdef\@title{}
\gdef\@pubdate{}\gdef\@pubnumber{}\gdef\@publabel{}
\gdef\@dpublabel{}
}
\def\@makepub{\vbox to \z@{\hbox to \textwidth{\hfill
\@publabel \hfill
\llap{\parbox[t]{0.25\textwidth}{\raggedleft\@pubnumber}}}%
\vss}}
\def\@maketitle{\vskip 60pt \begin{center}
 {\LARGE \@title \par}
 \end{center}}
\def\@makeauthor{{%
\def\and{\smallskip {\normalsize \rm and\smallskip }}
\def\And{\medskip {\normalsize \rm and\\}\medskip}
\long\def\address##1{{\def\and{\\and\\}\medskip
				{\small \it \\##1\\}
}}
{\centering
 \vskip 3em
 \large \lineskip .75em
 \@author}
 \par}} 
\def\@makedate{\vskip 1.5em 
 {\raggedright \small \noindent\@pubdate \par}}
\def\@makeabstract{\vskip 1.5em
{\small 
\begin{center}
{\bf ABSTRACT\vspace{-.5em}\vspace{0pt}} 
\end{center}
\quotation \@abstract \endquotation}}
\def\maketitle{\titlepage
\let\footnotesize\small \setcounter{page}{1}
\def\thefootnote{\arabic{footnote}}
\@makepub
\vfil
\@maketitle
\@makeauthor
\vfil
\@makeabstract
\@thanks
\vfil
\@makedate
\if@restonecol\twocolumn \else \eject \fi
\titlerelax \titleclean
\def\thefootnote{\alph{footnote}}
\setcounter{footnote}{0}
}

\ifcase\@ptsize
 \font\tenmsa=msam10
 \font\sevenmsa=msam7
 \font\fivemsa=msam5
 \font\tenmsb=msbm10
 \font\sevenmsb=msbm7
 \font\fivemsb=msbm5
\or
 \font\tenmsa=msam10 scaled \magstephalf
 \font\sevenmsa=msam8
 \font\fivemsa=msam6
 \font\tenmsb=msbm10 scaled \magstephalf
 \font\sevenmsb=msbm8
 \font\fivemsb=msbm6
\or
 \font\tenmsa=msam10 scaled \magstep1
 \font\sevenmsa=msam8
 \font\fivemsa=msam6
 \font\tenmsb=msbm10 scaled \magstep1
 \font\sevenmsb=msbm8
 \font\fivemsb=msbm6
\fi

\newfam\msafam
\newfam\msbfam
\textfont\msafam=\tenmsa  \scriptfont\msafam=\sevenmsa
  \scriptscriptfont\msafam=\fivemsa
\textfont\msbfam=\tenmsb  \scriptfont\msbfam=\sevenmsb
  \scriptscriptfont\msbfam=\fivemsb

\def\hexnumber@#1{\ifnum#1<10 \number#1\else
 \ifnum#1=10 A\else\ifnum#1=11 B\else\ifnum#1=12 C\else
 \ifnum#1=13 D\else\ifnum#1=14 E\else\ifnum#1=15 F\fi\fi\fi\fi\fi\fi\fi}

\def\msa@{\hexnumber@\msafam}
\def\msb@{\hexnumber@\msbfam}
\mathchardef\boxdot="2\msa@00
\mathchardef\boxplus="2\msa@01
\mathchardef\boxtimes="2\msa@02
\mathchardef\square="0\msa@03
\mathchardef\blacksquare="0\msa@04
\mathchardef\centerdot="2\msa@05
\mathchardef\lozenge="0\msa@06
\mathchardef\blacklozenge="0\msa@07
\mathchardef\circlearrowright="3\msa@08
\mathchardef\circlearrowleft="3\msa@09
\mathchardef\rightleftharpoons="3\msa@0A
\mathchardef\leftrightharpoons="3\msa@0B
\mathchardef\boxminus="2\msa@0C
\mathchardef\Vdash="3\msa@0D
\mathchardef\Vvdash="3\msa@0E
\mathchardef\vDash="3\msa@0F
\mathchardef\twoheadrightarrow="3\msa@10
\mathchardef\twoheadleftarrow="3\msa@11
\mathchardef\leftleftarrows="3\msa@12
\mathchardef\rightrightarrows="3\msa@13
\mathchardef\upuparrows="3\msa@14
\mathchardef\downdownarrows="3\msa@15
\mathchardef\upharpoonright="3\msa@16

\mathchardef\downharpoonright="3\msa@17
\mathchardef\upharpoonleft="3\msa@18
\mathchardef\downharpoonleft="3\msa@19
\mathchardef\rightarrowtail="3\msa@1A
\mathchardef\leftarrowtail="3\msa@1B
\mathchardef\leftrightarrows="3\msa@1C
\mathchardef\rightleftarrows="3\msa@1D
\mathchardef\Lsh="3\msa@1E
\mathchardef\Rsh="3\msa@1F
\mathchardef\rightsquigarrow="3\msa@20
\mathchardef\leftrightsquigarrow="3\msa@21
\mathchardef\looparrowleft="3\msa@22
\mathchardef\looparrowright="3\msa@23
\mathchardef\circeq="3\msa@24
\mathchardef\succsim="3\msa@25
\mathchardef\gtrsim="3\msa@26
\mathchardef\gtrapprox="3\msa@27
\mathchardef\multimap="3\msa@28
\mathchardef\therefore="3\msa@29
\mathchardef\because="3\msa@2A
\mathchardef\doteqdot="3\msa@2B

\mathchardef\triangleq="3\msa@2C
\mathchardef\precsim="3\msa@2D
\mathchardef\lesssim="3\msa@2E
\mathchardef\lessapprox="3\msa@2F
\mathchardef\eqslantless="3\msa@30
\mathchardef\eqslantgtr="3\msa@31
\mathchardef\curlyeqprec="3\msa@32
\mathchardef\curlyeqsucc="3\msa@33
\mathchardef\preccurlyeq="3\msa@34
\mathchardef\leqq="3\msa@35
\mathchardef\leqslant="3\msa@36
\mathchardef\lessgtr="3\msa@37
\mathchardef\backprime="0\msa@38
\mathchardef\risingdotseq="3\msa@3A
\mathchardef\fallingdotseq="3\msa@3B
\mathchardef\succcurlyeq="3\msa@3C
\mathchardef\geqq="3\msa@3D
\mathchardef\geqslant="3\msa@3E
\mathchardef\gtrless="3\msa@3F
\mathchardef\sqsubset="3\msa@40
\mathchardef\sqsupset="3\msa@41
\mathchardef\vartriangleright="3\msa@42
\mathchardef\vartriangleleft="3\msa@43
\mathchardef\trianglerighteq="3\msa@44
\mathchardef\trianglelefteq="3\msa@45
\mathchardef\bigstar="0\msa@46
\mathchardef\between="3\msa@47
\mathchardef\blacktriangledown="0\msa@48
\mathchardef\blacktriangleright="3\msa@49
\mathchardef\blacktriangleleft="3\msa@4A
\mathchardef\vartriangle="3\msa@4D
\mathchardef\blacktriangle="0\msa@4E
\mathchardef\triangledown="0\msa@4F
\mathchardef\eqcirc="3\msa@50
\mathchardef\lesseqgtr="3\msa@51
\mathchardef\gtreqless="3\msa@52
\mathchardef\lesseqqgtr="3\msa@53
\mathchardef\gtreqqless="3\msa@54
\mathchardef\Rrightarrow="3\msa@56
\mathchardef\Lleftarrow="3\msa@57
\mathchardef\veebar="2\msa@59
\mathchardef\barwedge="2\msa@5A
\mathchardef\doublebarwedge="2\msa@5B
\mathchardef\angle="0\msa@5C
\mathchardef\measuredangle="0\msa@5D
\mathchardef\sphericalangle="0\msa@5E
\mathchardef\varpropto="3\msa@5F
\mathchardef\smallsmile="3\msa@60
\mathchardef\smallfrown="3\msa@61
\mathchardef\Subset="3\msa@62
\mathchardef\Supset="3\msa@63
\mathchardef\Cup="2\msa@64

\mathchardef\Cap="2\msa@65

\mathchardef\curlywedge="2\msa@66
\mathchardef\curlyvee="2\msa@67
\mathchardef\leftthreetimes="2\msa@68
\mathchardef\rightthreetimes="2\msa@69
\mathchardef\subseteqq="3\msa@6A
\mathchardef\supseteqq="3\msa@6B
\mathchardef\bumpeq="3\msa@6C
\mathchardef\Bumpeq="3\msa@6D
\mathchardef\lll="3\msa@6E

\mathchardef\ggg="3\msa@6F

\mathchardef\circledS="0\msa@73
\mathchardef\pitchfork="3\msa@74
\mathchardef\dotplus="2\msa@75
\mathchardef\backsim="3\msa@76
\mathchardef\backsimeq="3\msa@77
\mathchardef\complement="0\msa@7B
\mathchardef\intercal="2\msa@7C
\mathchardef\circledcirc="2\msa@7D
\mathchardef\circledast="2\msa@7E
\mathchardef\circleddash="2\msa@7F
\def\ulcorner{\delimiter"4\msa@70\msa@70 }
\def\urcorner{\delimiter"5\msa@71\msa@71 }
\def\llcorner{\delimiter"4\msa@78\msa@78 }
\def\lrcorner{\delimiter"5\msa@79\msa@79 }
\def\yen{\mathhexbox\msa@55 }
\def\checkmark{\mathhexbox\msa@58 }
\def\circledR{\mathhexbox\msa@72 }
\def\maltese{\mathhexbox\msa@7A }
\mathchardef\lvertneqq="3\msb@00
\mathchardef\gvertneqq="3\msb@01
\mathchardef\nleq="3\msb@02
\mathchardef\ngeq="3\msb@03
\mathchardef\nless="3\msb@04
\mathchardef\ngtr="3\msb@05
\mathchardef\nprec="3\msb@06
\mathchardef\nsucc="3\msb@07
\mathchardef\lneqq="3\msb@08
\mathchardef\gneqq="3\msb@09
\mathchardef\nleqslant="3\msb@0A
\mathchardef\ngeqslant="3\msb@0B
\mathchardef\lneq="3\msb@0C
\mathchardef\gneq="3\msb@0D
\mathchardef\npreceq="3\msb@0E
\mathchardef\nsucceq="3\msb@0F
\mathchardef\precnsim="3\msb@10
\mathchardef\succnsim="3\msb@11
\mathchardef\lnsim="3\msb@12
\mathchardef\gnsim="3\msb@13
\mathchardef\nleqq="3\msb@14
\mathchardef\ngeqq="3\msb@15
\mathchardef\precneqq="3\msb@16
\mathchardef\succneqq="3\msb@17
\mathchardef\precnapprox="3\msb@18
\mathchardef\succnapprox="3\msb@19
\mathchardef\lnapprox="3\msb@1A
\mathchardef\gnapprox="3\msb@1B
\mathchardef\nsim="3\msb@1C
\mathchardef\napprox="3\msb@1D
\mathchardef\varsubsetneq="3\msb@20
\mathchardef\varsupsetneq="3\msb@21
\mathchardef\nsubseteqq="3\msb@22
\mathchardef\nsupseteqq="3\msb@23
\mathchardef\subsetneqq="3\msb@24
\mathchardef\supsetneqq="3\msb@25
\mathchardef\varsubsetneqq="3\msb@26
\mathchardef\varsupsetneqq="3\msb@27
\mathchardef\subsetneq="3\msb@28
\mathchardef\supsetneq="3\msb@29
\mathchardef\nsubseteq="3\msb@2A
\mathchardef\nsupseteq="3\msb@2B
\mathchardef\nparallel="3\msb@2C
\mathchardef\nmid="3\msb@2D
\mathchardef\nshortmid="3\msb@2E
\mathchardef\nshortparallel="3\msb@2F
\mathchardef\nvdash="3\msb@30
\mathchardef\nVdash="3\msb@31
\mathchardef\nvDash="3\msb@32
\mathchardef\nVDash="3\msb@33
\mathchardef\ntrianglerighteq="3\msb@34
\mathchardef\ntrianglelefteq="3\msb@35
\mathchardef\ntriangleleft="3\msb@36
\mathchardef\ntriangleright="3\msb@37
\mathchardef\nleftarrow="3\msb@38
\mathchardef\nrightarrow="3\msb@39
\mathchardef\nLeftarrow="3\msb@3A
\mathchardef\nRightarrow="3\msb@3B
\mathchardef\nLeftrightarrow="3\msb@3C
\mathchardef\nleftrightarrow="3\msb@3D
\mathchardef\divideontimes="2\msb@3E
\mathchardef\varnothing="0\msb@3F
\mathchardef\nexists="0\msb@40
\mathchardef\mho="0\msb@66
\mathchardef\thorn="0\msb@67
\mathchardef\beth="0\msb@69
\mathchardef\gimel="0\msb@6A
\mathchardef\daleth="0\msb@6B
\mathchardef\lessdot="3\msb@6C
\mathchardef\gtrdot="3\msb@6D
\mathchardef\ltimes="2\msb@6E
\mathchardef\rtimes="2\msb@6F
\mathchardef\shortmid="3\msb@70
\mathchardef\shortparallel="3\msb@71
\mathchardef\smallsetminus="2\msb@72
\mathchardef\thicksim="3\msb@73
\mathchardef\thickapprox="3\msb@74
\mathchardef\approxeq="3\msb@75
\mathchardef\succapprox="3\msb@76
\mathchardef\precapprox="3\msb@77
\mathchardef\curvearrowleft="3\msb@78
\mathchardef\curvearrowright="3\msb@79
\mathchardef\digamma="0\msb@7A
\mathchardef\varkappa="0\msb@7B
\mathchardef\hslash="0\msb@7D
\mathchardef\hbar="0\msb@7E
\mathchardef\backepsilon="3\msb@7F
\def\Bbb{\ifmmode\let\next\Bbb@\else
 \def\next{\errmessage{Use \string\Bbb\space only in math mode}}\fi\next}
\def\Bbb@#1{{\Bbb@@{#1}}}
\def\Bbb@@#1{\fam\msbfam#1}

\def\bk {{\hskip 0.2 cm}}

\def\com{{\hskip 0.2 cm},}
\def\pkt{{\hskip 0.2 cm}.}
\def\acknowledgements{\@startsection{section}{4}
{\z@}{-3.5ex plus -1ex minus -.2ex}{2.3ex plus .2ex}{\normalsize\bf}
{Acknowledgements}}

\def\half {\frac{1}{2}}           
\newcommand{\case}[1]{\makebox(0,0)[b]{
\footnotesize#1}}                 
\newcommand{\tab}[1]{{\sc Tab.}\,{\sf #1}}        
\newcommand{\eq}[1]{{\sc Eq.}\,{\sf (#1)}}        
\newcommand{\eqs}[1]{{\sc Eqs.}\,{\sf (#1)}}      
\newcommand{\refoth}[1]{{\sf #1}}                 
\newcommand{\eqoth}[1]{{\sf (#1)}}                

\def\bbbz {\Bbb{Z}}               

\def\bbbn {\Bbb{N}}               
\def\bbbno{\Bbb{N}_{0}}           
\def\bbbc {\Bbb{C}}
\def\bbbq {\Bbb{Q}}

\newtheorem{definition}{Definition}[section]
\newtheorem{theorem}[definition]{Theorem}
\newtheorem{conjecture}[definition]{Conjecture}
\newtheorem{lemma}[definition]{Lemma}
\newtheorem{proposition}[definition]{Proposition}
\newcounter{defs}[section]

\newcommand{\be}{\begin{equation}}
\newcommand{\ee}{\end{equation}}
\newcommand{\bea}{\begin{eqnarray}}
\newcommand{\eea}{\end{eqnarray}}

\newcommand{\bdf}{\stepcounter{defs}\begin{definition}}
\newcommand{\edf}{\end{definition}}
\newcommand{\bth}{\stepcounter{defs}\begin{theorem}}
\newcommand{\eth}{\end{theorem}}
\newcommand{\bcj}{\stepcounter{defs}\begin{conjecture}}
\newcommand{\ecj}{\end{conjecture}}
\newcommand{\blm}{\stepcounter{defs}\begin{lemma}}
\newcommand{\elm}{\end{lemma}}
\newcommand{\bpr}{\stepcounter{defs}\begin{proposition}}
\newcommand{\epr}{\end{proposition}}
\newcommand{\bprf}{Proof: }
\newcommand{\eprf}{\hfill $\Box$ \\}
\newcounter{pics}
\newcommand{\bpic}[4]{\begin{center}\begin{picture}(#1,#2)(#3,#4)
\refstepcounter{pics}}
\renewcommand{\thepics}{{\sf\roman{pics}}}
\newcommand{\epic}[1]{\end{picture}\\
{\small {\sc Fig.} \thepics \bk #1} \end{center}}
\newcommand{\epicspl}{\end{picture}\\           
\addtocounter{pics}{-1}\end{center}}            

\renewcommand{\thefootnote}{\rm{\alph{footnote}}}
\newcounter{tabs}
\newcommand{\btab}[1]{\refstepcounter{tabs}\begin{center}
\begin{tabular}{#1}}
\renewcommand{\thetabs}{{\sf\alph{tabs}}}
\newcommand{\etab}[1]{\end{tabular}\\[1.5ex]
{\small {\sc Tab.} \thetabs \bk #1} \end{center}}

\def\noi {\noindent}
\newcommand{\nn}{\nonumber}

\newcommand{\ket}[1]{\left| {#1} \right\rangle} 
\newcommand{\bra}[1]{\left\langle {#1} \right|} 
\newcommand{\spn}[1]{{\rm span}\{{#1}\}}        
\def\det{{\rm det}}                             
\newcommand{\vm}[1]{{\langle #1 \rangle}}       

\def\pmb#1{\setbox0=\hbox{#1}%
 \kern-.025em\copy0\kern-\wd0
 \kern.05em\copy0\kern-\wd0
 \kern-.025em\raise.0433em\box0 }

\def\cL{{\cal L}}

\makeatother

\def\ram{{\sf R}_1}      
\def\uram{{\sf R}^{\times}_1}      
\def\vm{{\cal V}}        
\def\ordering{{\cal O}}  
\def\bsvm{{\cal B}}      
\def\cset{{\cal C}}      
\def\sset{{\cal S}}      
\def\lset{{\sf L}}      
\def\tset{{\sf T}}      
\def\gset{{\sf G}}      
\def\ph{{\hat{p}}}      
\def\qh{{\hat{q}}}      
\def\pt{{\tilde{p}}}      
\def\qt{{\tilde{q}}}      
\def\cs{{\sf III}}      

\newcommand{\levelt}[1]{{{|}#1{|}_{L}}} 
\newcommand{\charget}[1]{{{|}#1{|}_{F}}}        
\newcommand{\length}[1]{{{\|}#1{\|}}}   
\newcommand{\osm}[1]{{{<}_{{}_{#1}}}}           
\newcommand{\nontrivial}[2]{{{\Omega}_{#1}^{#2}}}           

\title{Highest weight representations of the $N=1$ Ramond algebra}
\author{Matthias D\"{o}rrzapf\thanks{m.doerrzapf@damtp.cam.ac.uk}
\address{
Department of Applied Mathematics and Theoretical
Physics\\
University of Cambridge, Silver Street \\
Cambridge, CB3 9EW, UK
}}
\pubdate{May 1999}
\pubnumber{DAMTP-99-28 \\
hep-th/9905150
}

\abstract{We analyse the highest weight representations of the $N=1$
Ramond algebra and show that their structure is richer 
than previously suggested in the literature.
In particular, 
we show that certain Verma modules over the $N=1$ Ramond algebra contain
degenerate (2-dimensional) singular vector spaces and that in the 
supersymmetric case they can even contain subsingular vectors. 
After choosing a suitable ordering for the $N=1$
Ramond algebra generators we compute the ordering kernel, which turns
out to be two-dimensional for complete Verma modules and
one-dimensional for $G$-closed Verma modules. These two-dimensional
ordering kernels allow us to derive multiplication rules for singular
vector operators and lead to expressions for degenerate singular
vectors. Using these multiplication rules we study descendant singular vectors
and derive the Ramond embedding diagrams
for the rational models.
We give all explicit examples for singular vectors, degenerate
singular vectors, and subsingular vectors until level $3$.
We conjecture the ordering kernel coefficients of all (primitive) 
singular vectors and therefore identify these vectors uniquely.}

\begin{document}

\maketitle



\section{Introduction}

The extension of string theory to supersymmetric string theories
identified the $N=1$ Ramond algebra and the $N=1$ Neveu-Schwarz algebra as
the symmetry algebras of closed superstrings as first introduced by  
Ademollo et al.\cite{ademollo}. Therefore, the Ramond\footnote{For simplicity
we shall call the $N=1$ Ramond algebra simply the {\it Ramond algebra} and similarly for
the $N=1$ Neveu-Schwarz algebra.} algebra and the Neveu-Schwarz
algebra are the most important symmetry algebras with respect to superstring theory.
Even though this significance of the Ramond algebra is known since more than
a decade, the representations of the Ramond algebra are still very poorly 
understood
and results given in the literature on its representation theory are largely incomplete
as we shall show in this paper.  

Due to the conformal invariance on the
string world-sheet only superconformal extensions of the Virasoro
algebra are suitable symmetry algebras of superstrings. In the case of
one fermionic current one obtains only two possible super
extensions of the Virasoro algebra\cite{kac3}: 
the $N=1$ Ramond algebra and the
$N=1$ Neveu-Schwarz algebra. Whilst the latter corresponds to
antiperiodic boundary conditions of the fields living on the 
world-sheet of the closed superstrings, the Ramond
algebra corresponds to periodic boundary conditions.
The space of states of a superstring theory will therefore decompose in
irreducible highest weight representations of the Ramond algebra and
Neveu-Schwarz algebra. Certain aspects of the highest weight 
representations of both
algebras have been studied by many authors in the past. 
But so far only in the Neveu-Schwarz case one has achieved all the desirable
results about the structure of its representations\cite{ast1}.
The Kac-determinant formulae play a crucial r\^ole for the analysis of
these representations. 
In the Ramond case it has first been conjectured
by Friedan, Qiu, and Shenker\cite{friedan} whilst in the Neveu-Schwarz case
the formula has first been given by Kac\cite{kac7}. 
Both formulae have been proven
by Meurman and Rocha-Caridi\cite{meur}.

One can
construct all irreducible representations from Verma modules by
considering quotients of the Verma modules divided by their largest proper
submodules. Submodules of Verma modules are spanned by null-vectors
which themselves are generated by singular and subsingular
vectors. In this context, singular vectors are the vectors with lowest conformal
weight of such submodules. 
Once the Kac-determinant formulae are known, one can obtain
information on null-vectors and singular vectors by analysing the determinant's
roots.  The largest proper submodule and thus all null-vectors
are given by the kernel of the inner product matrix. Hence, 
null-vectors are orthogonal on the whole space of states and
therefore decouple from the string theory. In this sense they lead us to
differential equations for the $n$-point functions and therefore
describe the dynamics of the string theory. 
In the Virasoro case\cite{ff1} and in the Neveu-Schwarz case\cite{ast1}
subsingular vectors do not exist and therefore their highest weight
representations can solely be analysed using singular vectors. 
Embedding diagrams of singular vectors
have been conjectured for the
Ramond algebra only for the unitary cases by assuming the non-existence of subsingular
vectors\cite{kiritsis2}. However, for the Ramond algebra this property has never
been proven. Furthermore, it was assumed that singular vectors
with the same weight would always be linearly dependent, a fact that is
true for the Virasoro case and has been proven to be true for the
Neveu-Schwarz case\cite{ast1}. We will show that both assumptions fail for
the Ramond algebra. The $N=1$ Ramond algebra contains both degenerate
singular vectors (linearly independent singular vectors with the same weights)
as well as subsingular vectors. The latter occur only in the supersymmetric
case\footnote{The cases with ground states of conformal weight $\Delta=\frac{c}{24}$
are called supersymmetric since the global supersymmetry is unbroken\cite{friedan}.}
$\Delta=\frac{c}{24}$ whilst the former appear even for some of the minimal
models, however not for the unitary cases. We therefore show that
the general Ramond embedding diagrams do not follow the simple
rules as for the Virasoro case and consequently the connexion between the Verma
modules and the irreducible characters contains more complications.

Motivated by similar findings for the $N=2$ superconformal algebras
in Ref. \icite{p6sdim1} a method has been developed
to derive upper limits for the degree of
degeneracy of singular vectors. Using this method one easily
finds for the case of the 
Virasoro algebra and also for the $N=1$ Neveu-Schwarz algebra
that both algebras do not allow any degeneracy at 
all\cite{adrian1,p6sdim1}. 
However for the
$N=2$ algebras degenerate singular vector spaces have been found
for all $4$ types of
$N=2$ superconformal algebras\cite{p6sdim1,p9sdim2}.
The main result of this so-called {\it adapted ordering procedure} is the
ordering kernel. The number of elements of the ordering kernel sets
an upper limit on the linearly independent singular vectors with the
same weight.
Surprisingly enough, we will find
that the $N=1$ Ramond algebra also has an ordering
kernel of size $2$ and therefore also allows degenerate
levels for which we will give explicit examples. Degenerate levels
complicate very much the analysis of the embedding structure of submodules
- so-called embedding diagrams - since the
simple comparison of levels is not enough to decide whether or not two
singular vectors are proportional. This information, however, is crucial for
the corresponding character formulae.
The ordering kernel helps even in this matter as
has been shown in Ref. \icite{p6sdim1} that the elements of the ordering kernel are
sufficient to uniquely identify a singular vector. By comparing their coefficients
with respect to the ordering kernel it is hence easy to
identify linearly independent singular vectors. Furthermore, we can obtain
product expressions for singular vector operators that easily reveal
the vanishing of a descendant singular vector.
In Ref. \icite{p11vanish} it will be shown that the vanishing product of two
singular vector operators may lead to subsingular vectors which therefore
can also be found using the ordering kernel. 

The $N=1$ Ramond algebra is a subalgebra of the $N=2$ twisted
superconformal algebra. Its highest weight representations
inherit features that are
very similar to those of the latter one. The $N=2$ twisted
superconformal highest weight representations have been studied
in Ref. \icite{p9sdim2}.
Considering these strong similarities
of their representation theories it
is rather puzzling that the $N=1$ Ramond algebra is so extremely important for
string theory whilst the $N=2$ twisted superconformal algebra has so far
obtained very little attention in string theory.
Despite these similarities,
the representation theories also have
a few important differences which will become
clear in section \refoth{\ref{sec:classification}}. 

The multiplication rules
derived in the present paper give the necessary basis for the analysis of
descendant singular vectors and consequently the
derivation of
the Ramond embedding diagrams.
We discuss the most interesting embedding diagrams including the
minimal models and indicate the limitations of the embedding diagrams
of the unitary cases by Kiritsis\cite{kiritsis2}.
We point out the main differences of Ramond embeddings to Virasoro embeddings.
A full set of Ramond embedding diagrams will be given in Ref. \icite{progress}.

Some necessary introductory remarks about the Ramond algebra
will be given in section \refoth{\ref{sec:alg}}. In section
\refoth{\ref{sec:ordering}}, we first review the main ideas and implications 
of the adapted ordering method introduced
in Ref. \icite{p6sdim1}. We then define an adapted ordering for the Ramond algebra
and compute its ordering kernel. In section 
\refoth{\ref{sec:sdim}}, we thus deduce the singular dimensions and
derive multiplication rules for vectors identified by their ordering kernel coefficients. 
In section \refoth{\ref{sec:classification}}, we compute necessary conditions on the
ordering kernel coefficients of (primitive) singular vectors, which
allows
 us to conjecture
these coefficients for all (primitive) singular vectors.
We hence compute all cases of degenerate singular vectors. 
Using the multiplication rules of section \refoth{\ref{sec:sdim}} we
give in section \refoth{\ref{sec:desc}}
a first analysis of the most interesting embedding diagrams,
the rational models. A complete 
discussion of the Ramond embedding diagrams is beyond the scope
of this work and shall be given in a future 
publication\cite{progress}.   
$G$-closed
Verma modules 
and $G$-closed singular vectors 
are considered in section
\refoth{\ref{sec:GclosedVM}} and \refoth{\ref{sec:Gclosedsvecs}} 
respectively. $G$-closed Verma modules also lead to the simplest type of
subsingular vectors. In section \refoth{\ref{sec:GclosedVM}} we will also
show that there are no further subsingular vectors that are related to the
vanishing of singular vector operator products.
Examples of all singular vectors, degenerate singular vector spaces, and subsingular vectors 
for levels $\leq3$ will be given in the appendix.
We conclude the 
paper with some final remarks
in section \refoth{\ref{sec:conclusions}}.


\section{The $N=1$ Ramond algebra}
\label{sec:alg}

Periodic boundary conditions for the fields on the world-sheet of
a closed string require the
fermionic current to have modes with integer scaling dimensions. Therefore,
the symmetry algebra is the super-extension of the Virasoro algebra by integer moded
odd generators $G_m$.

\bdf 
The $N=1$ Ramond algebra consists of the Virasoro
algebra generators\footnote{We use the notation $\bbbn=\{1,2,3,\ldots\}$, 
$\bbbno=\{0,1,2,\ldots\}$, and 
$\bbbz=\{\ldots,-1,0,1,2,\ldots\}$.}
$L_{m}$, $m\in\bbbz$, and the fermionic current generators $G_n$, 
$n\in\bbbz$ with the commutation relations\footnote{The $N=1$ Ramond algebra satisfies the same
commutation relations as the $N=1$ Neveu-Schwarz. The only
difference is that for the latter algebra the fermionic current has
modes with half-integer indices.}
\bea
[L_{m},L_{n}] & = & (m-n) L_{m+n} + \frac{C}{12} \:(m^{3}-m)\:
\delta_{m+n,0} \;\; ,\nn \\
\ [L_{m},G_{n}] & = & (\frac{m}{2}-n) G_{m+n} \;\; , \label{eq:cr} \\
\ \{G_{m},G_{n}\} &=& 2 L_{m+n} + \frac{C}{3} (m^2-\frac{1}{4})
\delta_{m+n,0} \;\; , \nn \\
\ [L_{m},C] & = &  [G_{n},C]  \;\; = \;\; 0 \;\; , \nn
\eea
for $m,n\in\bbbz$.
\edf

Let us note that the Ramond algebra is a subalgebra of the twisted $N=2$
superconformal algebra. Comparing with the results obtained in
Ref. \icite{p9sdim2} we
shall see that this fact leads to strong
similarities between the Ramond highest weight representations and
the highest weight representations of the twisted $N=2$ superconformal
algebra. 
The squares of the fermionic operators
can be expressed in terms of Virasoro operators, as one can easily 
obtain from the commutation relations \eqs{\ref{eq:cr}}:
\bea
G_m^2 &=& L_{2m} - \delta_{m,0}\frac{C}{24} \;\;\;\com m\in\bbbz 
\pkt \label{eq:square} 
\eea

For applications in physics one usually extends the Ramond algebra
by the {\it (fermion) parity operator}\footnote{Note 
that due to \eq{\ref{eq:square}}
the fermion number $F$ is not well defined, however, the {\it 
parity} $(-1)^F$ is.} $(-1)^F$. It commutes
with the operators $L_m$ and $C$ but anticommutes with $G_n$
\bea
[(-1)^F,L_m] =[(-1)^F,C] &=& 0 \com m\in\bbbz \,, \nn \\
\{(-1)^F,G_n\} &=& 0 \com n\in\bbbz \,. \label{eq:f}
\eea
The operator $(-1)^F$ therefore distinguishes fermionic states from 
bosonic states in the space of states and is present in most
applications in physics. We will therefore in the following always consider the
extended Ramond algebra. Later in this section we will show that this does
not at all restrict our results to the extended algebra but they can easily be 
transferred to the unextended algebra. We hence give the following definition.
\bdf \label{def:extsalg}
The Ramond algebra \eqs{\ref{eq:cr}} extended by the fermion parity
operator $(-1)^F$ of \eq{\ref{eq:f}} shall simply be called the Ramond algebra $\ram$. 
We will explicitly say the unextended Ramond algebra, denoted by
$\uram$ whenever we want to refer to 
the algebra of  \eqs{\ref{eq:cr}} only.
\edf

The central term $C$ commutes with all other operators and can therefore
be fixed as $c\in\bbbc$. ${\cal H}_{\ram}=\spn{L_0, (-1)^F, C}$ 
defines a Cartan subalgebra
of $\ram$ which elements can hence be 
diagonalised simultaneously. 
Generators with positive
index span the set of {\it positive operators} $\ram^+$ of $\ram$
and likewise generators with negative index 
span the set of {\it negative operators} $\ram^-$ of $\ram$:
\bea
\ram^{+} &=& \spn{L_{m},G_{m}: m\in\bbbn} \,, \\
\ram^{-} &=& \spn{L_{-m},G_{-m}: m\in\bbbn} \,.
\eea 
\noi The {\it zero modes} are spanned by $\ram^0=\spn{L_0,G_0,(-1)^F,C}$ which contain
besides ${\cal H}_{\ram}$ also the operator $G_0$. 

For eigenvectors of ${\cal H}_{\ram}$ we denote the 
$L_0$-eigenvalue by $\Delta$
(the conformal weight), the $C$-eigenvalue by $c$, and
the $(-1)^F$ eigenvalue simply by $\pm$ (the parity) for $\pm 1$. The
$C$-eigenvalue $c$ (the conformal anomaly) 
shall for simplicity be suppressed in the notation. 
Highest weight vectors, Verma modules and singular vectors are defined
in the usual way\footnote{There is the usual historical confusion:
what physicists call highest weight vector is in fact a vector of lowest
weight in the Verma module.} .

\bdf
An eigenvector $\ket{\Delta}$ of 
${\cal H}_{\ram}$ with vanishing $\ram^{+}$ action is called 
a highest weight vector. By convention we define a highest weight vector $\ket{\Delta}$
to have positive parity.
Additional zero-mode vanishing conditions\footnote{c.f. Ref. \icite{p6sdim1}, definition \refoth{3.A}.} 
are possible with respect to the operator $G_0$.
We write $\ket{\Delta}^G$ if $G_0\ket{\Delta}^G=0$ and call it a $G$-closed highest weight vector.
For $\ket{\Delta}^G$ we necessarily have $\Delta=\frac{c}{24}$.
\edf

\bdf
For a given highest weight vector $\ket{\Delta}$ the Verma module $\vm_{\Delta}$
is the left-module
\bea
\label{eq:vms}
\vm_{\Delta} &=& U(\ram)\otimes_{{\cal H}_{\ram}\oplus\ram^+}\ket{\Delta} \,,
\eea
where $U(\ram)$ denotes the universal enveloping algebra of $\ram$.
For $G$-closed highest weight vectors $\ket{\frac{c}{24}}^G$ we define
$G$-closed Verma modules as
\bea
\vm_{\frac{c}{24}}^G &=& U(\ram)\otimes_{{\cal H}_{\ram}\oplus\ram^+\oplus
\spn{G_0}}\ket{\frac{c}{24}}^G \,.
\eea
\edf

For an eigenvector $\Psi^p_{l}$ of ${\cal H}_{\ram}$  
in $\vm_{\Delta}$ the conformal weight is $\Delta+l$
and the parity is $p$ with $l\in\bbbno$ and $p\in\{\pm\}$ which
we denote as {\it level} $\levelt{\Psi^p_{l}}=l$ and {\it parity} $\charget{\Psi^p_{l}}=p$. 
We can hence define the notion of singular vectors which - as we shall
explain later - correspond to
vectors of lowest conformal weight in a given subrepresentation
of a Verma module and conversely every proper subrepresentation
of a Verma module needs to contain at least one singular vector.

\bdf
An eigenvector $\Psi^{\pm}_{l}$ of ${\cal H}_{\ram}$  
in the Verma module $\vm_{\Delta}$ is called singular vector
if it is not proportional to $\ket{\Delta}$ but is still
annihilated by $\ram^+$. $G$-closed singular vectors are
denoted as $\Psi^{\pm G}_{l}$ and satisfy necessarily $\Delta+l=\frac{c}{24}$. 
For $\Psi^{\pm}_l$ singular in $\vm_{\Delta}$ there exists a unique operator 
$\theta_{l}^{\pm}\in U(\ram^-\oplus\{G_0\})$ such that  
$\Psi^{\pm}_l=\theta^{\pm}_l\ket{\Delta}$ called the singular vector operator.
Similarly for $\Psi^{\pm G}_l$ where $\theta^{\pm G}_l\in U(\ram^-)$.
If $\theta^{\pm}_l\ket{\Delta}$ can be written as 
$\theta_1\theta_2\ket{\Delta}$ with two singular vector operators
$\theta_1$ and $\theta_2$, $\theta_1$ not proportional to $G_0$,
then  $\Psi^{\pm}_l$ is called a secondary singular
vector or a descendant singular vector of $\theta_2\ket{\Delta}$. Otherwise it is
called a primitive singular vector. 
\edf

In the definition of secondary singular vectors we had to exclude operators $\theta_1$
that are multiples of $G_0$ simply because, for most singular vectors,
$G_0$ interpolates between two singular vectors at the same level but of different parity. The
only exceptions appear for levels $l$ with $\Delta+l=\frac{c}{24}$ for which the action of
$G_0^2$ vanishes and therefore $G_0$ cannot interpolate between two states. 
Surely, if $G_0$ interpolates
between a pair of singular vectors then both vectors are either secondary or
primitive.

Before we continue, we shall now compare Verma modules over the 
(extended) Ramond algebra $\ram$ with Verma modules over the unextended
Ramond algebra. For the unextended algebra $\uram$ one chooses 
the fermionic generator
$G_0$ in the Cartan subalgebra ${\cal H}_{\uram}$
together with $L_0$ and $C$. 
We hence define highest weight vectors of the unextended algebra
with respect to their $G_0$ eigenvalue.
\bdf \label{def:uhw}
An (unextended) highest weight vector $\ket{\lambda}^{\times}$
of the unextended algebra $\uram$ is an eigenvector of
$G_0$ that satisfies
\bea
G_0 \ket{\lambda}^{\times} &=& \lambda\ket{\lambda}^{\times} \com \\
L_0 \ket{\lambda}^{\times}  &=& (\lambda^2+\frac{c}{24}) \ket{\lambda}^{\times} \com \\
\ram^+ \ket{\lambda}^{\times} &=& 0 \pkt
\eea
An (unextended) Verma module $\vm_{\lambda}^{\times}$ is defined by
\bea
\vm_{\lambda}^{\times} &=& U(\ram)
\otimes_{{\cal H}_{\uram}\oplus\ram^+}\ket{\lambda}^{\times} \,,
\eea
\edf
Except for the {\it supersymmetric case}\cite{cohn2} $\Delta=\frac{c}{24}$,
a Verma module $\vm_{\Delta}$ is always reducible to
two Verma modules $\vm^{\times}_{\pm\lambda}$ of the unextended Ramond
algebra\cite{cohn2,gerard}, with $\pm\lambda=\pm \sqrt{\Delta-\frac{c}{24}}$.
We should note that for Lie superalgebras {\it Schur's lemma} does not hold
any more in the sense that the Cartan subalgebra can always be diagonalised 
on irreducible representations\cite{kac3}. Instead we can obtain grade spaces
for which only the square of the negative
parity operator $G_0$ is diagonal but not necessarily
$G_0$ itself. But except for the supersymmetric case $\Delta=\frac{c}{24}$
we can always build a $\ram$ Verma module on an eigenvector
$\ket{\lambda}^{\times}$ of $G_0$. Suppose that $\ket{\lambda}^{\times}$ was not
an eigenvector of $G_0$ but of $L_0$ with eigenvalue $\Delta=\lambda^{2}+\frac{c}{24}$,
then the combinations $G_0\ket{\lambda}^{\times}\pm\lambda\ket{\lambda}^{\times}$ 
are two eigenvectors of $G_0$ with eigenvalues $\pm\lambda$ that generate together
the module built on $\ket{\lambda}^{\times}$ and form themselves two $\uram$ 
Verma modules
built on highest weight vectors $\ket{\pm\lambda}^{\times}$ as we shall see.
Therefore, the above definition \refoth{\ref{def:uhw}} of (unextended) highest weight vectors is justified.
The fact that odd generators of the Cartan subalgebra may not be diagonal
in an irreducible representation is another reason why for applications in physics 
we may want to choose a Lie algebra as Cartan subalgebra rather than 
a Lie superalgebra. Therefore the extended algebra $\ram$ is a more
natural choice. Nevertheless, let us show here that we can easily relate
extended and unextended Verma modules and in most cases the unextended
Verma modules embedded in an extended Verma module diagonalise $G_0$. 
 
If $\Delta\not = \frac{c}{24}$ it is easy to see that the two vectors
$\ket{\pm\lambda}^{\times}=G_0\ket{\Delta}\pm\sqrt{\Delta-\frac{c}{24}}\ket{\Delta}$
are eigenvectors of $G_0$ with eigenvalues $\pm\lambda=\pm
\sqrt{\Delta-\frac{c}{24}}$. These are the highest weight vectors of the
two unextended Verma modules $\vm_{\pm\lambda}^{\times}$ embedded
in $\vm_{\Delta}$. The sum of $\vm_{+\lambda}^{\times}$
and $\vm_{-\lambda}^{\times}$ is direct as we shall show in the following theorem.
This is essentially because the operator $(-1)^F$ - which is not contained in the
unextended algebra - interpolates between the two vectors.
\bth
The Verma module $\vm_{\Delta}$ contains for $\Delta\not = \frac{c}{24}$ the
two (unextended) highest weight vectors $\ket{\pm\lambda}^{\times}$ with
$\Delta=\lambda^2+\frac{c}{24}$ (i.e. $\lambda\not =0$). The sum 
\bea
\vm_{\Delta} &=& \vm_{\lambda}^{\times} \oplus \vm_{-\lambda}^{\times}
\eea
is direct and we have $(-1)^F$ interpolating between the two highest weight
vectors:
$(-1)^F\ket{\pm\lambda}^{\times}=\ket{\mp\lambda}^{\times}$.
\eth
\bprf
The vectors $\ket{\pm\lambda}^{\times}$ are given by 
$G_0\ket{\Delta}\pm\sqrt{\Delta-\frac{c}{24}}\ket{\Delta}$ as previously
discussed. Assume $\Psi\in\vm_{\lambda}^{\times}$ and 
$\Psi\in\vm_{-\lambda}^{\times}$, we hence find operators $\theta^+,\theta^-,
\gamma^+,\gamma^-\in U(\ram^-)$ such that 
\bea
\Psi &=& (\theta^++\gamma^+G_0)\ket{\lambda}^{\times}
=(\theta^-+\gamma^-G_0)\ket{-\lambda}^{\times} \pkt
\eea
We can write this in the standard basis for $\vm_{\Delta}$ and 
easily obtain $(\theta^+-\theta^-+\lambda\gamma^++\lambda\gamma^-)
G_0\ket{\Delta}+
\lambda(\theta^++\theta^-+\lambda\gamma^+-\lambda\gamma^-)
\ket{\Delta}=0$. Taking into account that in the non-supersymmetric
case $\ket{\Delta}$ and $G_0\ket{\Delta}$ do not satisfy any vanishing
conditions in $\vm_{\Delta}$ we therefore find that the operators acting on 
these two basis vectors must vanish identically. However, this results
in $\theta^+\lambda\gamma^+=0$ and hence 
$\Psi=(\theta^+\lambda\gamma^+)\ket{\lambda}^{\times}=0$.
Therefore the sum is direct. Furthermore, the highest weight vector $\ket{\Delta}
=\frac{\ket{\lambda}-\ket{-\lambda}}{2\lambda}$ is contained in the sum
$\vm_{\lambda}^{\times} \oplus \vm_{-\lambda}^{\times}$ and therefore 
$\vm_{\Delta}\subset\vm_{\lambda}^{\times} \oplus \vm_{-\lambda}^{\times}$.
By construction the vectors $\ket{\pm\lambda}$ are contained in
$\vm_{\Delta}$ and thus also 
$\vm_{\lambda}^{\times} \oplus \vm_{-\lambda}^{\times}\subset\vm_{\Delta}$.
Finally, one easily checks that $(-1)^F\ket{\pm\lambda}^{\times}$ has
$G_0$ eigenvalue $\mp\lambda$.
\eprf

Let us now try to diagonalise $G_0$ on the $L_0$-grade spaces of
$\vm_{\lambda}^{\times}$. We define
\bea
\lambda_{l} &=& \sqrt{\Delta+l-\frac{c}{24}} \pkt
\eea
We assume that $\lambda_{l}\not =0$ and take any operator $\theta_{l}\in
U(\ram^-\cup\{G_0\})$. It is easy to see that the vectors
\bea
\Psi_{\theta_l,\lambda}^{\pm\lambda_l} &=& (G_0\theta_l \pm \lambda_l \theta_l) \ket{\lambda} \com
\label{eq:xfmn}
\eea
are - if  non-trivial - eigenvectors of $G_0$ with eigenvalues $\pm\lambda_l$.
Again, we have $(-1)^F$ interpolating between the vectors
$\Psi_{\theta_l,\lambda}^{+\lambda_l}$ and $\Psi_{\theta_l,\lambda}^{-\lambda_l}$
 for $\lambda\not =0$.

An $L_0$-grade space in $\vm_{\lambda}^{\times}$
 with $l$ such that $\lambda_l\not = 0$ has therefore also a diagonal
$G_0$ action. With respect to the Verma module $\vm_{\Delta}$
this change of basis involves the 4 vectors 
$\theta_l\ket{\Delta}$, $G_0\theta_l\ket{\Delta}$,  
$\theta_lG_0\ket{\Delta}$, and $G_0\theta_lG_0\ket{\Delta}$ transforming
them into the 4 
$G_0$-eigenvectors $\Psi_{\theta_l,\pm\lambda}^{\pm\lambda_l}$.
Surely, in the cases
$\Delta=\frac{c}{24}$ or $\Delta+l=\frac{c}{24}$ two of these vectors would
obviously be trivial and the basis transformation breaks down. 
The case where some of these 4 vectors are proportional plays a
very important r\^ole for our later considerations. If we assume that
$\theta_l G_0\ket{\Delta} = \kappa G_0\theta_l \ket{\Delta}$ then we
easily obtain $G_0\theta_l G_0\ket{\Delta} = \kappa \lambda_l^2
\theta_l \ket{\Delta}$. If the first relation is true on $\ket{\Delta}$, then
it is also true on $G_0\ket{\Delta}$ and thus
$\lambda^2\theta_l \ket{\Delta} = \kappa G_0\theta_l G_0 \ket{\Delta}$.
Comparing the two last conditions finally results in $\lambda_l^2\kappa^2=
\lambda^2$. Hence, only two proportionality factors are allowed
by grouping the vectors in two proportional pairs. In this case, however,
the set of 4 vectors  $\theta_l\ket{\Delta}$, $G_0\theta_l\ket{\Delta}$,  
$\theta_lG_0\ket{\Delta}$, and $G_0\theta_lG_0\ket{\Delta}$ 
defines only a 2-dimensional vector space whose basis
transforms to
\bea
\Psi_{\theta_l,\lambda}^{\pm\lambda_l} &=&
\pm(\kappa\lambda_l\pm\lambda) (G_0\theta_l\pm\lambda_l\theta_l)\ket{\Delta} \com \\
\Psi_{\theta_l,-\lambda}^{\pm\lambda_l} &=&
\pm(\kappa\lambda_l\mp\lambda) (G_0\theta_l\pm\lambda_l\theta_l)\ket{\Delta} \pkt
\eea
Hence, in either case of the two possible proportionality factors $\kappa$
one of the vectors $\Psi_{\theta_l,\lambda}^{\pm\lambda_l}$ would be trivial and
also one of the vectors
$\Psi_{\theta_l,-\lambda}^{\pm\lambda_l}$ would vanish. Therefore, the two-dimensional
space defined by $\theta_l\ket{\Delta}$, $G_0\theta_l\ket{\Delta}$,  
$\theta_lG_0\ket{\Delta}$, and $G_0\theta_lG_0\ket{\Delta}$ 
would be transformed in a one-dimensional $G_0$-eigenspace in $\vm_{\lambda}^{\times}$
and a one dimensional $G_0$-eigenspace in  $\vm_{-\lambda}^{\times}$, one of them
having $G_0$-eigenvalue $\lambda_l$ and $-\lambda_l$ for the other one.
In the case that $\lambda_l=0$ we find that half of the operators $\theta_l$
cannot be diagonalised with respect to $G_0$.
The corresponding
vectors $G_0\theta_l\ket{\lambda}^{\times}$ are hence eigenvectors of $G_0$ with
eigenvalue $0$. The vectors generated by the other half of operators cannot be
transformed into a basis of $G_0$-eigenvectors and only the square of $G_0$ is diagonal
on them. 
\bth
The operator $G_0$ can be diagonalised on the $L_0$-grade spaces of 
$\vm_{\lambda}^{\times}$ 
if $\lambda_l \not = 0$. The eigenvalues are $\pm\lambda_l$ which both appear
except for the case of proportionality among pairs of the vectors
$G_0\theta_l\ket{\Delta}$ and  
$\theta_lG_0\ket{\Delta}$ 
for some $\theta_l\in 
U(\ram^-\cup\{G_0\})$. In this case the eigenvalues $\pm\lambda_l$ will be shared
among the Verma modules $\vm_{\pm\lambda}$.
\noi
If $\lambda_l=0$ then $G_0$ cannot be diagonalised on the corresponding $L_0$-grade space of $\vm_{\pm\lambda}^{\times}$.   
\eth

The fact that $G_0$ cannot be diagonalised appears in the
supersymmetric case already at level $0$. For
$\Delta=\frac{c}{24}$ the Verma module $\vm_{\frac{c}{24}}$
contains one unextended Verma module
generated by $\ket{0}^{\times}=G_0\ket{\frac{c}{24}}$. $G_0$ can easily
be diagonalised on the whole Verma module $\vm_{0}^{\times}$ as explained above as
$\lambda_l\not =0$, ($l>0$). However,
$\vm_{\frac{c}{24}}$ consists of vectors outside $\vm_{0}^{\times}$,
$\ket{\frac{c}{24}}$ being one example.
\bth \label{th:supersym}
The Verma module $\vm_{\frac{c}{24}}$ contains an (unextended)
Verma module $\vm_{0}^{\times}$. The vectors in $\vm_{\frac{c}{24}}$ 
which lie outside $\vm_{0}^{\times}$ do not belong to any (unextended) 
Verma module built on an (unextended) highest weight vector (i.e. 
$G_0$-eigenvector). 
\eth

Conversely, if we are now given a Verma module of the unextended Ramond algebra
with (only one) highest weight vector $\ket{\lambda}^{\times}$ with $G_0$ eigenvalue
$\lambda\not = 0$ and hence $L_0$-eigenvalue $\Delta=\lambda^2+\frac{c}{24}$
we can easily obtain $\vm_{\Delta}$ by considering $\ket{\lambda}^{\times}$ together
with its partner $\ket{-\lambda}^{\times}=(-1)^F\ket{\lambda}^{\times}$ 
with $G_0$-eigenvalue $-\lambda$ and imposing
$\ket{\Delta}= \frac{\ket{\lambda}^{\times}-\ket{-\lambda}^{\times}}{2\lambda}$. 
This automatically implies
$G_0\ket{\Delta}= \frac{\ket{\lambda}^{\times}+\ket{-\lambda}^{\times}}{2}$.
In the case of $\lambda=0$ we simply have to add $\chi$ with $G_0\chi=
\ket{0}^{\times}$ in order to obtain\footnote{Note that $\vm_{\frac{c}{24}}^{\times}$ is also
a $\ram$ module, which is not the case for $\vm_{\Delta}^{\times}$ with $\Delta\not =\frac{c}{24}$.} 
$\vm_{\frac{c}{24}}$ from
$\vm_{0}^{\times}$.
The notion of singular vectors is the same in extended and unextended
Verma modules and furthermore $G_0$ applied to a singular vector
produces another singular vector if non-trivial. Therefore, if a highest weight
representation of the extended $\ram$ algebra is irreducible (i.e. it has no
singular vectors), then the
two subrepresentations of the unextended algebra are obviously also both
irreducible. 
Conversely, if we have an irreducible representation of
the unextended algebra built on $\ket{\lambda}^{\times}$, $\lambda\not =0$, 
then the Verma module of $\ram$ built on $\ket{\Delta}$ constructed
from $\ket{\lambda}^{\times}$ and $\ket{-\lambda}^{\times}$ is also irreducible.
Since it is rather simple to
transform highest weight representations of the extended
Ramond algebra to highest weight representations of the 
unextended algebra and conversely, it is justified to restrict
ourselves to the extended Ramond algebra $\ram$ which turns out
to be more suitable for our considerations as the Cartan subalgebra
is just a Lie algebra.
Furthermore, applications in physics usually require to distinguish
bosonic from fermionic states and therefore the extended
algebra is more important for physics.

In the Verma module one usually defines an inner product by defining
\bea
L^+_{-m}=L_{m} \com & G^+_{-m}=G_{m} & \forall m\in\bbbz \,,
\eea 
and obvious linear hermitian extensions to the universal enveloping algebra.
One sets $\bra{\Delta}L_0=\Delta\bra{\Delta}$ and $\bra{\Delta}\ram^-\equiv 0$.
This consequently defines an inner product of the vectors $X\ket{\Delta}$ and
$Y\ket{\Delta}$, $X,Y\in U(\ram)$ via
\bea
\bra{\Delta}X^+Y\ket{\Delta} \com \label{eq:ip}
\eea
where we define $\bra{\Delta}G_0\ket{\Delta}=0$ since $(-1)^F$ is diagonal
on the two states. We obtain a {\it (pseudo-)norm}
of the vector $X\ket{\Delta}$, $X\in U(\ram)$ as $\bra{\Delta}X^+X\ket{\Delta}$.
A singular vector and the whole embedded Verma module built on it
have obviously vanishing (pseudo-)norms. Furthermore, every vector
contained in the largest proper submodule of $\vm_{\Delta}$ has vanishing
(pseudo-)norm\cite{p11vanish}. These vectors form the kernel of the inner
product matrix formed via \eq{\ref{eq:ip}} as it is shown in Ref. \icite{p11vanish}.
They therefore decouple from the
field theory. They are usually called 
{\it null-vectors}\footnote{If the pseudo-norm is 
positive semi-definite, then all vectors with vanishing norm are null-vectors.
However, if the pseudo-norm is 
not positive semi-definite 
there are vectors with vanishing pseudo-norm that are not 
null-vectors\cite{p11vanish} and therefore do not decouple from the field theory.}.
The quotient module where
all null-vectors are set to $0$ is hence an irreducible highest weight 
representation.
Conversely, all irreducible highest weight representations can be 
constructed in this way. 
Singular vectors and their descendants do not necessarily span the
whole submodule of null-vectors. The quotient module of the Verma 
module divided by
the submodule spanned by all singular vectors may again contain new singular
vectors, called subsingular vectors. 
\bdf
Let $H_{\Delta}$ be the submodule generated by all singular vectors of a Verma
module $\vm_{\Delta}$. The quotient module 
\bea
Q_{\Delta} &=& \frac{\vm_{\Delta}}{H_{\Delta}}
\eea
may or may not be irreducible. In the case that $Q_{\Delta}$ is reducible it contains 
new singular vectors $\Upsilon$ that were not singular in $\vm_{\Delta}$. $\Upsilon$ is
called subsingular vector in $\vm_{\Delta}$. Continuing with this process one may 
find subsingular vectors in $Q_{\Delta}$ and further quotient modules until one 
obtains an irreducible representation. 
\edf

For the Virasoro algebra and for the $N=1$ Neveu-Schwarz algebra there are no subsingular
vectors\cite{ff1,ast1}, however, they have been discovered 
for the $N=2$ superconformal algebras by Gato-Rivera et al.
in Refs. \icite{beatriz2,beatriz1,p9sdim2}. 
At first, it seems that $N=1$ superconformal symmetry is not large enough to
allow subsingular vectors. However, that this assumption does not hold for
one particular case of the extended Ramond algebra will be shown in section
\refoth{\ref{sec:GclosedVM}} of this paper.
The fact that the largest proper submodule is equivalent to the kernel of the
inner product matrix assigns an important r\^ole to the determinant of the
inner product matrix, usually called the {\it Kac-determinant}.
If we choose an ${\cal H}_{\ram}$ graded basis, then the inner product
matrix is block diagonal and the Kac-determinant factorises in determinants
of these blocks. The lowest grade for which the determinant vanishes
indicates the existence of a singular vector.
For the $N=1$ Ramond algebra the determinant
formula has been first conjectured by Friedan, Qiu,
Shenker\cite{friedan} 
and has later been
proven by Meurman and Rocha-Caridi\cite{meur}. 
For
level zero the determinant formulae are $\det{M}^+_0=1$ and 
$\det{M}^-_0= \Delta-\frac{c}{24}$. 
For the central term $c$ at
level $l\in\bbbn$ and parity $\pm$
we use the parametrisation\cite{gerard,friedan} 
in $t\in\bbbc$, ($t\not=0$),
\bea
c(t) &=& \frac{15}{2}-\frac{3}{t} -3t \pkt \label{eq:cparam}
\eea
We can then give the determinant formulae as
\bea
\det{M}^{\pm}_l(\Delta,t) &=& (\Delta-\frac{c(t)}{24})^{\frac{P_R(l)}{2}} 
\prod_{1\leq pq\leq 2l \atop p,q\in\bbbn , \, p-q \, {\rm odd}}
[\Delta-\Delta_{p,q}(t)]^{P_R(l-\frac{pq}{2})} \pkt \label{detform}
\eea
The functions $\Delta_{p,q}$ can hence be written as
\bea
\Delta_{p,q} &=& \frac{(q-tp)^2-(t-1)^2}{8t} +\frac{1}{16} \;\; = \;\; \frac{(q-tp)^2}{8t} +\frac{c}{24} \com
\eea
where $p$ and $q$ are positive integers and $p-q$ is odd. 
Finally, the partition function 
$P_R(l)$ is given by 
\bea
\sum_{i=0} x^i P_R(i) &=& \prod_{k=1} \frac{1+x^k}{1-x^k} \com
\eea
and satisfies $P_R(0)=1$.
Let us note that $P_R(l)$ is always even\cite{cohn2} for $l\geq 1$. 
For most $c$ the parametrisation \eq{\ref{eq:cparam}} has two
solutions $t^1$ and $t^2$. In which case we obtain $t^1=1/t^2$. We
note that $\Delta_{p,q}(t)=\Delta_{q,p}(1/t)$ and $c(t)=c(1/t)$ and therefore 
the set of curves $\Delta_{p,q}(t)$ is invariant under the choice of solution $t^1$ and
$t^2$. Hence the parametrisation is well-defined. In fact, the Verma modules
$\vm_{\Delta_{p,q}}(t)$ and $\vm_{\Delta_{q,p}}(1/t)$ are identical.
From the determinant we know that following the curves
$\Delta_{p,q}(t)$ we obtain singular vectors which turn out to be
generically primitive.
\bth
Along the curve $\Delta_{p,q}(t)$ we find (at least) two singular vectors $\Psi^{\pm}_{p,q}(t)$
both at level $\frac{pq}{2}$ and with parity $\pm$, where $p,q\in\bbbn$ and
$p-q$ is odd. Both vectors are
generically primitive.
\eth
\bprf
For a given $\Delta_{p,q}(t)$ which does not intersect with any other curve
$\Delta_{p^\prime,q^\prime}(t)$ where $\frac{p^\prime
q^\prime}{2}\leq\frac{pq}{2}$ the claim is trivial.
Intersections of such curves are only given by discrete points
$t=\pm\frac{q-q^\prime}{p-p^\prime}$. Continuity arguments easily show
that $\Psi_{p,q}^{\pm}$ exists also for these discrete points (if it
happens to vanish identically, then at least one derivative would be
singular) and
these are the only points for which $\Psi_{p,q}^{\pm}$ could be
secondary. Hence $\Psi_{p,q}^{\pm}$ is generically primitive.
\eprf

$G_0$ interpolates between the 
two parity sectors, provided
$G_0$ does not act on any $G$-closed vector. Therefore, the two sectors
have exactly the same determinant expressions. 
Starting with a given singular vector 
$\Psi^+_{l}=\theta^+_l\ket{\Delta}\in\vm_{\Delta}$ with singular operator 
$\theta^+_l$ we can construct two negative parity singular vectors:
\bea
\Psi_{G,l}^- &=& \theta^+_lG_0\ket{\Delta} \com \label{eq:psiG1} \\ 
{}_G\Psi_{l}^- &=& G_0\theta^+_l\ket{\Delta} \pkt
\eea
Likewise, $G_0$ constructs two positive parity singular vectors from the 
singular vector
$\Psi^-_{l}=\theta^-_l\ket{\Delta}\in\vm_{\Delta}$:
\bea
\Psi_{G,l}^+ &=& \theta^-_lG_0\ket{\Delta} \com \\
{}_G\Psi_{l}^+ &=& G_0\theta^-_l\ket{\Delta} \pkt \label{eq:psiG4}
\eea
This seems as if the (primitive) vectors $\Psi_{p,q}^{\pm}$ always come at least
in pairs of $4$ unless the corresponding vectors of
\eqs{\ref{eq:psiG1}}-\eqoth{\ref{eq:psiG4}}
are proportional. It is the Ramond partition function that requires
the latter.
\bth \label{th:psiprop}
For the singular vectors $\Psi_{p,q}^{\pm}$ the vectors 
$\Psi_{G,p,q}^{+}$ and ${}_G\Psi_{p,q}^{+}$ are
proportional\footnote{In the case of $G$-closed
Verma modules or $G$-closed singular
vectors some of these vectors may be trivial.}
as well as $\Psi_{G,p,q}^{-}$ and ${}_G\Psi_{p,q}^{-}$.
\eth
\bprf
The entries of the inner product matrix are polynomials in $\Delta$
and $c$. It is easy to show\cite{p11vanish} that if a determinant with
polynomial entries vanishes for $\Delta=\Delta(t)$, then the multiplicity of
the factor $(\Delta-\Delta(t))$ in the determinant sets an upper limit
for the dimension of the kernel of the inner product
matrix. Generically, the multiplicity of $\Delta_{p,q}(t)$ at level
$\frac{pq}{2}$ is $P_R(0)=1$  for each parity sectors. As
$\Psi_{p,q}^{\pm}$ has to be in the kernel, there will generically be exactly one
singular vector for each parity at level $\frac{pq}{2}$
along the curve $\Delta_{p,q}(t)$ and hence the claimed proportionalities
have to be true generically. Due to continuity this extends all along
the curves $\Delta_{p,q}(t)$. Note that the fact that there is just one
singular vector along $\Delta_{p,q}(t)$ at level $\frac{pq}{2}$ for each
parity must be true generically but may not be true for the discrete
intersection points of the curves $\Delta_{p,q}(t)$.
\eprf

We conclude this section by introducing convenient notation that will
be used in the following section.
\bdf
For $Y\in U(\ram)$ of the form
\bea 
Y&=&L_{-m_L} \ldots L_{-m_1} 
G_{-n_G} \ldots G_{-n_1}L_{-1}^n G_{-1}^{r_1}
G_{0}^{r_2} , \label{eq:ybasis}
\eea
\noi where $r_1,r_2\in\{0,1\}$ and for all its reorderings
we define 
the level $\levelt{Y}=\sum_{j=1}^{L}m_j
+\sum_{j=1}^{G}n_j+n+r_1$, 
the parity $\charget{Y}=(-1)^{G+r_1+r_2}$, and
their {\it length} $\length{Y}=L+G$. For the trivial case we put
$\levelt{1}=\length{1}=0$ and $\charget{1}=+1$.
Further, we set ($l\in\bbbn$, $k\in\bbbno$)
\bea
\lset_l &=& \{ Y=L_{-m_L} \ldots L_{-m_1}: \; m_L\geq\ldots\geq m_1\geq 2,
\; \levelt{Y}=l\} \,, \\
\gset_l &=& \{ Y=G_{-n_G} \ldots G_{-n_1}: \; n_G > \ldots > n_1\geq 2,
\; \levelt{Y}=l\} \,, \\
\lset_0 &=& \tset_0 ~=~ \gset_0 ~=~ \{1\} \,,
\eea
\bea
\sset^{\pm}_{k} &=& 
\left\{Y=LG: \; L\in\lset_m, \; G\in\gset_n, 
\right. \nn
\\ 
&& \left. \levelt{Y}=k=m+n, \;\charget{Y}=\pm 1=\charget{G} ,\; 
\,m,\in\bbbno ,\, n\in\bbbno\right\} \,,
\eea
\bea
\cset^{\pm}_{k} 
&=& \left\{S_{m,p} \, L_{-1}^{k-m-r_1}
G_{-1}^{r_1}G_{0}^{r_2}: \;S_{m,p}\in\sset^p_{m},
\;m\in\bbbno,\;r_1,r_2\in\{0,1\}, \right. \nn \\
&& \left. k-m-r_1\geq 0, \, p(-1)^{r_1+r_2}=\pm 1 \right\} \,.  
\label{eq:cset} 
\eea
\edf
\noi 
Elements of $\cset^{\pm}_{k}$ are therefore of the form of \eq{\ref{eq:ybasis}} with
$r_1,r_2\in\{0,1\}$, $\levelt{Y}=k$, and $\charget{Y}=\pm 1$.
We will always use $\cset^{\pm}_{k}$ in order to define the
basis of a Verma module $\vm_{\Delta}$:
\bdf
The standard basis for $\vm_{\Delta}$ is 
\bea
{\cal B}_{\Delta} &=& \left\{ X\ket{\Delta}: 
\; X\in\cset^{\pm}_{k} \;, k\in\bbbno \; \right\} \,,
\label{eq:basis1}
\eea
\noi 
which is naturally
$\bbbno\otimes\{\pm 1\}$ graded with respect to their $L_0$ and $(-1)^F$
eigenvalues relative to the highest weight vector. This grading is inherited
from the grading on $U(\ram)$.
The decomposition of $\Psi_l^p\in\vm_{\Delta}$ with respect to the
standard basis
\bea
\Psi^p_{l} &=& \sum_{X\in\cset^p_{l}} c_X X\ket{\Delta} \com
\eea 
is called the normal form of $\Psi_l^p$. 
$X\in\cset^p_{l}$ are the terms of 
$\Psi^p_{l}$ and the coefficients $c_X$ its coefficients. 
Terms $X$ with non-trivial coefficients
$c_X$ are called non-trivial terms of $\Psi^p_{l}$.
\edf

In the case of 
$G$-closed Verma modules the basis is obviously smaller and will be
defined similarly through
\bea
{\cal B}^G_{\frac{c}{24}} &=& \left\{ X\ket{\frac{c}{24}}: 
\; X\in\cset^{\pm G}_{k} \;, k\in\bbbno, \; \right\} \,,
\label{eq:basisG}
\eea
\noi with
\bea
\cset^{\pm G}_{k} 
&=& \left\{S_{m,p} \, L_{-1}^{k-m-r_1}
G_{-1}^{r_1}: \;S_{m,p}\in\sset^p_{m},
\;m\in\bbbno,\;r_1\in\{0,1\}, \right. \nn \\
&& \left. k-m-r_1\geq 0, \, p(-1)^{r_1}=\pm 1 \right\} \,, \;\;\; k\in\bbbno \,.
\label{eq:csetG} 
\eea
Whilst Verma modules $\vm_{\Delta}$ may contain both, singular vectors
and $G$-closed singular vectors, $G$-closed Verma modules
$\vm^G_{\frac{c}{24}}$ can only contain singular vectors but not
$G$-closed singular vectors. Similar phenomena are true for the
$N=2$ algebras as shown in Refs. \icite{beatriz2,p9sdim2}. To be
precise, for the topological and Neveu-Schwarz $N=2$ algebras
there are no chiral singular vectors in chiral Verma modules whereas
for the twisted $N=2$ algebra there are no $G$-closed singular vectors
in $G$-closed Verma modules. 


\section{Ordering kernels for Ramond Verma modules}
\label{sec:ordering}

Verma modules of the Virasoro algebra contain at most one singular
vector at a given level. This is a consequence of the embedding diagrams 
proven by Feigin and Fuchs\cite{ff1}. On the other hand, if one knows
in advance that there can be at most one singular vector for
a given level, then the analysis of descendant singular
vectors comes down to a detailed analysis of the roots
of the determinant formula. Whenever one finds that the level
of two secondary singular vectors agree then these singular
vectors need to be the same and the generated submodules would be
identical. This also allows us to derive the character formulae
of the irreducible highest weight representations. 
However, for Verma modules where such a {\it uniqueness}
of singular vectors at a given level is not true the procedure of deriving
the embedding patterns is much more complicated. Consequently 
the derivation of the characters via embedding diagrams
also needs more information in this case.
This has been demonstrated in Ref. \icite{npb1} for the minimal models
of the $N=2$ Neveu-Schwarz algebra. To solve such problems there 
are hence two questions of major interest. First, which are the maximal 
dimensions a space of singular vectors can have for a given weight 
- the so-called singular dimensions - and secondly, if the
singular dimension is greater than 1, how can we decide if two
singular vectors at the same weight are proportional? 
Both questions can be answered with a rather simple but powerful
method, the {\it adapted ordering method}, introduced in Ref. \icite{p6sdim1}.
There, this method has first been used for the $3$ isomorphic
$N=2$ superconformal algebras (the topological $N=2$ algebra, the $N=2$ 
Neveu-Schwarz algebra, and the $N=2$ Ramond algebra)
and just recently also for the twisted $N=2$ algebra in Ref. \icite{p9sdim2}.
In the latter case the adapted ordering method
shows strong similarities to what we will find for $\ram$.
{\it A priori} the connexion between the following definition and the answers
to the questions above does not seem very obvious. The power of this
method, however, lies in two theorems which were proven in Ref. \icite{p6sdim1}
and which we will review after introducing the notion of adapted orderings
in the case of the Ramond algebra. For convenience, let us first define the following
notation.
\bdf
For $X\in\cset^p_l$ and $\Gamma\in\ram$ we set
\bea
\nontrivial{X}{\Gamma}(\Delta) &=& \Bigl\{ {\rm non-trivial \: terms \: of\:\:} \Gamma \, X\ket{\Delta} \Bigr\}.
\eea
As the set of all non-trivial terms of $\Gamma \, X\ket{\Delta}$ in its normal form,
$\nontrivial{X}{\Gamma}(\Delta)$ is - if non-trivial - a subset of 
$\cset^{p\charget{\Gamma}}_{l+\levelt{\Gamma}}$.
\edf

In this notation we can give the definition of an adapted ordering in a particularly
suitable form: 
\bdf \label{def:adapt}
Let $\ordering$ be a total ordering on $\cset^p_{l}$. 
${\cal K}=\ram^+$ is the set of annihilation operators.
$\ordering$ is said to be adapted to the
subset $\cset^{pA}_{l}\subset\cset^p_{l}$, 
in $\vm_{\Delta}$
with ${\cal K}=\ram^+$,
if for any $X\in\cset^{pA}_{l}$ at least one 
$\Gamma \in{\cal K}$ exists for which 
\bea
\nontrivial{X}{\Gamma}(\Delta) \not \subset 
\bigcup_{Y\in\cset^p_{l}\atop X \osm{\ordering} Y, \, Y\not = X}
\nontrivial{Y}{\Gamma}(\Delta) \pkt
\eea
The complement of $\cset^{pA}_{l}$, 
${\ }\cset^{pK}_{l}=\cset^p_{l}\setminus
\cset^{pA}_{l}$ is the (ordering) kernel with respect to
the ordering $\ordering$ in the Verma module $\vm_{\Delta}$.
\edf

In other words, for each $X\in\cset^{pA}_{l}$ there exists an annihilation operator
$\Gamma$ such that $\Gamma \, X\ket{\Delta}$ contains at least one non-trivial term
that is not contained in $\Gamma \, Y\ket{\Delta}$ for all $Y$ that are 
strictly $\ordering$-larger than $X$. The significance of this definition lies in the fact that
if we know that  $\Gamma \Psi_l^p\equiv 0$ for $\Psi_l^p\in\vm_{\Delta}$ and we assume that 
$X$ is the $\ordering$-smallest non-trivial term in $\Psi_l^p$, then there will be at least one 
non-trivial term in $\Gamma \Psi_l^p$ which contradicts $\Gamma \Psi_l^p\equiv 0$.
Hence only elements of the ordering kernel $\cset^{pK}_{l}$ can be $\ordering$-smallest 
non-trivial terms of a singular vector $\Psi_l^p$.  These rather simple thoughts
result in the following two theorems that were proven in a more general setting
in Ref. \icite{p6sdim1}.

\bth \label{th:dims}
If the ordering kernel $\cset^{pK}_{l}$ of an adapted ordering at level $l$, parity $p$,
in $\vm_{\Delta}$ and with annihilation
operators ${\cal K}=\ram^+$ has $n$ elements,
then there are at most $n$ linearly independent singular vectors
$\Psi^p_{l}$ in $\vm_{\Delta}$ at level $l$ with parity $p$.
\eth

\bth \label{th:kernel}
Let $\ordering$ denote an adapted ordering
at level $l$, parity $p$, with ordering kernel $\cset^{pK}_{l}$ 
for a given Verma module $\vm_{\Delta}$ and annihilation operators 
${\cal K}=\ram^+$. If the normal form of two vectors $\Psi^{p,1}_{l}$
and $\Psi^{p,2}_{l}$ in $\vm_{\Delta}$ at the same level $l$ and parity
$p$, both satisfying the highest weight conditions, 
have $c_{X}^1=c_{X}^2$ for all terms $X\in\cset^{pK}_{l}$, then
\bea
\Psi^{p,1}_{l} &\equiv & \Psi^{p,2}_{l}\,.
\eea
\eth

Surely, every total ordering on $\cset_l^p$
would be adapted at least to the empty set $\emptyset\subset\cset_l^p$.
This, however, is not at all useful as the adapted kernel would be the whole
set  $\cset_l^p$. The aim is to find orderings that lead to very small 
ordering kernels. Coming back to the two questions set in
the introduction of this section, theorem \refoth{\ref{th:dims}} answers
the first one: the size of the adapted kernel sets an upper limit
for the dimension of a space of singular vectors for a given weight.
Theorem \refoth{\ref{th:kernel}} answers the second question: in order to
decide if two singular vectors with the same weight are proportional we
do not need to compare all their coefficients, in fact we do not even need
to know all their coefficients, it is already enough to compare 
the few coefficients
with respect to the ordering kernel. These coefficients
completely identify a singular vector. This fact
has been used for the $N=2$ algebra in Refs. \icite{p6sdim1,p9sdim2}
and in a simpler version already in Ref. \icite{cmp1} for the $N=2$
Neveu-Schwarz case. As the trivial vector has vanishing coefficients
for all terms it follows that a singular vector needs to have
at least one non-trivial term of the ordering kernel.
In the following we will compute ordering kernels
for the Ramond algebra $\ram$ which will prove to be the smallest 
possible by explicit examples given in the appendix.

We obtain the corresponding definition of adapted orderings for
$G$-closed Verma modules by replacing in definition \refoth{\ref{def:adapt}} 
$\, \cset^p_{l}$ by $\cset^{pG}_{l}$, 
$\cset^{pA}_{l}$ by $\cset_{l}^{pGA}$,  $\vm_{\Delta}$ by 
$\vm_{\frac{c}{24}}^G$, and
$\bsvm_{\frac{c}{24}}$ by $\bsvm_{\frac{c}{24}}^G$.
Theorem \refoth{\ref{th:dims}} and theorem \refoth{\ref{th:kernel}}
hold accordingly.
The same is true for $G$-closed singular vectors for which we simply have to
extend the set of annihilation operators to
${\cal K}=\ram^+\oplus\{G_0\}$. 

The adapted ordering we are now going to introduce coincides formally with
the adapted ordering given in Ref. \icite{p6sdim1} for the
topological $N=2$ superconformal algebra, restricted to the fewer
fermionic and bosonic generators in the $N=1$ Ramond case. Even though
our later results will be similar to the twisted $N=2$ case, the ordering
for the twisted $N=2$ case\cite{p9sdim2} 
is very different in the sense that the
special r\^ole of the translation operator $L_{-1}$ has been taken over by
another $N=2$ bosonic operator.
We first introduce an ordering on the
sets\footnote{Examples of definition \refoth{\ref{def:lorder}} can be
found in Refs. \icite{p6sdim1,p9sdim2}.} $\lset_n$ and $\gset_n$.

\bdf \label{def:lorder}
We take $Y_i\in\lset_{n^i}$ for $i=1,2$ which we call case $\Upsilon=L$ 
or $Y_i\in\gset_{n^i}$ for $i=1,2$ which we denote by case
$\Upsilon=G$. Hence $Y_i$ is of the form
\bea
Y_i &=& Z^i_{-m_{\length{Y_i}}^i}\ldots Z^i_{-m^i_1} \com
\eea
with $\levelt{Y_i}=n^i$ ($n^i\in\bbbno$), or
$Y_i=1$, $i=1,2$, with $Z^i_{-m^i_j}$
being either $L^i_{-m^i_j}$ in case $L$ or $G^i_{-m^i_j}$ in case $G$.
We compute the index\footnote{For subsets of $\bbbn$ we define 
$\min\emptyset=0$.} $j_0=\min\{j:m^1_j-m^2_j\neq 0,
j=1,\ldots,\min(\length{Y_1},\length{Y_2})\}$. If non-trivial,
$j_0$ is the index for which the levels of the generators 
in $Y_1$ and $Y_2$ first disagree when read from the right to the left. 
For $j_0>0$ we then define
\bea
Y_1\osm{\Upsilon}Y_2 & {\rm if} & m^1_{j_0}<m^2_{j_0} \,.
\eea
If, however, $j_0=0$, we set
\bea
Y_1\osm{\Upsilon}Y_2 & {\rm if} & \length{Y_1}>\length{Y_2} \,.
\eea
Throughout this definition, $\Upsilon$ either stands for $L$ or for $G$.
\edf

\noi 
With the help of definition \refoth{\ref{def:lorder}} we now introduce
a suitable adapted ordering for Ramond Verma modules.
It is easy to see that the following
definition really defines a total ordering
on $\cset^p_{l}$
with global minimum $L_{-1}^{l}$ for $\cset^+_{l}$
and
$L_{-1}^{l}G_{0}$ for $\cset^-_{l}$.
$L_{-1}^{l-1}G_{-1}G_0$ and 
$L_{-1}^{l-1}G_{-1}$ are the smallest elements after the minimum
elements respectively.

\bdf \label{def:cord}
On the set $\cset^p_{l}$, $l\in\bbbn$, $p\in\{\pm\}$ we introduce the 
total ordering $\ordering$. 
For two elements $X_1,X_2\in\cset^p_{l}$, $X_1\neq X_2$
with $X_i=L^i G^i L_{-1}^{n^i}G_{-1}^{r^i_1}
G_{0}^{r^i_2}$, $L^i\in\lset_{m_i}$,
$G^i\in\gset_{k_i}$ for
some $m_i,n^i,k^i\in\bbbno$, $\, r_1^i,r_2^i\in\{0,1\}$, 
$i=1,2$, we define
\bea
X_1\osm{\ordering}X_2 & {\rm if} & n^1>n^2 \,. \label{eq:cord1}
\eea
For $n^1=n^2$ we set
\bea
X_1\osm{\ordering}X_2 & {\rm if} & r^1_1>r^2_1 \,.  \label{eq:cord2}
\eea
If $r^1_1=r^2_1$ then we set
\bea
X_1\osm{\ordering}X_2 & {\rm if} & G^1\osm{G} G^2 \,. \label{eq:cord3}
\eea
In the case that $G^1=G^2$ we then define
\bea
X_1\osm{\ordering}X_2 & {\rm if} & L^1\osm{L} L^2 \,. \label{eq:cord4}
\eea
For $X_1=X_2$ we define
$\, X_1\osm{\ordering}X_2$ and $\, X_2\osm{\ordering}X_1$.
\edf 

As explained before, the main result we intend to extract from an
adapted ordering is the ordering kernel as the number of elements of
the ordering kernel defines an upper limit for the singular
dimensions. Secondly, the terms in the ordering kernel identify
singular vectors uniquely according to theorem \refoth{\ref{th:dims}}
and theorem \refoth{\ref{th:kernel}}. We find the following result.

\bth \label{th:adkernel}
The ordering kernels of $\ordering$ on $\cset^p_{l}$
for the Verma module $\vm_{\Delta}$ are given for
$l\in\bbbn$, $p\in\{ \pm\}$ and
for all central terms $c\in\bbbc$ in the following tables.

\btab{|r|l|}
\hline \label{tab:adkern1}
 $(-1)^F$ & ordering kernel \\
\hline
$+$ &  $\{L_{-1}^{l}, L_{-1}^{l-1}G_{-1}G_0\}$ \\
\hline
$-$ &   $\{L_{-1}^{l}G_0, L_{-1}^{l-1}G_{-1}\}$ \\
\hline
\etab{Ordering kernels for $\ordering$,
annihilation operators $\ram^+$.}

\btab{|r|l|}
\hline \label{tab:adkern2}
 $(-1)^F$ & ordering kernel \\
\hline
$+$ &  $\{L_{-1}^{l}\}$ \\
\hline
$-$ &   $\{L_{-1}^{l}G_0\}$ \\
\hline
\etab{Ordering kernels for $\ordering$,
annihilation operators $\ram^+$ and $G_0$.}
\eth

For the proof of theorem \refoth{\ref{th:adkernel}}
we follow the lines of the proof for the topological $N=2$
algebra given in Ref. \icite{p6sdim1}. 

\bprf
The most general term $X_0$ at level $l$, parity $p$,
in $\cset^p_{l}$, $l\in\bbbn$, $p\in\{\pm\}$ is
\bea
X_0 &=& L^0 G^0 L_{-1}^{n} G_{-1}^{r_1} G_0^{r_2} \; 
\in\cset^{p}_{l} \com \label{eq:x0}
\eea
with 
\bea
L^0 & = & L_{-m_{\length{L^0}}}\ldots L_{-m_1}\in\lset_m \,, \nn \\
G^0 & = & G_{-k_{\length{G^0}}}\ldots G_{-k_1}\in\gset_k \,,
\eea
$m,n,k\in\bbbno$, $r_1,r_2\in\{0,1\}$, such that
$l=m+n+k+r_1$ and $p=(-1)^{\length{G^0}+r_1+r_2}$. 
We then construct
the vector $\Psi^0=X_0\ket{\Delta}$.

In the case $G^0\not =1$ we look at $G_{-k_1}$ which necessarily has
$k_1>1$ and consider
the positive operator $G_{k_1-1}$. The commutation relations of
$G_{-k_1}$ and $G_{k_1-1}$ create a generator $L_{-1}$. 
Therefore,  
$\nontrivial{X_0}{G_{k_1-1}}$
contains the term
\bea
X^G &=& L^0 \tilde{G}^0 L_{-1}^{n+1} G_{-1}^{r_1}G_0^{r_2} \com \nn \\
\tilde{G}^0 &=& G_{-k_{\length{G^0}}}\ldots G_{-k_2} \com
\eea
or $\tilde{G}^0=1$ in the case $\length{G^0}=1$. 
For any other term 
$Y\in\cset^{p}_{l}$ which also contains $X^G$ 
in $\nontrivial{Y}{G_{k_1-1}}$
the action of
$G_{k_1-1}$ also creates at least one generator $L_{-1}$ in order
to obtain $X^G$ or $Y$ has more generators $L_{-1}$ than $X_0$.
In the latter case $Y\osm{\ordering}X_0$.
In the first case, however,
the action of one positive operator can according to the commutation
relations create at most
one new generator. We can therefore restrict ourselves to terms 
$Y$ of the form
\bea
Y &=& L^Y G^Y L_{-1}^{n} G_{-1}^{r^Y_1} G_0^{r^Y_2} \; 
\in\cset^{p}_{l} \com
\eea
with the same number $n$ of generators $L_{-1}$ than $X_0$. 
$G_{k_1-1}$ commuting through operators of $L^Y$ can only create
operators of the form $G_{-k}$ with $k\leq k_1$. Hence, this
additional operator $L_{-1}$ can only be created by commutation of
$G_k$ with operators in $G^Y$ or with $G_{-1}^{r_1}$.
In the first case the operator with smallest (absolute) index in $G^Y$
must have an index $k_1^Y\leq k$ and consequently $Y\osm{\ordering}X_0$
(note that in this case $r_i^Y=r_i$). In the latter case, however, we
necessarily obtain $r_1^Y=1>r_1$ and again $Y\osm{\ordering}X_0$. Hence, we
have shown that
$X^G\in\nontrivial{Y}{G_{k_1-1}}$ implies $Y\osm{\ordering}X_0$ and therefore
\bea
X^G \not \in
\bigcup_{Y\in\cset^p_{l}\atop X_0 \osm{\ordering} Y, \, Y\not = X_0}
\nontrivial{Y}{G_{k_1-1}} \com
\eea
but
\bea
X^G\in\nontrivial{X_0}{G_{k_1-1}}\pkt 
\eea

Let us now continue with terms $X_0$ of the form
\bea
X_0 &=& L^0 L_{-1}^{n} G_{-1}^{r_1} G_0^{r_2} \; 
\in\cset^{p}_{l} \pkt \label{eq:x03}
\eea
If $L^0\not =1$ we consider the positive operator $L_{m_1-1}$. 
$\nontrivial{X_0}{L_{m_1}-1}$
contains
\bea
X^L &=& \tilde{L}^0 L_{-1}^{n+1} G_{-1}^{r_1} G_0^{r_2} \com \nn \\
\tilde{L}^0 &=& L_{-m_{\length{L^0}}}\ldots L_{-m_2} \com
\eea
or $\tilde{L}=1$ in the case $\length{L^0}=1$.
For any other term 
$Y\in\cset^{p}_{l}$ which also contains $X^L$ 
in $\nontrivial{Y}{L_{m_1-1}}$
we again find that the action of $L_{m_1-1}$ creates exactly one
additional $L_{-1}$ in order to obtain $X^L$ or we have automatically
$Y\osm{\ordering}X_0$. If the additional $L_{-1}$ is created by commutation
with $G_{-1}^{r^Y_1}$ we would as before find $Y\osm{\ordering} X_0$.
We can hence concentrate on terms $Y$ of the form
\bea
Y &=& L^Y L_{-1}^{n} G_{-1}^{r_1} G_0^{r_2} \; 
\in\cset^{p}_{l} \com
\eea
where also $r^Y_2=r_2$ due to parity equality with $X_0$. 
The only way of
creating $L_{-1}$ from $Y$ under the action of $L_{m_1-1}$ 
is via commutation with
generators in $L^Y$. Therefore $L^Y$ needs to contain
an operator $L_{-m}$ with $m\leq m_1$ and therefore again 
$Y\osm{\ordering} X_0$. Thus
\bea
\nontrivial{X_0}{L_{m_1 -1}} \not \subset
\bigcup_{Y\in\cset^p_{l}\atop X_0 \osm{\ordering} Y, \, Y\not = X_0}
\nontrivial{Y}{L_{m_1-1}} \com
\eea
This completes the proof of the ordering kernels
of table \tab{\ref{tab:adkern1}}.

Finally, if we also have $G_0$ as annihilation operator, we can then 
act with $G_0$ on terms $X_0$ of the form
\bea
X_0 &=& L_{-1}^{n} G_{-1} G_0^{r_2} \; \in\cset^{p}_{l} \com 
\label{eq:x05}
\eea
which creates a non-trivial term
\bea
X^{\prime} &=& L_{-1}^{n+1} G_0^{r_2} \;\;\;\in\cset^{-p}_{l} \pkt 
\eea
Obviously, the only way of creating $L_{-1}$ by commuting 
$G_0$ with products of generators in $\cset_l^p$
is by commuting $G_0$ directly with $G_{-1}$. Therefore any other 
term $Y$ that creates $X^{\prime}$
under the action of $G_0$ and is not $\ordering$-smaller 
than $X_0$ would also need to create one
$L_{-1}$ and thus also have $r^Y_1=r_1=1$, besides
$n^Y=n$. Trivially, we find in this case that $Y=X_0$. 
This proves the ordering kernel of table \tab{\ref{tab:adkern2}} 
which finally completes our proof of theorem \refoth{\ref{th:adkernel}}.
\eprf

Obviously $\ordering$ also defines a total ordering on 
$\cset^{pG}_{l}$.
We can set $r_2\equiv 0$ everywhere in the above proof and obtain the
same reasoning for the $G$-closed Verma modules  
$\vm_{\frac{c}{24}}^G$. Hence, we can already give the ordering
kernels for $G$-closed Verma modules in the following theorem.

\bth \label{th:adkernel2}
The ordering kernels of $\ordering$ on $\cset^{pG}_{l}$
for the Verma module $\vm_{\frac{c}{24}}^G$ are given for
$l\in\bbbn$, $p\in\{ \pm\}$ and
for all central terms $c\in\bbbc$ by:

\btab{|r|l|}
\hline \label{tab:adkern3}
 $(-1)^F$ & ordering kernel \\
\hline
$+$ &  $\{L_{-1}^{l} \}$ \\
\hline
$-$ &   $\{L_{-1}^{l-1}G_{-1}\}$ \\
\hline
\etab{Ordering kernels for $\ordering$ on $\cset^{pG}_{l}$,
annihilation operators $\ram^+$.}

\eth

As mentioned earlier, $G$-closed Verma modules do not contain any 
$G$-closed singular vectors. Therefore,
theorem \refoth{\ref{th:adkernel2}} does not consider orderings
on $\cset^{pG}_{l}$ with annihilation operators $\ram^+$ and $G_0$.
Knowing the ordering kernels, we can now use our powerful theorems 
\refoth{\ref{th:dims}} and \refoth{\ref{th:kernel}} to find upper
limits for singular dimensions and to identify singular vectors. Both
will be demonstrated in the following section.


\section{Singular dimensions and singular vectors in Verma modules
over the $N=1$ Ramond algebra}
\label{sec:sdim}

Theorem \refoth{\ref{th:dims}} tells us that the dimension of spaces
of singular vectors for a given level $l$ and parity $p$ is bounded by
the number of elements of the corresponding ordering kernel found in
the previous section. In appendix \refoth{\ref{app:a}} we give
all the singular vectors until level $3$. In particular, at
level $3$ we give explicit examples of two-dimensional singular vector
spaces in (complete) Verma modules. The upper limits for the
singular dimensions given by the ordering kernels turn out to be
maxima. We can therefore state that the ordering we have chosen is
the best possible since the results we have found cannot be improved
by choosing any other ordering. We easily deduce the following theorem from
\refoth{\ref{th:adkernel}} and \refoth{\ref{th:adkernel2}} which is
surprisingly similar to the corresponding results for the twisted $N=2$
superconformal algebra\cite{p9sdim2}.

\bth
For singular vectors $\Psi^p_{l}$ and $G$-closed singular vectors
$\Psi^{pG}_{l}$ in the Verma module
$\vm_{\Delta}$, or singular vectors $\Psi^p_{l}$ in the
$G$-closed Verma module $\vm_{\frac{c}{24}}^G$, one finds the
following maximal singular dimensions
for singular vectors at the same level $l\in\bbbno$ and with the same 
parity $p\in\{\pm\}$.
\btab{|l|c|c|}
\hline \label{tab:dim1}
 & $p=+$ & $p=-$ \\
\hline 
$\Psi^p_{l}\in\vm_{\Delta}$ & 
$2$ & $2$ \\
\hline 
$\Psi^p_{l}\in\vm_{\frac{c}{24}}^G$ & 
$1$ & $1$ \\
\hline 
$\Psi^{pG}_{\frac{c}{24}-\Delta}\in\vm_{\Delta}$ & 
$1$ & $1$ \\
\hline 
$\Psi^{pG}_{0}\in\vm_{\frac{c}{24}}^G$ & 
$0$ & $0$ \\
\hline
\etab{Maximal dimensions for singular spaces of the $N=1$ Ramond algebra.}
\eth

Using theorem \refoth{\ref{th:kernel}} every singular vector in its
normal form can be
uniquely identified by its components with respect to elements in the
ordering kernel. All the other coefficients of a singular vector 
with respect to the
standard basis are hence fixed once the coefficients on the ordering
kernel are known. However, it may be extremely hard to actually compute
them especially at very high levels. For the Ramond algebra
the so-called fusion method has been used by Watts\cite{gerard} to
compute certain classes of Ramond singular vectors explicitly. 

\bdf  \label{def:ab}
If $\Psi^{\pm}_{l}=\theta^{\pm}_{l}\ket{\Delta}$ is a 
singular vector\footnote{Note that this definition only makes sense if 
$\Psi^{\pm}_{l}$ is a singular vector. Certainly for most 
pairs $a,b\in\bbbc$ singular vectors do not exist.} in $\vm_{\Delta}$ at level
$l\in\bbbno$ with parity $\{\pm\}$ then the normal form of $\Psi^+_{l}$
is completely determined by the coefficients of $L_{-1}^{l}$ and 
$L_{-1}^{l-1}G_{-1}G_0$. Alternatively
the normal form of $\Psi^-_{l}$
is determined by the coefficients of $L_{-1}^{l}G_0$ and
$L_{-1}^{l-1}G_{-1}$. 
We therefore introduce the following notation
\bea
(a,b)_{l}^+=\theta^+_{l}=aL_{-1}^{l}   + 
bL_{-1}^{l-1}G_{-1}G_0 + \ldots  \com \\
(a,b)_{l}^-=\theta^-_{l}= aL_{-1}^{l}G_0   + 
bL_{-1}^{l-1}G_{-1} + \ldots  \com
\eea
for the singular vector operators $\theta^{\pm}_{l}$ given in 
their normal form, where $a,b\in\bbbc$. 
Similarly, $G$-closed singular vectors 
$\Psi^{\pm G}_{l}=\theta^{\pm G}_{l}\ket{\frac{c}{24}-l}$ are
identified by their $L_{-1}^{l}$ or $L_{-1}^{l}G_{0}$
component for parity $+$ or $-$ respectively. We use the notation
\bea
(a)_{l}^{+G}=\theta^{+G}_{l}=aL_{-1}^{l} + \ldots  \com \\
(a)_{l}^{-G}=\theta^{-G}_{l}= aL_{-1}^{l}G_0   +  \ldots  \pkt
\eea
Finally, singular vectors
$\Psi^{\pm}_{l}=\theta^{\pm}_{l}\ket{\frac{c}{24}}^G$
in $G$-closed Verma modules $\vm_{\frac{c}{24}}^G$ can be identified
in the following way:
\bea
(a)_{l}^{+}=\theta^{+}_{l}=aL_{-1}^{l} + \ldots  \com \\
(a)_{l}^{-}=\theta^{-}_{l}= aL_{-1}^{l-1}G_{-1}   +  \ldots  \pkt
\eea
\edf

In the introduction we mentioned that in the Virasoro case it was
sufficient to compare the level of two singular vectors in the same
Verma module in order to
deduce whether they are proportional. This important feature simplified
the derivation of the embedding structure of submodules\cite{ff1} and
finally the connexion to the Virasoro irreducible characters. Our
examples in the appendix show that in the Ramond case the simple 
comparison of weights is not sufficient in order to decide if
singular vectors are proportional and therefore the derivation of the
embedding structure and character formulae seems almost unmanageable if we
do not know the singular vectors explicitly. 
However, theorem \refoth{\ref{th:kernel}} reduces this problem to the
knowledge of the coefficients with respect to the ordering kernel
only. In
the case of $G$-closed singular vectors or singular vectors in
$G$-closed Verma modules this single coefficient can always be
normalised to $1$ - if the vector is non-trivial - and therefore for these cases we can
almost proceed as in the Virsoro case. For singular vectors that are
not $G$-closed in complete Verma modules we need to find the two
leading coefficients. This will be done in the following section. 
However, in
order to analyse secondary singular vectors we need to find
product formulae for singular vector operators. This can easily be
achieved by taking into account all possible contributions to leading
terms. For example
\bea
\lefteqn{(a_1L_{-1}^{l_1} + b_1L_{-1}^{l_1-1}G_{-1}G_0+\ldots)  
(a_2L_{-1}^{l_2} + b_2L_{-1}^{l_2-1}G_{-1}G_0+\ldots)  
\;\;=} \nn \\
&& a_1a_2L_{-1}^{l_1+l_2}   + (a_1b_2+a_2b_1
+2b_1b_2) L_{-1}^{l_1+l_2-1}G_{-1}G_0+\ldots  \com
\eea
leads to the multiplication rule for $(a_1,b_1)_{l_1}^+(a_2,b_2)_{l_2}^+$ 
provided the conformal weights satisfy $\Delta_2+l_2=\Delta_1$. In the same
way we can easily find similar rules for other types of singular
vector operators.
\bth \label{th:multab}
$\theta^{p_1}_{l_1}$ and 
$\theta^{p_2}_{l_2}$ denote two singular vector operators
for the Verma modules $\vm_{\Delta_1}$ and $\vm_{\Delta_2}$ with 
$\Delta_2+l_2=\Delta_1$ then
$\theta^{p_1}_{l_1}\theta^{p_2}_{l_2}\ket{\Delta_2}$ is either trivial 
or singular in $\vm_{\Delta_2}$
at level $l_1+l_2$ with parity $p_1p_2$. Depending on the parities
the resulting singular vector 
operator is:
\bea
(a_1,b_1)^+_{l_1} \, (a_2,b_2)_{l_2}^+ \ket{\Delta_2} &=& 
(a_1a_2, \, a_1b_2+a_2b_1+2b_1b_2)^+_{l_1+l_2} 
\ket{\Delta_2} \,, \\
(a_1,b_1)^+_{l_1} \, (a_2,b_2)_{l_2}^- \ket{\Delta_2} &=& 
(a_1a_2, \, a_1b_2+a_2b_1[\Delta_2-\frac{c}{24}] 
+2b_1b_2)^-_{l_1+l_2} 
\ket{\Delta_2} \,, \\
(a_1,b_1)^-_{l_1} \, (a_2,b_2)_{l_2}^+ \ket{\Delta_2} &=& 
(a_1a_2+2a_1b_2,\half l_2a_1a_2+a_2b_1
-a_1b_2[\Delta_2-\frac{c}{24}])^-_{l_1+l_2} 
\ket{\Delta_2} \,, \\
(a_1,b_1)^-_{l_1} \, (a_2,b_2)_{l_2}^- \ket{\Delta_2} &=& 
(a_1a_2[\Delta_2-\frac{c}{24}]+2a_1b_2, \,
a_2b_1+\half l_2a_1a_2-a_1b_2)^+_{l_1+l_2} 
\ket{\Delta_2} \,.
\eea
If, however, 
$\theta^{p_2}_{l_2}$ denotes a singular vector in the $G$-closed
Verma module $\vm_{\frac{c}{24}}^G$ and 
$\Delta_1=\frac{c}{24}+l_2$ then
$\theta^{p_1}_{l_1}\theta^{p_2}_{l_2}\ket{\frac{c}{24}}^G$ is either trivial 
or singular in $\vm_{\frac{c}{24}}^G$
at level $l_1+l_2$ with parity $p_1p_2$. The multiplication rules for
these cases are:
\bea
(a_1,b_1)^+_{l_1} \, (a_2)_{l_2}^+ \ket{\Delta_2}^G &=& 
(a_1a_2)^+_{l_1+l_2} 
\ket{\Delta_2}^G \,, \\
(a_1,b_1)^+_{l_1} \, (b_2)_{l_2}^- \ket{\Delta_2}^G &=& 
(a_1b_2+2b_1b_2)^-_{l_1+l_2} 
\ket{\Delta_2}^G \,, \\
(a_1,b_1)^-_{l_1} \, (a_2)_{l_2}^+ \ket{\Delta_2}^G &=& 
(\half l_2a_1a_2+a_2b_1)^-_{l_1+l_2} 
\ket{\Delta_2}^G \,, \\
(a_1,b_1)^-_{l_1} \, (b_2)_{l_2}^- \ket{\Delta_2}^G &=& 
(2a_1b_2)^+_{l_1+l_2} 
\ket{\Delta_2}^G \,.
\eea
Finally, if $\theta^{p_1G}_{l_1}$
denotes a $G$-closed singular vector operator
of $\ket{\frac{c}{24}-l_1}$ and $\theta^{p_2}_{l_2}$
a singular vector operator of $\ket{\Delta_2}$ with 
$\Delta_2+l_1+l_2=\frac{c}{24}$ then
$\theta^{p_1G}_{l_1}\theta^{p_2}_{l_2}\ket{\Delta_2}$ is either trivial 
or a $G$-closed singular in $\vm_{\frac{c}{24}-l_1-l_2}$
at level $l_1+l_2$ with parity $p_1p_2$. 
\bea
(a_1)^+_{l_1} \, (a_2,b_2)_{l_2}^+ \ket{\Delta_2} &=& 
(a_1a_2)^+_{l_1+l_2} 
\ket{\Delta_2} \,, \\
(a_1)^+_{l_1} \, (a_2,b_2)_{l_2}^- \ket{\Delta_2} &=& 
(a_1a_2)^-_{l_1+l_2} 
\ket{\Delta_2} \,, \\
(a_1)^-_{l_1} \, (a_2,b_2)_{l_2}^+ \ket{\Delta_2} &=& 
(a_1a_2+2a_1b_2)^-_{l_1+l_2} 
\ket{\Delta_2} \,, \\
(a_1)^-_{l_1} \, (a_2,b_2)_{l_2}^- \ket{\Delta_2} &=& 
(-a_1a_2(l_1+l_2)+2a_1b_2)^+_{l_1+l_2} 
\ket{\Delta_2} \,.
\eea
\eth

In exactly the same way, we can also find the leading coefficients of
the vectors $\Psi^{\pm}_{G,l}=(a,b)^{\pm}_{G,l}\ket{\Delta}$ 
and ${}_{G}\Psi^{\pm}_{l}={}_G(a,b)^{\pm}_{l} \ket{\Delta}$, introduced
in \eqs{\ref{eq:psiG1}}-\eqoth{\ref{eq:psiG4}}. For example, the
product
\bea
(aL_{-1}^{l}   + bL_{-1}^{l-1}G_{-1}G_0+\ldots)  G_0  &=& 
a L_{-1}^{l} G_0   
+ bL_{-1}^{l-1}G_{-1}(L_0-\frac{c}{24})+\ldots \,,
\eea 
leads to $(a,b)^-_{G,l}$. Using similar products one easily obtains
the following theorem.
\bth \label{th:Gprod}
If the vector $\Psi^{\pm}_{l}$ is given by $(a,b)^{\pm}_l$, then the
vectors $\Psi^{\pm}_{G,l}=(a,b)^{\pm}_{G,l}\ket{\Delta}$ 
and ${}_{G}\Psi^{\pm}_{l}={}_G(a,b)^{\pm}_{l} \ket{\Delta}$
are:
\bea
{}_G(a,b)^-_l \ket{\Delta} &=& (a+2b,
-b[\Delta-\frac{c}{24}]+\half la)^-_l \ket{\Delta} \com \label{eq:thetaG1} \\
(a,b)^-_{G,l} \ket{\Delta} &=& (a,b[\Delta-\frac{c}{24}])^-_l 
\ket{\Delta} \com \\
{}_G(a,b)^+_l \ket{\Delta} &=& (a[\Delta-\frac{c}{24}]
+2b,\half la-b)^+_l \ket{\Delta} \com \\
(a,b)^+_{G,l} \ket{\Delta} &=& (a[\Delta-\frac{c}{24}],b)^+_l 
\ket{\Delta} \pkt \label{eq:thetaG4}
\eea
\eth

For $(a,b)^{\pm}_{l}$ being $G$-closed we require
${}_G(a,b)^{\mp}_{l}\equiv 0$, which is necessary and sufficient. From
the above multiplication formulae we easily compute that in both cases
non-trivial solutions exist only for the known condition
$\Delta+l=\frac{c}{24}$.
For these solutions we obtain $(a,b)^{\pm}_l$ is
$G$-closed as given in the following theorem.
\bth
$(a,-\frac{a}{2})^+_{l}=(a)^{+G}_{l}$ and $(\frac{2}{l}a,a)^-_{l}=(a)^{-G}_{l}$
($l\not =0$) are the only possibilities for $G$-closed singular vectors in
$\vm_{\Delta}$ with $\Delta=\frac{c}{24}-l$.
\eth



\section{All Ramond singular vectors.}
\label{sec:classification}

In the previous section, we were able to derive the ordering kernel coefficients
for the vectors $\Psi_{G,l}^{\pm}$ and ${}_G\Psi_{l}^{\pm}$ provided
the ones of $\Psi_l^{\pm}$ were known. In theorem \refoth{\ref{th:psiprop}}, we showed that
in the case of the primitive singular vectors $\Psi_{p,q}^{\pm}$ the corresponding
singular vectors $\Psi_{G,l}^{\mp}$ and ${}_G\Psi_{l}^{\mp}$ are proportional.
This proportionality requirement restricts the coefficients of $\Psi_{p,q}^{\pm}$.

\bth \label{th:idsols}
The singular vectors $\Psi_{p,q}^{+}$ in $\vm_{\Delta}$ at level $\frac{pq}{2}$
are either given by
\bea
(q-tp,-q)^+_l  & \; {\rm or} \; & (q-tp,tq)^+_l \pkt  \label{eq:factor1}
\eea 
Similarly, the singular vectors $\Psi_{p,q}^{-}$ in $\vm_{\Delta}$ at level $\frac{pq}{2}$
are either
\bea
(-\frac{8t}{q},q-tp)^-_l  & \; {\rm or} \; & (\frac{8}{p},q-tp)^-_l \pkt  \label{eq:factor2}
\eea 
\eth
\bprf
Requiring $(a,b)^{-}_{G,l}\propto {}_G(a,b)_{l}^{-}$ has for $\Delta\not =\frac{c}{24}$ exactly
two solutions as one can easily see: $\frac{a}{b}=-\frac{q}{q-tp}$ or $\frac{a}{b}=\frac{tp}{q-tp}$.
The cases for which $\Delta=\frac{c}{24}$ intersects with $\Delta=\Delta_{p,q}$ are only 
the discrete
points at $t=\frac{q}{p}$. Continuity arguments for the coefficients of 
polynomial type of the singular 
vectors lead to the claim for parity $+$. One easily verifies the parity $-$ case in the same way.   
\eprf

We note that the pairs of possible solutions given in theorem \refoth{\ref{th:idsols}} are 
symmetric 
under the operation $p \leftrightarrow q$ , $t \leftrightarrow \frac{1}{t}$. Since the
Verma modules $\vm_{\Delta_{p,q}(t)}$ and $\vm_{\Delta_{q,p}(\frac{1}{t})}$ are identical\footnote{Note that
$c(t)=c(\frac{1}{t})$.} we find the following identity of singular vectors using the normalisation
of equations \eqs{\ref{eq:factor1}}-\eqoth{\ref{eq:factor2}}:
\bea
\Psi^{\pm}_{p,q}(t) &=& -\frac{1}{t} \Psi^{\pm}_{q,p}(\frac{1}{t}) \pkt
\eea 
Due to this identity, all (primitive) singular vectors have been found once the vectors
$\Psi_{p,q}^{\pm}(t)$ for $p,q\in\bbbn$, $q$ even and $p$ odd have been given.

The examples in the appendix and computer exploration\cite{maple} until level $6$ show
which of the two possibilities can be assigned to the primitive singular vectors
$\Psi_{p,q}^{\pm}(t)$ depending on $p$ or $q$ being even or odd. We can therefore
give the following conjecture.
\bcj  \label{cj:ab}
The singular vectors $\Psi_{p,q}^{\pm}(t)$ at level $\frac{pq}{2}$ are given by
\bea
\Psi_{p,q}^{+}(t) &=& (q-tp,-q)^+_{\frac{pq}{2}} \ket{\Delta_{p,q}(t)} \;\;, q \; {\rm even} \com \label{eq:qeven} \\
\Psi_{p,q}^{+}(t) &=& (q-tp,tp)^+_{\frac{pq}{2}} \ket{\Delta_{p,q}(t)} \;\;, p \; {\rm even} \com \label{eq:peven} \\
\Psi_{p,q}^{-}(t) &=& (-\frac{8t}{q},q-tp)^-_{\frac{pq}{2}} \ket{\Delta_{p,q}(t)} \;\;, q \; {\rm even} \com \\
\Psi_{p,q}^{-}(t) &=& (\frac{8}{p},q-tp)^-_{\frac{pq}{2}} \ket{\Delta_{p,q}(t)} \;\;, p \; {\rm even} \com
\eea
for $p,q\in\bbbn$, $p-q$ odd.
\ecj

According to theorem \refoth{\ref{cj:ab}} the singular vectors $\Psi_{p,q}^{\pm}$ 
only have two possible choices for their coefficients, therefore explicit computer calculations
are a sufficiently convincing indication for conjecture \refoth{\ref{cj:ab}}. 
A final proof could follow in the line
of Watts\cite{gerard}, using the fusion of conformal fields in order to derive 
expressions for singular vectors. But for our purpose only two coefficients
are needed and therefore the construction of Watts should be applicable
in a simplified form and may only be needed for a particular limit of the parameter $t$. 

The expressions of \refoth{\ref{cj:ab}} reveal once more striking similarities to the 
twisted $N=2$ superconformal algebra. For the twisted $N=2$ algebra
the singular vectors also follow a choice of two possible 
ordering kernel coefficients\cite{p9sdim2}.
They are also labelled by two integers $r$, $s$ where $r$ plays the r\^ole of $p$
and $s$ the one of $q$, however $s$ has to be odd. In the twisted $N=2$ case, the  
observed expressions for singular vectors always take an expression similar to
\eq{\ref{eq:peven}} whilst the second possibility similar to \eq{\ref{eq:qeven}}
never appears. Considering the fact that $s$ has to be odd for the twisted $N=2$ case
and taking into account that $q$ is odd for \eq{\ref{eq:peven}}, we observe 
a rather close connexion between the $N=1$ Ramond case and the twisted $N=2$ case.

Let us note that the singular vectors of  conjecture \refoth{\ref{cj:ab}} never vanish identically.
This reveals a significant difference with respect to the singular vectors of the
three isomorphic $N=2$ algebras,
where discrete vanishings of the primitive singular vector expressions lead to the fact that
the two-dimensional tangent space is spanned by two secondary\footnote{For example, 
in the case of the $N=2$ Neveu-Schwarz algebra the spaces of linearly
independent singular vectors are always uncharged and are secondary singular vectors
of charged singular vectors at lower levels.} linearly independent
singular vectors\cite{cmp1}. In contrast, the primitive singular vector expressions for the 
twisted $N=2$ algebra never vanish\cite{p9sdim2}, like for the $N=1$ Ramond algebra.

Based on conjecture \refoth{\ref{cj:ab}} it is easy to compute 
the Verma modules with degenerate singular vectors, i.e. with singular 
spaces of dimension $2$. For a given level $l\in\bbbn$ there are
normally multiple solutions to the number theoretical problem $pq=2l$, 
$p,q\in\bbbn$, $p-q$ odd. Solutions with $p$ even are said to
be of {\it type} $\epsilon=+$ whilst for $q$ even we assign type $\epsilon=-$.
In the general case one factorises the integer $2l$ in its prime 
factors  
\bea
2l &=& 2^n \prod_{i=1}^{P} \gamma_i^{n_i} \com
\eea
with $\gamma_i>2$ distinct primes and $n\in\bbbn$, $n_i\in\bbbno$. 
The type $\epsilon=-$ ($q$ even) solutions to $pq=2l$ are then 
given by
\bea
p_{\pi^{-}} &=& \prod_{i=1}^P \gamma_i^{k_i} \com \nn \\
q_{\pi^{-}} &=& 2^n \prod_{i=1}^P \gamma_i^{n_i-k_i} \com \label{eq:rspi1}
\eea
whilst for $\epsilon=+$ ($p$ even) we have
 \bea
p_{\pi^{+}} &=& 2^n \prod_{i=1}^P \gamma_i^{k_i} \com \nn \\
q_{\pi^{+}} &=& \prod_{i=1}^P \gamma_i^{n_i-k_i} \com \label{eq:rspi2}
\eea
for each $P$-tuple $\pi^{\pm}=(k_1,k_2,\ldots,k_P)\in\bbbno^P$, 
with $k_i \leq n_i, \, \forall i=1,\ldots,P$ where the type of the solution is
indicated with a superscript $\pm$. 
Note that for each level $l\in\bbbn$ there exist at 
least two solutions: 
$p=2l$, $q=1$ and $p=1$, $q=2l$. The number of solutions to $pq=2l$, $p,q\in\bbbn$, 
$p-q$ odd is thus determined by the number
of $P$-tuples $\pi^{\pm}=(k_1,\ldots,k_P)$ with $\pi^{\pm}\in\bbbno^P$ and 
$\pi^{\pm}\leq(n_1,\ldots,n_P)$ for the prime factorisation of $2l$. 
Hence, there are\footnote{We define the empty product as 
$\prod_{i=1}^{0}=1$.} $\Pi_l=2\prod_{i=1}^P (n_i+1)$ solutions to 
$pq=2l$, $p,q\in\bbbn$, $p-q$ odd.

The proof of theorem \refoth{\ref{th:psiprop}} sheds light on the fact 
that there are generically 
only one-dimensional singular spaces defined by the singular 
vectors $\Psi^+_{p,q}$ (or $\Psi^-_{p,q}$) at level $l=\frac{pq}{2}$ for 
all $\Pi_l$ solutions $p$ and $q$. The
most interesting Verma modules are those, where some of these $\Pi_l$ 
solutions intersect and may hence lead to two-dimensional singular spaces 
provided the corresponding singular vectors are not proportional. 
Using conjecture \refoth{\ref{cj:ab}} 
we shall now investigate such cases of degeneration. Let us remark again 
that the singular dimension cannot be bigger than $2$ (the size of the ordering
kernel) even though a
priori we would expect that even more than $2$ solutions $\pi_i$ could
intersect. 

Let us analyse the intersections of the conformal
weights $\Delta_{p,q}(t)$ of the $\Pi_l$ solutions 
corresponding to 
$pq=2l$, $p,q\in\bbbn$, $p-q$ odd.
For fixed $l\in\bbbn$ we find the prime factorisation as 
given above. Let us now take two $P$-tuples
$\pi^{\epsilon_1}_1$ and $\pi^{\epsilon_2}_2$ of assigned
types $\epsilon_1$ and $\epsilon_2$. We write $p_i$ for
$p_{\pi_i^{\epsilon_i}}$ and $q_i$ for
$q_{\pi_i^{\epsilon_i}}$ which satisfy $p_1q_1=p_2q_2=2l$. 
Let us further assume that 
the corresponding conformal weights intersect:
$\Delta_{p_1,q_1}= \Delta_{p_2,q_2}$. 
This easily implies $(q_1-p_1t)^2=(q_2-p_2t)^2$ for
the intersection points
which results in
\bea
t &=& \pm \frac{q_2}{p_1} \;=\; \pm \frac{q_1}{p_2} \pkt \label{eq:intsec}
\eea
Using the prime factorisation the intersection values of the central parameter
$t$ can be given as
\bea
t &=& \pm \beta_n^{\epsilon_1,\epsilon_2} \prod_{i=1}^P \gamma_i^{n_i-k_i^1-k_i^2} \com 
\label{eq:intersect}
\eea
with
\bea
\beta_n^{\epsilon_1,\epsilon_2} &=& \left\{ \begin{array}{l}
2^n \:\:\: \;\;\; {\rm for} \;\;\; (\epsilon_1,\epsilon_2)=(-,-) \\ 
1 \:\:\: \;\;\;\;\; {\rm for} \;\;\; (\epsilon_1,\epsilon_2)=(-,+), (+,-)  
\\ \frac{1}{2^n} \:\:\: \;\;\; {\rm for} \;\;\; (\epsilon_1,\epsilon_2)=(+,+) 
\end{array} \right.
\eea
Therefore, the conformal weights 
$\Delta_{p_1,q_1}$ and
$\Delta_{p_2,q_2}$ have exactly $2$ intersection 
points
where the two singular vectors $\Psi^+_{p_1,q_1}$ and 
$\Psi^+_{p_2,q_2}$ both exist as well as $\Psi^-_{p_1,q_1}$ and 
$\Psi^-_{p_2,q_2}$. The key question is 
whether these singular vectors are different and hence
define two-dimensional singular spaces, or if they are proportional 
and thus lead to one-dimensional singular spaces instead.

For this purpose, let us take again two $P$-tuples 
$\pi_1^{\epsilon_1}$ and $\pi_2^{\epsilon_2}$. 
The corresponding conformal weights intersect for 
$t=\pm \beta_n^{\epsilon_1,\epsilon_2} \prod_{i=1}^P \gamma_i^{n_i-k_i^1-k_i^2}
=\pm\frac{q_1}{p_2}=\pm\frac{q_2}{p_1}$. 
From \refoth{\ref{cj:ab}} we obtain the corresponding 
singular vectors
$\Psi^{\pm}_{p_j,q_j}$, $j=1,2$ as:
\bea
\Psi^+_{p_j,q_j} (t=\pm\frac{q_1}{p_2}) &=&
(q_j\mp q_{\bar{j}},-q_j)^+_l\ket{\Delta_{p_j,q_j}} \com
\epsilon_j=- \com  (q_j \;{\rm even}) \com\;\;
\label{eq:int1} \\
\Psi^+_{p_j,q_j} (t=\pm\frac{q_1}{p_2}) &=&
(q_j\mp q_{\bar{j}},\pm q_{\bar{j}})^+_l\ket{\Delta_{p_j,q_j}} \com \;\; \epsilon_j=+ 
\com  (p_j \;{\rm even}) \\
\Psi^-_{p_j,q_j} (t=\pm\frac{q_1}{p_2}) &=&
(\mp\frac{8}{p_{\bar{j}}},q_j\mp q_{\bar{j}})^+_l\ket{\Delta_{p_j,q_j}} \com \;\; \epsilon_j=- 
\com  (q_j \;{\rm even}) \\
\Psi^-_{p_j,q_j} (t=\pm\frac{q_1}{p_2}) &=&
(\frac{8}{p_j},q_j\mp q_{\bar{j}})^+_l\ket{\Delta_{p_j,q_j}} \com  
\epsilon_j=+ \com  (p_j \;{\rm even}) \;\; \com \label{eq:int4} 
\eea 
with $\bar{1}=2$ and $\bar{2}=1$.
By considering the determinant of these coefficients we
easily obtain the following result.

\bth \label{th:prsvecs}
For a given level $l\in\bbbn$ the singular vectors $\Psi^{\pm}_{p,q}$ 
with $l=\frac{pq}{2}$ 
define generically one-dimensional singular spaces. 
There are two intersection points of $\Delta_{p_1,q_1}=\Delta_{p_2,q_2}$
for $p_1q_1=p_2q_2=2l$
which are
$t=\pm \beta_n^{\epsilon_{1},\epsilon_{2}} \frac{1}{2^n}\prod_{i=1}^P \gamma_i^{n_i-k_i^1-k_i^2}
=\pm\frac{q_{\pi^1}}{p_{\pi^2}}=\pm\frac{q_{\pi^2}}{p_{\pi^1}}$. 
At these points degenerate singular spaces of dimension $2$
occur as given in the following table
\btab{|l|l|l|l|l|l|}
\hline \label{tab:degeneration}
 $\epsilon_1$ & $\epsilon_2$ & $p_1$ & $p_2$ & 
$t$ & dimension \\
\hline
- & - & odd & odd & $\pm2^n\prod_{i=1}^P \gamma_i^{n_i-k_i^1-k_i^2}$ &  $2$ \\ 
\hline
- & + & odd & even & $\pm\prod_{i=1}^P \gamma_i^{n_i-k_i^1-k_i^2}$ &  $1$ \\ 
\hline
+ & - & even & odd & $\pm\prod_{i=1}^P \gamma_i^{n_i-k_i^1-k_i^2}$ &  $1$ \\ 
\hline
+ & + & even & even & $\pm\frac{1}{2^n}\prod_{i=1}^P 
\gamma_i^{n_i-k_i^1-k_i^2}$ & $2$ \\ 
\hline
\etab{Degenerate $N=1$ Ramond cases.}
The values for $t$ at the intersection points are real and rational.
However, at most 2 of these conformal weight curves
can intersect in any common values of $t$. 
\eth 
\bprf
The determinant of the ordering kernel coefficients of 
\eqs{\ref{eq:int1}}-\eqoth{\ref{eq:int4}} in the different cases
are $\pm(q_2^2-q_1^2)$ for $(\epsilon_1,\epsilon_2)=(-,-)$ and
$\pm(q_1^2-q_2^2)$ for $(\epsilon_1,\epsilon_2)=(+,+)$ by considering
only the positive parity case. In the case of 
$(\epsilon_1,\epsilon_2)=(+,-)$ and $(\epsilon_1,\epsilon_2)=(-,+)$
the determinant always vanishes. Similar relations can easily be found
for the negative parity cases. Hence for $\epsilon_1\epsilon_2=+$
(equal type singular vectors)
we obtain $2$-dimensional spaces, whilst for 
$\epsilon_1\epsilon_2=-$ (different type singular vectors)
we have just $1$-dimensional singular spaces
defined by the two singular vectors.
Let us assume now that the 
conformal weights of $\pi^{\epsilon_1}_1$ and 
$\pi^{\epsilon_2}_2$ intersect at
$t=\pm \beta_{n}^{\epsilon_1,\epsilon_2}
\frac{1}{2^n}\prod_{i=1}^P
\gamma_i^{n_i-k_i^1-k_i^2}=\pm\frac{q_1}{p_2}$. 
In the same way 
the conformal weights of $\pi^{\epsilon_2}_2$ 
and $\pi^{\epsilon_3}_3$
intersect at
$\bar{t}=\pm \beta_{n}^{\epsilon_2,\epsilon_3}
\frac{1}{2^n}\prod_{i=1}^P \gamma_i^{n_i-k_i^2-k_i^3}
=\pm\frac{q_3}{p_2}$. 
Assuming $t=\bar{t}$ obviously leads to $q_1=q_3$ and hence also $p_1=p_3$.
Therefore, there are no values of $t$ for which more than two conformal 
weights $\Delta_{p,q}$ intersect at the same level $pq=2l$.
\eprf

Intersections of equal type singular vectors occur for the first time at
level $3$ when $\Delta_{1,6}$ intersects $\Delta_{3,2}$ and
for $\Delta_{6,1}$ intersecting with $\Delta_{2,3}$ and therefore lead
to degenerate primitive singular vectors. All other
intersections at level $3$ and below are intersections of different
type of singular vectors and therefore do not lead to any degenerate levels.
In appendix \refoth{\ref{app:a}} we give all primitive singular
vectors until level $3$ and indicate explicitly the degenerate
singular vectors. So far, degenerate singular vector spaces had only been
discovered for the $N=2$ superconformal 
algebras\cite{cmp1,beatriz2,p9sdim2}, and the fact that also the
$N=1$ Ramond algebra leads to degenerate singular vector spaces in its highest weight
representations has been overlooked in the past.
Let us stress that for the $N=2$ isomorphic algebras
(Neveu-Schwarz, topological and Ramond) the degenerate singular
vectors found so far are always secondary whereas for the $N=2$
twisted algebra and for the $N=1$ Ramond algebra the degenerate
singular vectors are generically primitive\footnote{We are very grateful 
to B. Gato-Rivera for pointing this out to us.}.

The multiplication rules of theorem \refoth{\ref{th:multab}} together with
the identification of the singular vectors with their ordering kernel
coefficients supply the fundamental tools needed to investigate the
structure of embedded singular vectors and descendant singular vectors
and hence the embedding diagrams. In the following section, we show
that theorem \refoth{\ref{th:multab}} allows us to 
follow the lines of Feigin and Fuchs\cite{ff1}, 
who discussed the Virasoro case, to extend their approach to the Ramond
algebra using the results of this section. 
In a first analysis we will comment on the main similarities
for the rational models ($t\in\bbbq$)  
with the Virasoro case but we will also reveal certain major
differences. A complete discussion and derivation of all 
Ramond embedding diagrams will be given in a forthcoming 
paper\cite{progress}.


\section{Descendant singular vectors.}
\label{sec:desc}

In the previous sections we have discussed that a singular vector 
$\theta_{l_1}\ket{\Delta_1}$ of the Verma module $\vm_{\Delta_1}$
generates a submodule homomorphic to $\vm_{\Delta_1+l_1}$.
On the other hand $\vm_{\Delta_1+l_1}$ may again contain
a singular vector $\theta_{l_2}\ket{\Delta_1+l_1}$. In this section
we want to focus on the resulting vector
$\theta_{l_2}\theta_{l_1}\ket{\Delta_1}$, which - if non-trivial - is again
a singular vector, called a {\it descendant or secondary singular vector},
as was explained before. Descendant
singular vectors that are proportional are considered to be equal. 
The relative structure of the singular vectors and their descending patterns
are usually given in so-called {\it embedding diagrams}. Embedding
diagrams are important to find out the dimension of the null submodule 
of a Verma module and are hence crucial for the character formulae
of the irreducible highest weight representations. Embedding diagrams
include all singular vectors and usually - if applicable - also subsingular
vectors. The diagrams indicate which singular vectors appear as descendant
singular vectors of others and proportional singular vectors are indicated
as one singular vector only.

For the Virasoro algebra the derivation of the embedding diagrams\cite{ff1}
is merely a matter of solving the number theoretical problem for which
levels the roots of the Kac determinant formula predict a singular
vector and of comparing the relative levels of these singular
vectors to reveal the embedding pattern. This
procedure is rather simple due to the properties of Virasoro algebra that
does not allow two or more Virasoro singular vectors at the same
level in the same Verma module. 
Hence Virasoro singular vectors at the same level in the same
Verma module are 
always proportional. As a remark, let us add
that also for the Virasoro case for a given level
$l\in\bbbno$ we have different one-parameter families 
of singular vectors $\xi_{p,q}(t)$ that intersect
for certain values of $t$, pairs $(p_1,q_1)$, and $(p_2,q_2)$ at the
same level $p_1q_1=p_2q_2$. However, for Virasoro Verma modules 
these singular vectors at such
intersection points are always linearly dependent\cite{ff1,adrian1,p6sdim1}. 
Furthermore, for the Virasoro algebra the products of singular
vector operators never vanish 
and also Virasoro Verma modules do not contain any
subsingular vectors. All these non-trivial properties make the
derivation of Virasoro embedding diagrams straightforward.

For superconformal algebras most of these remarkable properties of the Virasoro 
algebra may not hold. Products of singular vector operators 
may vanish identically, a very common 
consequence of Lie superalgebras, singular spaces may be bigger than
just one-dimensional and subsingular vectors may also arise. Such
features have recently been discussed for the $N=2$ superconformal
algebras\cite{beatriz1,beatriz2,p6sdim1,p9sdim2,cmp1,thesis}.
In this section we will show that the Ramond algebra also shares some
of these difficulties and we will discuss the derivation of the most
interesting embedding diagrams, the {\it rational models}. A full consideration
of the Ramond embedding diagrams shall be given elsewhere\cite{progress}.
The multiplication rules of theorem \refoth{\ref{th:multab}}
and the expressions of \refoth{\ref{cj:ab}} will prove to be the main
tools for the analysis of the Ramond embedding diagrams. With their
help it will be possible to find out which descendant singular vectors are
identical and which are trivial.

Like in the Virasoro case\cite{ff1} and the $N=2$ Neveu-Schwarz case\cite{thesis}
we first compute for a given
Verma module $\vm_{\Delta}$ and a central parameter $t$ the {\it intercept}
$a$ as
\bea
a^2 &=& 8t\Delta- \frac{5t}{2}+1+t^2 \pkt  
\eea
The sign of $a$ is not relevant but for real $a$ we
choose $a$ positive.
In the case that $\Delta=\Delta_{p,q}$ for $p,q\in\bbbn$, $p-q$ odd,
we obviously find $(q-pt)^2=a^2$. A vanishing $a$-value corresponds
to the supersymmetric case $\Delta=\frac{c}{24}$ which will be analysed in
much detail in the following section. Therefore, $a=0$ shall be
excluded for the rest of this section. Conversely, if for given values of
$t$ and $a$ we are looking for the integers $p$, $q$ that solve 
$\Delta=\Delta_{p,q}$, then we find as solutions all $\ph$, $\qh$
with
\bea
\qh = \ph t -a \com \;\; \ph, \qh \in \bbbz\com\;\;
\ph-\qh \;{\rm odd}\; \com \;\; \ph\qh >0 .
\eea
For these values of $(\ph,\qh)$ we have the required
$\Delta_{| \ph |, | \qh |}=\Delta$ with
$p= | \ph |$, $q= | \qh |$ and hence $p-q$ odd.
Note that we distinguish $p$, $q$ from $\ph$, $\qh$ as the
latter are also allowed to be negative as long as their product stays
positive. Just like in the Virasoro case\cite{ff1} the problem reduces
to finding the integer pair solutions $(\ph,\qh)$ on the straight line
$\cL$: $\qh = \ph t -a$ with positive product $\ph\qh$. In contrast
to the Virasoro case we require here the difference $\ph-\qh$ to be odd.
The cases of irrational $t$ are simple and will be included in a full
discussion on the embedding diagrams in a forthcoming publication\cite{progress}, the
cases of rational $t$ are however very interesting. These so-called {\it rational
models} shall be analysed here. We therefore assume $t\in\bbbq$
and find the unique pair of integers $(u,v)$ with $u$, $v$ coprime
and $v$ positive such that $t=\frac{u}{v}$.

If $\cL$ has no integer pair solutions at all or only
integer pair solutions with $\ph-\qh$ even, then the
Verma module $\vm_{\Delta(a,t)}$ is obviously irreducible as the 
determinant formula is non-trivial for all levels. In the case that
$\cL$ has at least one integer pair solution with $\ph-\qh$ odd, 
then it obviously has infinitely
many. In the notation of Feigin and Fuchs\cite{ff1} these cases are
denoted by $\cs_{-}$ for positive $t$ and $\cs_{+}$
for negative\footnote{Note the opposite definition of the index of $\cs_{\pm}$ and
the sign of $t$ for historical reasons.} values of  $t$. 
Our task is now to describe among these infinitely many solutions those
solutions $(\ph,\qh)$
with $\ph\qh >0$ and $\ph-\qh$ odd. For this purpose it is essential
to distinguish two different cases for which we shall add the label $e$
for {\it even} or $o$ for {\it odd} to $\cs_{\pm}$. Thus $\cs_{e\pm}$ denotes the case 
where one of $u$ or $v$ is even (but obviously not both) whilst
$\cs_{o\pm}$ stands for the cases with both $u$ and $v$ odd.

Let us first concentrate on $t$ positive. The integer pair
solution of $\cL$ with $\ph -\qh$ odd, both $\ph$ and $\qh$ positive,
and the smallest for $\ph$ is denoted by $(p_0,q_0)$. Likewise,
the integer pair
solution of $\cL$ with $\ph-\qh$ odd, both $\ph$ and $\qh$ negative,
and the smallest value of $| \ph |$ is denoted by 
$(-\tilde{p}_0,-\tilde{q}_0)$. We can now easily give all other allowed
solutions:
\bea
p_n =  p_0 + n v &\com &q_n =  q_0 + n u \com \\
-\tilde{p}_n =  -\tilde{p}_0 - n v &
\com & -\tilde{q}_n =  -\tilde{q}_0 - n u \com
\eea
with $n\in\bbbn_0$ for the case of $\cs_{o-}$. However, for $\cs_{e-}$ we
only find every other point as a solution and hence $n\in 2\bbbn_0$.
Therefore, we can already state that the Verma module $\vm_{p_0,q_0}$
contains singular vectors at the levels $\frac{p_nq_n}{2}$ and
$\frac{\pt_n\qt_n}{2}$ for all $n\in\bbbn_0$ in the case $\cs_{o-}$
but only for $n\in 2\bbbn_0$ for $\cs_{e-}$.  For all the solutions of type
$(p_n,q_n)$ we have $q-tp=-a$ whilst for $(\pt_n,\qt_n)$ we obtain $q-tp=a$
which fixes already one of the coefficients of \refoth{\ref{cj:ab}} for each case.
We shall now investigate descendant singular vectors of these primitive singular
vectors and shall use the multiplication rules of theorem \refoth{\ref{th:multab}}
to compare their ordering kernel coefficients. 

Along this line, we first look at the singular vector $\Psi^+_{p_0,q_0}$. Its highest weight
$\Delta_{p_0,q_0}+\frac{p_0q_0}{2}$ leads to a parameter $a^{\prime}$, which
is related to the original $a$ for $\Delta_{p_0,q_0}$ through:
$a^{\prime}=a+2q_0$. For the descendant singular vectors we hence have
to solve exactly the same number theoretical problem, namely finding
integer solutions on the straight line $\qh =\ph t-a^{\prime}$. This again
resembles the Virasoro case\cite{ff1}, however, a main difference in our case
is that we have to take into account the implications of the two
subcases $\cs_{e-}$ and $\cs_{o-}$. 

Figure \refoth{\ref{fig:IIImlines}} shows that
in the case of $\cs_{o-}$ the two lowest level
solutions of this shifted straight line are $(p_0+v,-q_0+u)$ (if $q_0\not = u$)
and $(-\pt_0,
-\qt_0-2q_0)$ which again both satisfy that the difference of the $p$-value and 
the $q$-value is odd. In the case that $q_0=u$ the former solution
changes to $(p_0+2v,u)$. This case shall be denoted by $\cs_{o-}^{0(u)}$. 
In the same way we can analyse 
the shifted straight line for the singular vector  $\Psi^+_{\pt_0,\qt_0}$
which results in the two lowest level solutions\footnote{For convenience we invert 
the line in the origin.}
$(p_0+2\pt_0,q_0)$ and $(\pt_0-v,-\qt_0-u)$ (for $\pt_0\not = v$). In the case
of $\pt_0 = v$ the latter changes to $(-v,-\qt_0-2u)$ which shall be denoted by
$\cs_{o-}^{0(v)}$. 

\vbox{
\bpic{400}{200}{-200}{-100}
\put(-200,0){\vector(1,0){400}}
\put(0,-100){\vector(0,1){200}}
\put(195,-10){\scriptsize $\ph$}
\put(-10,90){\scriptsize $\qh$}
\put(155,100){\line(-5,-2){355}}
\put(-195,-100){\line(5,2){395}}
\put(-145,-100){\line(5,2){340}}
\put(-120,-90){\circle*{2}}
\put(-70,-70){\circle*{2}}
\put(-20,-50){\circle*{4}}
\put(30,-30){\circle*{2}}
\put(80,-10){\circle*{2}}
\put(130,10){\circle*{4}}
\put(180,30){\circle*{2}}
\put(-170,-90){\circle*{2}}
\put(-120,-70){\circle*{2}}
\put(-70,-50){\circle*{2}}
\put(-20,-30){\circle*{4}}
\put(30,-10){\circle*{2}}
\put(80,10){\circle*{4}}
\put(130,30){\circle*{2}}
\put(180,50){\circle*{2}}
\put(-170,-30){\circle*{2}}
\put(-120,-10){\circle*{4}}
\put(-70,10){\circle*{2}}
\put(-20,30){\circle*{2}}
\put(30,50){\circle*{4}}
\put(80,70){\circle*{2}}
\put(130,90){\circle*{2}}
\put(2,-45){\tiny $-(a+2q_{0})$}
\put(2,-25){\tiny $-a$}
\put(-36,40){\tiny $-(a-2\qt_{0})$}
\put(85,7){\tiny $(p_{0},q_{0})$}
\put(-36,-41){\tiny $(-\pt_{0},-\qt_{0})$}
\put(139,9){\tiny $(p_{0}+v,-q_{0}+u)$}
\put(135,3){\tiny or $(p_0+2v,u)$ for $q_0=u$}
\put(-35,-60){\tiny $(-\pt_{0},-\qt_0-2q_0)$}
\put(35,45){\tiny $(-\pt_0+v,\qt_0+u)$}
\put(35,39){\tiny or $(v,\qt_0+2u)$ for $\pt_0=v$}
\put(-115,-15){\tiny $(-p_0-2\pt_0,-q_0)$}
\put(80,7){\vector(0,-1){14}}
\put(-20,-27){\vector(0,1){54}}
\put(-32,18){\tiny $2\qt_{0}$}
\put(64,-8){\tiny $2q_{0}$}
\label{fig:IIImlines}
\epic{$\cs_{o-}$ solutions of $\qh = \ph t -a$ at low levels.}
}

\noi
Therefore, we find for the case $\cs_{o-}$ the descendant singular vectors
\bea
\Psi^+_a &=& \theta_{p_0+v,u-q_0}^+ \theta_{p_0,q_0}^+ \ket{\Delta_{p_0,q_0}} \com \label{eq:desc1} \\
\Psi^+_b &=& \theta_{\tilde{p}_0,\tilde{q}_0+2q_0}^+ \theta_{p_0,q_0}^+ \ket{\Delta_{p_0,q_0}} \com \\
\Psi^+_c &=& \theta_{p_0+2\pt_0,q_0}^+ \theta_{\pt_0,\qt_0}^+ \ket{\Delta_{p_0,q_0}} \com \\
\Psi^+_d &=& \theta_{v-\pt_0,\qt_0+u}^+ \theta_{\pt_0,\qt_0}^+ \ket{\Delta_{p_0,q_0}} \pkt
\eea
The vectors given are for the positive parity sector which can easily be transformed into the
negative parity sector. We observe that the vectors $\Psi^+_a$ and $\Psi^+_d$ are at the same level
and the same is true for the pair $\Psi^+_b$ and $\Psi^+_c$.
Using the multiplication rules of \refoth{\ref{th:multab}} and taking into account
the different possible types for the singular vector operators\footnote{Note that not all combinations
of types are possible. If for example $\theta_{p_0,q_0}^+$ is of type $\epsilon=+$, then  $\theta_{p_0+v,u-q_0}^+$
has to be of type $\epsilon=-$. Indeed, wrong type combinations do not lead to proportional 
singular vectors in some of the following cases.} we easily find by explicit calculations
that the singular vectors 
$\Psi^+_a$ and $\Psi^+_d$ are proportional as well as $\Psi^+_b$ and $\Psi^+_c$.
Performing the same procedure on the singular vectors $\Psi^+_a$ and $\Psi^+_b$ using exactly the
same method of shifting the straight line $\ph=\qh t-a$ appropriately,
we consequently obtain a second list of descendant
singular vectors of $\Psi_{p_0,q_0}$ (descendant in second order)
for the case $\cs_{o-}$.
\bea
\Psi^+_e &=& \theta_{p_0+2v,q_0}^+ \theta_{p_0+v,u-q_0}^+ \theta_{p_0,q_0}^+ \ket{\Delta_{p_0,q_0}} \com \\
\Psi^+_f &=& \theta_{\pt_0,\qt_0+2u}^+ \theta_{p_0+v,u-q_0}^+ \theta_{p_0,q_0}^+ \ket{\Delta_{p_0,q_0}} \com \\
\Psi^+_g &=& \theta_{p_0+2\pt_0+v,u-q_0}^+ 
\theta_{\pt_0,\qt_0+2q_0}^+ \theta_{p_0,q_0}^+ \ket{\Delta_{p_0,q_0}} \com \\
\Psi^+_h &=& \theta_{v-\pt_0,\qt_0+2q_0+u}^+ \theta_{\pt_0,\qt_0+2q_0}^+ 
\theta_{p_0,q_0}^+ \ket{\Delta_{p_0,q_0}} \pkt \label{eq:desc8}
\eea

Again, the multiplication rules \refoth{\ref{th:multab}} easily prove that $\Psi^+_e$ and
$\Psi^+_h$ are proportional and also the pair $\Psi^+_f$ and $\Psi^+_g$. Furthermore,
the results of the multiplication show that the latter pair of singular vectors is
proportional to $\Psi^+_{\pt_1,\qt_1}$ whilst the former is proportional to $\Psi^+_{p_1,q_1}$.
Therefore, after having reached singular vectors of the original straight line $\cL$
($\Psi^+_{p_1,q_1}$ and $\Psi^+_{\pt_1 ,\qt_1}$) the whole procedure
starts again and ultimately exhausts all (infinitely many) singular vectors\footnote{The fact
that these are all singular vectors will be explained in Ref. \icite{progress}}.
So far, this reminds us very much of the Virasoro case\cite{ff1} but we should keep in mind that
differently to the Virasoro case 
the multiplication rules \refoth{\ref{th:multab}} were crucial to prove this 
embedding structure. We will summarise this structure in the {\it embedding diagrams}
given below in theorem \refoth{\ref{th:emb1}}.
Before continuing to the case $\cs_{e-}$ we shall discuss what happens in
the special cases $\cs_{o-}^{0(u)}$ and $\cs_{o-}^{0(v)}$. According to the solutions on
the shifted straight lines given above for the case $\cs_{o-}^{0(u)}$ we find that the singular
vector $\Psi^+_a$ has to be replaced by
\bea
\Psi^+_{a^{\prime}} &=& \theta_{p_0+2v,u}^+ \theta_{p_0,q_0}^+ \ket{\Delta_{p_0,q_0}} \pkt
\eea  
Taking into account that $p_0$ is even for $\cs_{o-}^{0(u)}$ ($q_0=u$) we hence find
that $\Psi^+_d$ becomes proportional to $\Psi^+_{p_0,q_0}$ whilst $\Psi^+_{a^{\prime}}$
is already proportional to $\Psi^+_{p_1,q_1}$, again by using our multiplication rules. 
Therefore the above embedding pattern collapses in the way indicated in theorem 
\refoth{\ref{th:emb1}}. Similarly, for the cases $\cs_{o-}^{0(v)}$ the vectors
$\Psi^+_a$ and $\Psi^+_{\pt_0,\qt_0}$ become proportional whilst $\Psi^+_d$ is already
proportional to $\Psi^+_{\pt_1,\qt_1}$ resulting in almost the same embedding patterns as
for $\cs_{o-}^{0(u)}$. Finally, if both conditions hold, $q_0=u$ and $\pt_0 =v$, we obtain
the proportionalities of both cases. This case shall be denoted by $\cs_{o-}^{00}$ 
following the notation of Feigin and Fuchs\cite{ff1}.
Let us finally mention that for the cases $\cs_{o+}$ with negative, rational $t$ we find
exactly the same patterns except that the embedding patterns come to an end
just like in the Virasoro case. In figure \refoth{\ref{fig:IIImlines}} this corresponds to
solutions in the regions with $\ph \qh< 0$ and therefore the number of
solutions has to be finite. 
We have hence proven that the descendant singular vectors of the 
primitive singular vectors have the following embedding patterns.

\bth \label{th:emb1}
For the subsets $\cs_{o\pm}$, $\cs_{o\pm}^{0(u)}$, $\cs_{o\pm}^{0(v)}$, and $\cs_{o\pm}^{00}$
of rational models ($\Delta\not =\frac{c}{24}$)
we find the following embedding patterns for descendant singular 
vectors of the type $\Psi^+_{p,q}$ in the positive parity sector. The negative parity sector
leads to exactly the same diagrams. The zero mode $G_0$ interpolates between
the embedding diagrams of the positive parity sector and the negative parity sector.
\bpic{400}{230}{30}{-5}
\put(78,220){\framebox(4,3){}}
\put(55,170){\circle*{3}}
\put(105,170){\circle*{3}}
\put(55,120){\circle{3}}
\put(105,120){\circle{3}}
\put(55,70){\circle*{3}}
\put(105,70){\circle*{3}}
\put(80,220){\line(1,-2){25}}
\put(80,220){\line(-1,-2){25}}
\put(55,169){\line(0,-1){48}}
\put(105,169){\line(0,-1){48}}
\put(56,169){\line(1,-1){48}}
\put(104,169){\line(-1,-1){48}}
\put(55,119){\line(0,-1){48}}
\put(105,119){\line(0,-1){48}}
\put(56,119){\line(1,-1){48}}
\put(104,119){\line(-1,-1){48}}
\put(53,55){$\vdots$}
\put(103,55){$\vdots$}
\put(53,40){$\vdots$}
\put(103,40){$\vdots$}
\put(84,220){\tiny $|\Delta_{p_0,q_0}>$}
\put(30,170){\tiny $\Psi^+_{p_{0},q_{0}}$}
\put(107,170){\tiny $\Psi^+_{\pt_{0},\qt_{0}}$}
\put(43,120){\tiny $\Psi^+_{a}$}
\put(107,120){\tiny $\Psi^+_{b}$}
\put(30,70){\tiny $\Psi^+_{p_{1},q_{1}}$}
\put(107,70){\tiny $\Psi^+_{\pt_{1},\qt_{1}}$}
\put(80,5){\case{$\cs_{o-}$}}
\put(148,220){\framebox(4,3){}}
\put(150,170){\circle*{3}}
\put(150,120){\circle*{3}}
\put(150,70){\circle*{3}}
\put(148,55){$\vdots$}
\put(148,40){$\vdots$}
\put(150,170){\line(0,-1){50}}
\put(150,120){\line(0,-1){50}}
\put(150,220){\line(0,-1){50}}
\put(154,220){\tiny $|\Delta_{p_0,q_0}>$}
\put(152,170){\tiny $\Psi^+_{\pt_{0},\qt_{0}}$}
\put(152,120){\tiny $\Psi^+_{p_0,q_0}$}
\put(152,70){\tiny $\Psi^+_{\pt_{1},\qt_{1}}$}
\put(150,5){\case{$\cs_{o-}^{0(u)}$}}
\put(198,220){\framebox(4,3){}}
\put(200,170){\circle*{3}}
\put(200,120){\circle*{3}}
\put(200,70){\circle*{3}}
\put(198,55){$\vdots$}
\put(198,40){$\vdots$}
\put(200,170){\line(0,-1){50}}
\put(200,120){\line(0,-1){50}}
\put(200,220){\line(0,-1){50}}
\put(204,220){\tiny $|\Delta_{p_0,q_0}>$}
\put(202,170){\tiny $\Psi^+_{p_{0},q_{0}}$}
\put(202,120){\tiny $\Psi^+_{\pt_0,\qt_0}$}
\put(202,70){\tiny $\Psi^+_{p_{1},q_{1}}$}
\put(200,5){\case{$\cs_{o-}^{0(v)}$}}
\put(248,220){\framebox(4,3){}}
\put(250,170){\circle*{3}}
\put(250,120){\circle*{3}}
\put(250,70){\circle*{3}}
\put(248,55){$\vdots$}
\put(248,40){$\vdots$}
\put(250,170){\line(0,-1){50}}
\put(250,120){\line(0,-1){50}}
\put(250,220){\line(0,-1){50}}
\put(254,220){\tiny $|\Delta_{p_0,q_0}>$}
\put(252,170){\tiny $\Psi^+_{p_0,q_0}$}
\put(252,120){\tiny $\Psi^+_{p_1,q_1}$}
\put(252,70){\tiny $\Psi^+_{p_{2},q_{2}}$}
\put(250,5){\case{$\cs_{o-}^{00}$}}
\put(328,220){\framebox(4,3){}}
\put(305,170){\circle*{3}}
\put(355,170){\circle*{3}}
\put(305,120){\circle{3}}
\put(355,120){\circle{3}}
\put(305,70){\circle{3}}
\put(355,70){\circle{3}}
\put(330,20){\circle*{3}}
\put(330,220){\line(1,-2){25}}
\put(330,220){\line(-1,-2){25}}
\put(305,169){\line(0,-1){48}}
\put(355,169){\line(0,-1){48}}
\put(306,169){\line(1,-1){48}}
\put(354,169){\line(-1,-1){48}}
\put(330,21){\line(1,2){24}}
\put(330,21){\line(-1,2){24}}
\put(303,105){$\vdots$}
\put(353,105){$\vdots$}
\put(303,80){$\vdots$}
\put(353,80){$\vdots$}
\put(330,5){\case{$\cs_{o+}$}}
\put(398,220){\framebox(4,3){}}
\put(400,170){\circle*{3}}
\put(400,120){\circle*{3}}
\put(400,70){\circle*{3}}
\put(400,20){\circle*{3}}
\put(398,105){$\vdots$}
\put(398,80){$\vdots$}
\put(400,220){\line(0,-1){50}}
\put(400,170){\line(0,-1){50}}
\put(400,70){\line(0,-1){50}}
\put(400,5){\case{$\cs_{o+}^{0(u)}$, $\cs_{o+}^{0(v)}$, $\cs_{o+}^{00}$}}
\label{pic:virembed}
\epic{$N=1$ Ramond embedding diagrams $\cs_{o}$, positive parity sector.}
\eth

In the following we analyse in the same way the cases $\cs_{e-}$
which means that either $u$ or $v$ (but not both) is odd. In this case not every integral
point pair on the straight line $\cL$ leads to a singular vector because the 
addition of $u$ and $v$ to $q_0$ and $p_0$ respectively changes $p-q$ 
from odd to even. Hence, only every other integral point pair on the straight line 
$\cL$ satisfies the condition $p-q$ odd. At first, this seems to complicate
our analysis. However, it turns out that the only difference is that $v$ and
$u$ need to be replaced by $2v$ and $2u$ respectively. Analysing 
descending solutions of $\Psi_{p_0,q_0}$ and $\Psi_{\pt_0,\qt_0}$
we obtain the following figure.

\vbox{
\bpic{450}{200}{-200}{-100}
\put(-200,0){\vector(1,0){400}}
\put(0,-100){\vector(0,1){200}}
\put(195,-10){\scriptsize $\ph$}
\put(-10,90){\scriptsize $\qh$}
\put(155,100){\line(-5,-2){355}}
\put(-195,-100){\line(5,2){395}}
\put(-145,-100){\line(5,2){340}}
\put(-120,-90){\circle*{2}}
\put(-70,-70){\circle*{2}}
\put(-20,-50){\circle*{4}}
\put(30,-30){\circle*{2}}
\put(80,-10){\circle*{2}}
\put(130,10){\circle*{2}}
\put(180,30){\circle*{4}}
\put(-170,-90){\circle*{2}}
\put(-120,-70){\circle*{2}}
\put(-70,-50){\circle*{2}}
\put(-20,-30){\circle*{4}}
\put(30,-10){\circle*{2}}
\put(80,10){\circle*{4}}
\put(130,30){\circle*{2}}
\put(180,50){\circle*{2}}
\put(-170,-30){\circle*{2}}
\put(-120,-10){\circle*{4}}
\put(-70,10){\circle*{2}}
\put(-20,30){\circle*{2}}
\put(30,50){\circle*{2}}
\put(80,70){\circle*{4}}
\put(130,90){\circle*{2}}
\put(2,-45){\tiny $-(a+2q_{0})$}
\put(2,-25){\tiny $-a$}
\put(-36,40){\tiny $-(a-2\qt_{0})$}
\put(85,7){\tiny $(p_{0},q_{0})$}
\put(-36,-41){\tiny $(-\pt_{0},-\qt_{0})$}
\put(165,19){\tiny $(p_{0}+2v,-q_{0}+2u)$}
\put(150,13){\tiny or $(p_0+4v,2u)$ for $q_0=2u$}
\put(-35,-60){\tiny $(-\pt_{0},-\qt_0-2q_0)$}
\put(85,65){\tiny $(-\pt_0+2v,\qt_0+2u)$}
\put(85,59){\tiny or $(2v,\qt_0+4u)$ for $\pt_0=2v$}
\put(-115,-15){\tiny $(-p_0-2\pt_0,-q_0)$}
\put(80,7){\vector(0,-1){14}}
\put(-20,-27){\vector(0,1){54}}
\put(-32,18){\tiny $2\qt_{0}$}
\put(64,-8){\tiny $2q_{0}$}
\label{fig:IIImlinese}
\epic{$\cs_{e-}$ solutions of $\qh = \ph t -a$ at low levels.}
}

It is worth noting that in the cases $\cs_{e}$ the singular vectors of one embedding diagram
are always of the same type whilst for $\cs_{o}$ they can be of different type.
As before, we can easily define the descendant singular vectors of first and second order
analogous to the \eqs{\ref{eq:desc1}}-\eqoth{\ref{eq:desc8}} simply by replacing $v$ and
$u$ by $2v$ and $2u$ respectively everywhere in the index of the vectors.
In this way we obtain for example the singular vectors
\bea
\Psi^+_{\hat{a}} &=& \theta_{p_0+2v,2u-q_0}^+ \theta_{p_0,q_0}^+ \ket{\Delta_{p_0,q_0}} \com \\
\Psi^+_{\hat{b}} &=& \theta_{\tilde{p}_0,\tilde{q}_0+2q_0}^+ \theta_{p_0,q_0}^+ \ket{\Delta_{p_0,q_0}} \pkt
\eea 
We can again use the multiplication rules of \refoth{\ref{th:multab}} to obtain exactly the same
proportionalities as above. The main difference, however, is that the embedding pattern
does not collapse for $q_0=u$ and neither for $\pt_0=v$ as it does for $\cs_{o\pm}$ and also
for the Virasoro case. Nevertheless, the embedding pattern does simplify to the single
line pattern for $q_0=2u$ and for $\pt_0=2v$. For example, in the former case
the vector
\bea
\Psi^+_{\hat{a}^{\prime}} &=& \theta_{p_0+4v,2u}^+ \theta_{p_0,q_0}^+ \ket{\Delta_{p_0,q_0}} \com
\eea  
becomes proportional
to $\Psi^+_{p_2,q_2}$ and $\Psi^+_{\hat{d}}$ is proportional to $\Psi_{p_0,q_0}^+$.
We shall denote these special cases by $\cs_{e-}^{0(2u)}$ and $\cs_{e-}^{0(2v)}$ correspondingly.
Let us note that $q_0=2u$ and $\pt_0=2v$ results
in $\frac{2u-\qt_0}{2v-p_0}=\frac{u}{v}$
and therefore either $\qt_0$ or $p_0$ must be even. This is related to the fact that
the singular vector solutions for $\cs_{e\pm}$ are always of the same
type. But since $q_0$ and
$\pt_0$ are both
even this leads to a contradiction. Therefore, cases of the type $\cs_{e\pm}^{00(2u)(2v)}$
do not exist.
We have therefore proven the results contained in the following theorem.

\bth \label{th:emb2}
For the subsets $\cs_{e\pm}$, $\cs_{e\pm}^{0(2u)}$, and $\cs_{e\pm}^{0(2v)}$
of rational models 
($\Delta\not = \frac{c}{24}$)
we find the following embedding patterns for descendant singular 
vectors of the type $\Psi^+_{p,q}$ in the positive parity sector. The negative parity sector
leads to exactly the same diagrams. The zero mode $G_0$ interpolates between
the embedding diagrams of the positive parity sector and the negative parity sector.
\bpic{400}{230}{30}{-5}
\put(78,220){\framebox(4,3){}}
\put(55,170){\circle*{3}}
\put(105,170){\circle*{3}}
\put(55,120){\circle{3}}
\put(105,120){\circle{3}}
\put(55,70){\circle*{3}}
\put(105,70){\circle*{3}}
\put(80,220){\line(1,-2){25}}
\put(80,220){\line(-1,-2){25}}
\put(55,169){\line(0,-1){48}}
\put(105,169){\line(0,-1){48}}
\put(56,169){\line(1,-1){48}}
\put(104,169){\line(-1,-1){48}}
\put(55,119){\line(0,-1){48}}
\put(105,119){\line(0,-1){48}}
\put(56,119){\line(1,-1){48}}
\put(104,119){\line(-1,-1){48}}
\put(53,55){$\vdots$}
\put(103,55){$\vdots$}
\put(53,40){$\vdots$}
\put(103,40){$\vdots$}
\put(84,220){\tiny $|\Delta_{p_0,q_0}>$}
\put(30,170){\tiny $\Psi_{p_{0},q_{0}}$}
\put(107,170){\tiny $\Psi^+_{\pt_{0},\qt_{0}}$}
\put(43,120){\tiny $\Psi^+_{\hat{a}}$}
\put(107,120){\tiny $\Psi^+_{\hat{b}}$}
\put(30,70){\tiny $\Psi^+_{p_{2},q_{2}}$}
\put(107,70){\tiny $\Psi^+_{\pt_{2},\qt_{2}}$}
\put(80,5){\case{$\cs_{o-}$}}
\put(148,220){\framebox(4,3){}}
\put(150,170){\circle*{3}}
\put(150,120){\circle*{3}}
\put(150,70){\circle*{3}}
\put(148,55){$\vdots$}
\put(148,40){$\vdots$}
\put(150,170){\line(0,-1){50}}
\put(150,120){\line(0,-1){50}}
\put(150,220){\line(0,-1){50}}
\put(154,220){\tiny $|\Delta_{p_0,q_0}>$}
\put(152,170){\tiny $\Psi^+_{\pt_{0},\qt_{0}}$}
\put(152,120){\tiny $\Psi^+_{p_0,q_0}$}
\put(152,70){\tiny $\Psi^+_{\pt_{2},\qt_{2}}$}
\put(150,5){\case{$\cs_{o-}^{0(2u)}$}}
\put(198,220){\framebox(4,3){}}
\put(200,170){\circle*{3}}
\put(200,120){\circle*{3}}
\put(200,70){\circle*{3}}
\put(198,55){$\vdots$}
\put(198,40){$\vdots$}
\put(200,170){\line(0,-1){50}}
\put(200,120){\line(0,-1){50}}
\put(200,220){\line(0,-1){50}}
\put(204,220){\tiny $|\Delta_{p_0,q_0}>$}
\put(202,170){\tiny $\Psi^+_{p_{0},q_{0}}$}
\put(202,120){\tiny $\Psi^+_{\pt_0,\qt_0}$}
\put(202,70){\tiny $\Psi^+_{p_{2},q_{2}}$}
\put(200,5){\case{$\cs_{o-}^{0(2v)}$}}
\put(328,220){\framebox(4,3){}}
\put(305,170){\circle*{3}}
\put(355,170){\circle*{3}}
\put(305,120){\circle{3}}
\put(355,120){\circle{3}}
\put(305,70){\circle{3}}
\put(355,70){\circle{3}}
\put(330,20){\circle*{3}}
\put(330,220){\line(1,-2){25}}
\put(330,220){\line(-1,-2){25}}
\put(305,169){\line(0,-1){48}}
\put(355,169){\line(0,-1){48}}
\put(306,169){\line(1,-1){48}}
\put(354,169){\line(-1,-1){48}}
\put(330,21){\line(1,2){24}}
\put(330,21){\line(-1,2){24}}
\put(303,105){$\vdots$}
\put(353,105){$\vdots$}
\put(303,80){$\vdots$}
\put(353,80){$\vdots$}
\put(330,5){\case{$\cs_{o+}$}}
\put(398,220){\framebox(4,3){}}
\put(400,170){\circle*{3}}
\put(400,120){\circle*{3}}
\put(400,70){\circle*{3}}
\put(400,20){\circle*{3}}
\put(398,105){$\vdots$}
\put(398,80){$\vdots$}
\put(400,220){\line(0,-1){50}}
\put(400,170){\line(0,-1){50}}
\put(400,70){\line(0,-1){50}}
\put(400,5){\case{$\cs_{o+}^{0(2u)}$, $\cs_{o+}^{0(2v)}$}}
\label{pic:virembede}
\epic{$N=1$ Ramond embedding diagrams $\cs_{e}$, positive parity sector.}
\eth

As explained, the main difference to the embedding patterns of the Virasoro case
is that the cases $\cs_{e\pm}$ do not simplify to the single line embedding patterns
for $q_0=u$ or $\pt_0=v$. By grouping these cases together with the types
$\cs_{e\pm}$, we have defined our notation $\cs_{e\pm}^{0}$ in such a way
that the resulting embedding patterns seem similar to the Virasoro case.
Let us now analyse some of these special cases further.  $q_0=u$ and $\pt_0=v$
easily implies $(a\frac{v}{u}+v,u)$ and $(-v,-a-u)$ as the two lowest level solutions.
The corresponding singular vectors are hence both at the level $av+uv$.
Conversely, $p_0q_0=\pt_0\qt_0$ implies that the two intercepts $-a$ and $\frac{a}{t}$
are both integers. Therefore, we have only two possibilities for each of
the points $(p_0,q_0)$ and $(\pt_0,\qt_0)$ which for the former are for example
$(a\frac{v}{u}+v,u)$ or $(a\frac{v}{u}+2v,2u)$. However, equality of the levels is only possible
for  $q_0=u$ and $\pt_0=v$. Therefore the subset of these special cases which satisfy both
$q_0=u$ and $\pt_0=v$ correspond exactly to the cases where two curves of primitive singular
vectors intersect at the same level. The values of $t$ for these cases were given in \eq{\ref{eq:intsec}}
as $t=\pm\frac{q_1}{p_2}=\pm\frac{q_2}{p_1}$. If the two intersecting primitive singular
vectors are of different type $\epsilon$, then $u$ and $v$ would obviously
be both odd and therefore the embedding diagram would go over to the straight line
embedding pattern in accordance with our earlier result that intersections of
singular vectors of different type do not lead to degenerate singular states.
However, if the two intersecting singular vectors are of the same type then consequently
$t=\pm\frac{q_1}{p_2}=\pm\frac{q_2}{p_1}$ implies that $u$ and $v$ cannot both be odd
and therefore we obtain case $\cs_{e\mp}$. The corresponding embedding diagram
does not simplify to the straight line pattern as in the Virasoro case, which clearly shows 
the degenerate singular vectors.

\bth \label{th:IIIspec}
The Verma modules with degenerate singular vectors given by the intersection of 2 curves of primitive
singular vectors of the same type at the same level are exactly the cases $\cs_{e\pm}$ with
$q_0=u$ and $\pt_0=v$. The corresponding embedding diagrams do not simplify to a straight line pattern but 
stay as the crossed embedding pattern given in figure \refoth{\ref{pic:virembede}}.
\eth

The embedding diagrams given above are equally valid for the positive parity sector and the negative
parity sector with the only exception of the special case $\Delta=\frac{c}{24}$. As we will see in
the following section, for $\Delta=\frac{c}{24}$ the singular vector operators of the primitive singular
vectors $\Psi_{p,q}^{\pm}$ always annihilate $G_0\ket{\frac{c}{24}}$. Therefore, these modules do not contain
two parallel embedding structures of singular vectors of the crossed type, but only one. However, as we shall
also show in the following section, we then obtain subsingular vectors in $\vm_{\frac{c}{24}}$. 
Therefore one part of
the parallel embedding structure is not given by
singular vectors but by subsingular vectors. Let us also mention that the operator $G_0$
always interpolates between the two corresponding vectors of the parallel embedding structures
except for the cases when some of the singular vectors become $G$-closed. The conditions
for $G$-closed singular vectors will be analysed in section \refoth{\ref{sec:Gclosedsvecs}}. A complete set
of Ramond embedding patterns with the interpolation of the two parity sectors will be given
in Ref. \icite{progress}

Let us finally identify the unitary minimal models
among the cases of our analysis. 
The conditions for the Ramond unitary minimal
models were first given by Friedan et al.\cite{friedan}. They
correspond to the central parameter $t=\frac{m+2}{m}$ with $m\in\bbbn$
but $m\not = 1$ and $\Delta=\Delta_{p,q}$ with the integer pair $(p,q)$, $0<p\leq m-1$,
$0<q\leq m+1$, $p-q$ odd. Obviously, the highest common factor of $m$
and $m+2$ is $2$ for $m$ even and $1$ for $m$ odd. For $m$ even we
therefore have $u=\frac{m}{2}+1$ and $v=\frac{m}{2}$ whilst for $m$
odd we obtain $u=m+2$ and $v=m$. Consequently, we find for $m$
even that the solutions $(p_0,q_0)$ and $(\pt_0,\qt_0)$
satisfy the conditions $0<p_0,\pt_0\leq m-1<2v$,
$0<q_0,\qt_0\leq m+1<2u$ and for $m$ odd $0<p_0,\pt_0\leq m-1<v$,
$0<q_0,\qt_0\leq m+1<u$. Therefore, the unitary minimal models 
for $\Delta\not =\frac{c}{24}$ ($a\not = 0$) are of
type $\cs_{e-}$ for $m$ even and of type $\cs_{o-}$ for $m$ odd. 
The corresponding embedding diagrams are always of the crossed type.
This finally proves for $\Delta\not = \frac{c}{24}$
the only $N=1$ Ramond embedding diagrams 
that had been conjectured in the
literature so far\cite{kiritsis2} for the unitary minimal models.
Note that none of these cases has degenerate singular
levels since $t=\frac{q_0}{p_0}=\frac{q_0}{\pt_0}=\frac{\qt_0}{p_0}$
implies $p_0=\pt_0$ and $q_0=\qt_0$ and thus $a=0$.
Therefore, we should stress that the conditions of the
critical cases of $\cs_{e-}$ of
theorem \refoth{\ref{th:IIIspec}} lead for the unitary minimal models
to the excluded case $a=0$. Hence, embedding diagrams of the unitary
minimal supersymmetric cases, i.e. with $\Delta=\frac{c}{24}$ 
do not have the embedding diagrams 
of singular vectors given
in the literature\cite{kiritsis2}. Their embedding patterns change and
in addition
we need to incorporate subsingular vectors
as described in the following section.
We summarise these results in the following theorem.

\bth
The unitary minimal models with $\Delta \not =\frac{c}{24}$ ($a\not =0$) are of type
$\cs_{o-}$ for $m$ odd and of type $\cs_{e-}$ for $m$ even.
\eth


\section{The supersymmetric case and subsingular vectors.}
\label{sec:GclosedVM}

The supersymmetric case $\Delta= \frac{c}{24}$ corresponds to $a=0$ which we
had to exclude in our analysis in the previous section.
In theorem \refoth{\ref{th:supersym}} we have explained that 
for the supersymmetric case $\Delta=\frac{c}{24}$ the 
structure of the Verma module $\vm_{\Delta}$ changes compared to other
values of $\Delta$. In this case $\vm_{\frac{c}{24}}$ contains just
one (unextended) submodule $\vm^{\times}_{0}$, which in the language of
the extended algebra can be written as the $G$-closed Verma module
$\vm_{\frac{c}{24}}^G=\vm^{\times}_{0}$. Note that $G_0$ can be
diagonalised on the whole $G$-closed module $\vm_{\frac{c}{24}}^G$.
Besides $\vm_{\frac{c}{24}}^G$, the module $\vm_{\frac{c}{24}}$
contains vectors outside $\vm_{\frac{c}{24}}^G$ that are not contained
in any (unextended) Verma module built on a $G_0$-eigenvector.
The submdoule $\vm_{\frac{c}{24}}^G$ is generated by the singular
vector $\Psi^-_{0} = G_0 \ket{\frac{c}{24}}$ at level $0$ with parity
$-$. The quotient module
\bea
\vm_{\frac{c}{24}}^G &=& \frac{\vm_{\frac{c}{24}}}{U(\ram)\Psi^-_{0}} \com
\eea
is again the $G$-closed Verma module $\vm_{\frac{c}{24}}^G$. 
This process of constructing the quotient module shall be considered
in more detail in this section. At first, it seems that 
this quotient module is only one of many quotient modules
one may want to consider. We could also consider quotients
with respect to modules generated by singular vectors $\Psi_{p,q}^{\pm}$ 
at levels $\frac{pq}{2}$. However, the supersymmetric quotient module
is special as we shall explain at the end of this section.

In a forthcoming publication\cite{p11vanish} it will be shown that the
existence of singular vector operators with vanishing product leads to
the appearance of subsingular vectors. In the supersymmetric case
we have
$G_0^2\ket{\frac{c}{24}}\equiv 0$ and therefore we expect the
possibility of having subsingular vectors in $\vm_{\frac{c}{24}}$.
Via the canonical map defined for quotient spaces, singular vectors from
$\vm_{\frac{c}{24}}$ are either also singular in $\vm_{\frac{c}{24}}^G$ or
they lie in the kernel of the quotient and hence vanish in the
quotient space. The latter happens if and only if the 
singular vector in question is
a descendant vector of the singular vector $\Psi_{0}^-$ in 
$\vm_{\frac{c}{24}}$.
Thus, singular vectors of $\vm_{\frac{c}{24}}^G$ can either be 
inherited from
$\vm_{\frac{c}{24}}$ via the canonical quotient map or they 
appear for the first time in the quotient
$\vm_{\frac{c}{24}}^G$ as vectors that were not singular in 
$\vm_{\frac{c}{24}}$.
In the latter case these singular vectors of 
$\vm_{\frac{c}{24}}^G$ are called subsingular vectors
in $\vm_{\frac{c}{24}}$. Whilst for the Virasoro algebra\cite{ff1}
and for the $N=1$ Neveu-Schwarz algebra\cite{ast1} there are no
subsingular vectors, they have been found for all $N=2$ superconformal
algebras by Gato-Rivera et al.\cite{beatriz1,beatriz2,p9sdim2}.
The fact that the $N=1$ Ramond algebra also contains subsingular
vectors - at least in the supersymmetric case - shall be demonstrated here.
This may at first seem surprising and confirms the obvious problems
encountered so far with $N=1$ Ramond theories. However, we 
shall argue that
subsingular vectors are likely to be
restricted to the (extended) supersymmetric case which is though a very
particular and highly interesting case.

Singular dimensions for $G$-closed Verma modules are less
than or equal to $1$ as shown in section \refoth{\ref{sec:sdim}}.
Hence, there cannot be any degenerate singular vectors in
$G$-closed Verma modules and furthermore, two singular vectors
in a $G$-closed Verma module at the same level and
with the same parity must be proportional.
A singular vector of a $G$-closed Verma module
at level $l$ can therefore
be identified by one coefficient only, which is
the coefficient with respect to $L_{-1}^{l}$ or
$L_{-1}^{l-1}G_{-1}$ depending on the parity.
\bth \label{th:Gvanish}
Two singular vectors in the $G$-closed Verma module 
$\vm_{\frac{c}{24}}^G$ at the same 
level and with the same parity are always proportional. 
If a vector satisfies the highest weight
conditions at level $l$, with parity $+$ or parity $-$, 
and its coefficient for the term
$L_{-1}^{l}$ or $L_{-1}^{l-1}G_{-1}$ 
vanishes, respectively, when written in the normal form,
then this vector is trivial.
\eth

Besides the singular vector $\psi_0^-=G_0\ket{\frac{c}{24}}$
the Verma module $\vm_{\frac{c}{24}}$ contains
also the singular vectors $\Psi^{\pm}_{p,q}$ at level
$\frac{pq}{2}$ whenever $\Delta_{p,q}(t)=\frac{c(t)}{24}$.
The latter equation has the solutions
\bea
t_{p,q} &=& \frac{q}{p} \;\;\; \com p,q\in\bbbn \com p-q \;{\rm odd} \pkt
\eea
For these values of $t$ the corresponding singular vectors
are given by $\Psi_{p,q}^+(t_{p,q})=(0,1)_{\frac{pq}{2}}^+$ and
by $\Psi_{p,q}^-(t_{p,q})=(1,0)_{\frac{pq}{2}}^-$. In the quotient module
$\vm_{\frac{c}{24}}^G = \frac{\vm_{\frac{c}{24}}}{U(\ram)\Psi^-_{0}}$
which is isomorphic to the $G$-closed module $\vm_{\frac{c}{24}}^G$
these singular vectors hence vanish due to theorem
\refoth{\ref{th:Gvanish}}. Therefore, the singular vectors
$\Psi_{p,q}^+(t_{p,q})$ and $\Psi_{p,q}^-(t_{p,q})$ are both descendants
of $\Psi_0^-$. Hence, these vectors can be written
as $\Psi_{p,q}^+(t_{p,q})=\tilde{\theta}^-_{p,q}G_0\ket{\frac{c}{24}}$ and
$\Psi_{p,q}^-(t_{p,q})=\tilde{\theta}^+_{p,q}G_0\ket{\frac{c}{24}}$
with singular vector operators $\tilde{\theta}^{\mp}_{p,q}G_0$
where $\tilde{\theta}^{\mp}_{p,q}$ does not contain any operators $G_0$. 
We can now construct the vectors 
$\Upsilon_{p,q}^{\pm}=
\tilde{\theta}^{\pm}_{p,q}\ket{\frac{c}{24}}$. Clearly, the
vectors $\Upsilon_{p,q}^{\pm}(\frac{q}{p})$ 
are singular in the quotient module 
$\frac{\vm_{\frac{c}{24}}}{U(\ram)G_0\ket{\frac{c}{24}}}$.
In principle $\Upsilon_{p,q}^{\pm}(\frac{q}{p})$ 
could also be singular in 
$\vm_{\frac{c}{24}}$, a simple consideration, however, shows that this
is not the case. Theorem \refoth{\ref{th:Gvanish}} shows that
$\Upsilon_{p,q}^{\pm}(\frac{q}{p})$ contains the non-trivial term
$L_{-1}^{\frac{pq}{2}}\ket{\frac{c}{24}}$ or
$L_{-1}^{\frac{pq}{2}-1}G_{-1}\ket{\frac{c}{24}}$
 depending
on its parity. Terms 
like $L_{-1}^{\frac{pq}{2}-2}G_{-2}G_0\ket{\frac{c}{24}}$ and
$L_{-1}^{\frac{pq}{2}-3}G_{-2}G_{-1}G_0\ket{\frac{c}{24}}$ 
are not contained in
$\Upsilon_{p,q}^{\pm}(\frac{q}{p})$ by definition of 
$\tilde{\theta}^{\pm}_{p,q}$. Therefore, the action of $G_1$ on
$L_{-1}^{\frac{pq}{2}}\ket{\frac{c}{24}}$ creates a non-trivial term 
$L_{-1}^{\frac{pq}{2}-1}G_0\ket{\frac{c}{24}}$ which obviously
cannot be created by any other non-trivial term of 
$\Upsilon_{p,q}^{+}(\frac{q}{p})$. Hence, 
$G_1\Upsilon_{p,q}^{+}(\frac{q}{p})$ is non-trivial in
$\vm_{\frac{c}{24}}$. Similar considerations are valid for
$G_1\Upsilon_{p,q}^{-}(\frac{q}{p})$. Therefore, the vectors  
$\Upsilon_{p,q}^{\pm}(t_{p,q})$ are both subsingular vectors in 
$\vm_{\frac{c}{24}}$ for $t_{p,q}=\frac{q}{p}$. We summarise our results
in the following theorem.

\bth \label{th:subsvecs}
For $t_{p,q}=\frac{q}{p}$ the singular vectors $\Psi^{\pm}_{p,q}(t)$ are
contained in $\vm_{\frac{c}{24}}$ and are descendants of the singular vector
$\Psi_0^-=G_0\ket{\frac{c}{24}}$. They can therefore be written as 
$\Psi^{\pm}_{p,q}(t_{p,q})=\tilde{\theta}^{\mp}_{p,q}G_0\ket{\frac{c}{24}}$
with $\tilde{\theta}^{\mp}_{p,q}\in U(\ram^-)$. The vectors
\bea
\Upsilon_{p,q}^{\pm} &=&
\tilde{\theta}^{\pm}_{p,q}\ket{\frac{c}{24}}
\eea
are then subsingular vectors in $\vm_{\frac{c}{24}}$ for $t_{p,q}$
at level $\frac{pq}{2}$
becoming singular in the quotient module
$\vm_{\frac{c}{24}}^G = \frac{\vm_{\frac{c}{24}}}{U(\ram)\Psi^-_{0}}$.
\eth

Theorem \refoth{\ref{th:subsvecs}} does not come as a surprise
considering the fact that $U(\ram)\Psi_0^-$ and
$\vm_{\frac{c}{24}}^G = \frac{\vm_{\frac{c}{24}}}{U(\ram)\Psi^-_{0}}$
are in fact both isomorphic to the $G$-closed Verma module
$\vm_{\frac{c}{24}}^G$ and therefore show the same structure. Singular
vectors appear in the quotient therefore at exactly the grades where
singular vectors disappear in the kernel of the quotient. 
Explicit examples of the subsingular vectors $\Upsilon_{p,q}^{\pm}$
are given in appendix \refoth{\ref{app:b}}. The symmetry of
$\Psi_{p,q}^{\pm}(t)$ and $\Psi_{q,p}^{\pm}(\frac{1}{t})$ leads to
the proportionality of the operators
$\tilde{\theta}^{\pm}_{p,q}(\frac{q}{p})$ and
$\tilde{\theta}^{\pm}_{q,p}(\frac{p}{q})$. Taking into account that
$c(t)=c(\frac{1}{t})$ consequently leads to $\Upsilon_{p,q}^{\pm}$
and $\Upsilon_{q,p}^{\pm}$ being proportional.

We have shown that the singular vectors $\Psi_{p,q}^{\pm}$ disappear
in the kernel of the quotient module for the $G$-closed case. However,
based on their singular vector operators we can construct subsingular
vectors in $\vm_{\frac{c}{24}}$ and thus the singular vectors in the
$G$-closed Verma module $\vm_{\frac{c}{24}}^G$. 
For the case that $\Psi_{p,q}^{\pm}$ leads to a two-dimensional
singular space one would most likely obtain a two-dimensional singular
space in the corresponding $G$-closed Verma module which would
contradict theorem \refoth{\ref{th:Gvanish}}. Therefore,
$2$-dimensional singular spaces defined by the intersection of
vectors $\Psi_{p,q}^{\pm}$ are not allowed for $t=t_{p,q}$. This can
easily be seen by requiring $t=\frac{q_1}{p_1}$ and
$t=\frac{q_1}{p_2}=\frac{q_2}{p_1}$. The first condition takes care
of the $G$-closed case and the second condition arises from the
intersection of two such vectors. This leads to $(p_1,q_1)=(p_2,q_2)$
and therefore excludes the existence of degenerate cases for the $G$-closed
Verma modules. Even more interestingly due to $p-q$ odd
exactly one of the coprime factors $(u,v)$ of $t=\frac{u}{v}$ has to be even.
Furthermore, since $a=0$ the straight line $\cL$ is point symmetric in the
origin and the negative solutions $(\pt_n,\qt_n)$ are exactly the same
as the positive solutions $(p_n,q_n)$. They are simply given by $(p_n,q_n)=
(v+nv,u+nu)$ with $n\in2\bbbn_0$. Since $(0,0)$ is on the straight line $\cL$ these are
the only possible reducible Verma modules having $a=0$. Thus, reducible $G$-closed 
Verma modules always belong to type $\cs_{e-}$. However, their structure is
quite different from the structure of the Verma modules of type $\cs_{e-}$ with 
$a\not =0$. Therefore, we shall use the notation $\cs_{e-}^{\frac{c}{24}}$. 
The reason why their structure is different to $\cs_{e-}$ is mainly due to the
fact that the positive and negative solutions are not only at the same level but they
are the same and therefore do not lead to degenerate singular vectors as we would
expect it from the cases $\cs_{e-}$. The degeneracy of null-vectors is for
$\cs_{e-}^{\frac{c}{24}}$ obtained through the subsingular vectors of theorem 
\refoth{\ref{th:subsvecs}}. 

In Ref. \icite{p11vanish}, it will be shown that for superconformal algebras 
the existence of subsingular vectors is often a consequence of
additional discrete vanishing conditions on an embedded singular
vector. To understand this, let us assume 
that we have given a singular vector $\Psi_l(t)$ at level $l$ along
the curve $\Delta=\Delta(t)$. This singular vector generates an
embedded Verma module built on $\Psi_l$. If for a discrete point $t_0$
the singular vector satisfies an additional vanishing condition
$\theta_{l_0}\Psi_l\equiv 0$ at level $l_0$, then the embedded Verma module
has to be smaller at this level compared to the generic cases along 
$\Delta=\Delta(t)$. Since null-vectors correspond to the kernel of the
inner product matrix and the dimension of this kernel is
upper semi-continuous, there must exist for this discrete case an
additional null-vector at level $l+l_0$ 
outside the embedded module built on $\Psi_l$
in order to compensate for this embedded module shrinking at level $l+l_0$.
This additional null-vector is usually a subsingular vector. As shown
for the $N=2$ superconformal algebras\footnote{See appendix \refoth{B}
of Ref. \icite{p8det}.} in some cases they may, however, 
also be ordinary singular vectors\cite{p8det,p11vanish}. 
This argument is explained in more detail in Ref. \icite{p11vanish},
where it is also shown that only singular vector operators need to be
considered in order to find possible vanishing conditions on singular
vectors. We therefore conclude this section by analysing the products
of singular vector operators and the modules and grades for which the
product of two singular vector operators may vanish in the Ramond case.

We consider products of the form $\Psi^{\mu}_{\frac{p_1q_1+p_2q_2}{2}}=
\theta_{p_1,q_1}^{\pm}
\theta_{p_2,q_2}^{\pm}\ket{\Delta_{p_2,q_2}}$ for
$\Delta_{p_2,q_2}+\frac{p_2q_2}{2}=\Delta_{p_1,q_1}$. There are four cases to be
considered depending on the parities of the two singular vector operators indicated
by
$\mu$ as $++$, $+-$, $-+$, or $--$ for the parities of  $\theta_{p_1,q_1}^{\pm}$
and $\theta_{p_2,q_2}^{\pm}$. We write $\theta_{p_i,q_i}^{\pm}=
(a_i,b_i)^{\pm}_{\frac{p_iq_i}{2}}$ in the notation of definition \refoth{\ref{def:ab}}
and use the product expressions of \refoth{\ref{th:multab}}. Requiring that one of
these product expressions
vanishes identically, which means that both coefficients with respect to
the ordering kernel vanish for the product, leads to the conditions
$a_1=0$ and $a_2+2b_2=0$ or $a_2=0$ and $a_1+2b_1=0$ for the case
of $\mu=++$. One easily obtains similar 
relations for the other cases of $\mu$. Using the actual coefficients for the singular vector operators
given in \refoth{\ref{cj:ab}} we again encounter four subcases for each value
of $\mu$ depending on the type $\epsilon_i$ of the singular
vector operators. The resulting equations need to be solved for $t$. 
In the case of $\mu=++$ we easily find that the first condition
given above ($a_1=0$) leads to $t=\frac{q_1}{p_1}>0$, whilst the second
($a_2+2b_2=0$) results in $t=-\frac{q_2}{p_2}<0$ independent of the types 
$\epsilon_i$. The second set of conditions for $\mu=++$
leads to a similar contradiction on $t$. Hence, there are no vanishing products
of singular vector operators in the case $\mu=++$.
We determine the solutions to all other cases of $\mu$ in exactly the same
way. We easily obtain the following result.
\bth \label{th:van1}
The vector  $\Psi^{\mu}_{\frac{p_1q_1+p_2q_2}{2}}=
\theta_{p_1,q_1}^{\pm}
\theta_{p_2,q_2}^{\pm}\ket{\Delta_{p_2,q_2}}$ for
$\Delta_{p_2,q_2}+\frac{p_2q_2}{2}=\Delta_{p_1,q_1}$, $\mu\in\{++,+-,-+,--\}$
is always non-trivial.
\eth
Finally, we can also allow one of the singular vector operators $\theta_i$
to be the operator $G_0$ which is a singular vector operator for 
$\Delta_i=\frac{c}{24}$. Inserting $\theta_i=(1,0)^-_0$ in the conditions above
leads to the following cases for vanishing products.
\bth \label{th:van2}
The vector  $\Psi^{\mu}_{l_1+l_2}=
\theta_{1}^{\pm}
\theta_{2}^{\pm}\ket{\Delta_{2}}$ for
$\Delta_{2}+l_2=\Delta_{1}$, $\mu\in\{++,+-,-+,--\}$, one of $\theta_i^{\pm}$ being
of the form $\theta_{p,q}^{\pm}$ and the other one equals $G_0$,
is trivial if and only if
\bea
\theta^{\pm}_1=\theta_{p,q}^{\pm} \com \theta^{-}_2=G_0 \com \Delta_2=\frac{c}{24} \com 
{\rm and} \;\; t=\frac{q}{p} \label{eq:van1} \com
\eea
or
\bea
\theta^{-}_1=G_0 \com \theta_2^{\pm}=\theta_{p,q}^{\pm} \com \Delta_2+\frac{pq}{2}=\frac{c}{24} \com 
{\rm and} \;\; t=-\frac{q}{p} \label{eq:van2} \pkt
\eea
\eth
Obviously the cases of \eq{\ref{eq:van1}} correspond exactly to the supersymmetric cases $\Delta_2=
\frac{c}{24}$.
Hence, in this case the operator $\theta_{p,q}^{\pm}$ vanishes on $\Psi_0^-$. 
As expected, these are exactly
the cases for which we have found the subsingular vectors given above
in this section. Our earlier explanation that an additional vanishing condition on $\Psi_0^-$ is responsible for
these subsingular vectors does, however, not apply
in this case since the vanishing of $\theta_{p,q}^{\pm}\Psi_0^-$
is not an additional vanishing condition on $\Psi_0^-$ but arises simply from $G_0 \Psi_0^-
\equiv 0$ and the fact that $\theta_{p,q}^{\pm}$ can be written as $\tilde{\theta}_{p,q}^{\mp}G_0$ 
as shown earlier. Nevertheless, exactly these cases contain
subsingular vectors. The other cases in theorem \refoth{\ref{th:van2}} are 
cases when vectors $\Psi_{p,q}^{\pm}$ become $G$-closed and as we shall
show in the following section that these are all such cases. 
In this context, one automatically
raises the question whether there are subsingular vectors appearing for the case that 
$\Psi_{p,q}^{\pm}$ becomes $G$-closed. This interesting question will be analysed in the
following section. Finally, let us remark that theorem \refoth{\ref{th:van1}} states that besides
the $G$-closed cases and the supersymmetric cases there are no trivial products
of singular vector operators given by \refoth{\ref{cj:ab}}. However, if these
singular vector operators form a 2-dimensional space it could well be that a linear
combination of two such operators may lead to a trivial product by acting on another singular
vector. By noting that the operators $(a,b)_l^{+}$ are elements of a (larger) algebra
with operations \refoth{\ref{th:multab}} 
in which all operators are invertible except for the $G$-closed and 
supersymmetric cases, we find that such vanishing products are only possible
for these cases. We therefore do 
not have any further subsingular vectors arising in the way described above.
Furthermore we can state that there are no subsingular vectors in Verma modules $\vm_{\Delta}$
other than the ones found in $\vm_{\frac{c}{24}}$. A rigorous proof for this claim will be
presented in Ref. \icite{progress} using a detailed analysis of the multiplicities of
the roots of the determinant together with the results of theorem \refoth{\ref{th:van1}} and
\refoth{\ref{th:van2}}.

For the embedding structure of $\cs_{e-}^{\frac{c}{24}}$ we find the singular vectors corresponding
to $(p_n,q_n)$, $n\in\bbbn_0$ in a single line diagram as descendants of $\Psi_0^-$.
At first, it seems as if the descendant singular vectors of $\Psi_{p_0,q_0}^+$ would form
a two-dimensional singular space as the shifted line $\cL$ has the intercept
$a^{\prime}\not =0$ and is of type $\cs_{e-}$ with negative and positive solutions being at the
same levels. However, as explained above, one dimension of this 2-dimensional
space is being projected out in the product on $\Psi_{p_0,q_0}^+$ and the resulting
singular vector equals $\Psi_{p_2,q_2}^+$. This single line embedding pattern for the 
singular vectors is, however, complemented by the subsingular vectors and their 
descendants.  Again, $G_0$ interpolates between identical embedding patterns for
the different parity sectors.

To conclude this subsection we note that if we had started off with (unextended) Verma modules
$\vm_{\lambda}^{\times}$, then we would have not found these subsingular vectors
of the complete Verma module in the supersymmetric case $\lambda =0$ as
$\vm_{0}^{\times}$ is simply the above quotient module. On the other hand,
in our search for irreducible representations we might as well start off using
this quotient module rather than the full (extended) Verma module for
$\lambda =0$. Hence we have shown that for the irreducible highest weight
representations also in the supersymmetric case $\Delta=\frac{c}{24}$ the
highest weight vector can be chosen to be an eigenvector of $G_0$. We can
therefore state that for all irreducible highest weight representations of
the Ramond algebra $\ram$ we may choose highest weight vectors that
are eigenvectors of $G_0$ in both the extended and the unextended case. In the former
case we find the condition $(-1)^F\ket{\lambda}=\ket{-\lambda}$. However, $G_0$
may not be completely diagonalisable on the 
highest weight module for $\lambda-\frac{c}{24}\in\bbbn$. 


\section{$G$-closed singular vectors.}
\label{sec:Gclosedsvecs}

$G$-closed singular vectors $\Psi_l^{\pm G}$
in $\vm_{\Delta}$ can only appear at levels $l$ that
satisfy $\Delta+l=\frac{c}{24}$. The embedded Verma module
generated by $\Psi_l^{\pm G}$ is homomorphic to $\vm_{\frac{c}{24}}^G$.
This homomorphism is an isomorphism if there are no additional
vanishing conditions on $\Psi_l^{\pm G}$ except for $G_0\Psi_l^{\pm G}
\equiv 0$ and surely for all descendants of $G_0$. In theorem
\refoth{\ref{th:van2}} we have analysed vanishing conditions coming
from the singular vector operators $\theta_{p,q}^{\pm}$. We have found
$\theta_{p,q}^{\pm}G_0\ket{\frac{c}{24}}\equiv 0$ for $t=\frac{q}{p}$.
In these cases however $\theta_{p,q}^{\pm}$ can be written as 
$\tilde{\theta}_{p,q}^{\mp}G_0$ as shown in the previous section.
Therefore, these vanishing conditions simply arise from the
vanishing condition for the operator $G_0$.
The fact that we only need to look at singular vector operators and their
descendants whilst
dealing with additional vanishing conditions is proven in Ref. \icite{p11vanish}.
Hence, the embedded modules generated by $\Psi_{l}^{\pm G}$ are 
isomorphic\footnote{Strictly speaking, we would first need to prove that
we have really found all null-vectors which will become clear in our
work on the embedding diagrams\cite{progress}.} to $\vm_{\frac{c}{24}}$. 

We shall now analyse for which cases singular vectors become
$G$-closed. Starting with the singular vectors $\Psi_{p,q}^{\pm}(t)$ we find that they
can only become $G$-closed for $\Delta_{p,q}+\frac{pq}{2}=\frac{c}{24}$ which
leads to:
\bea
\tilde{t}_{p,q} &=& -\frac{q}{p} \pkt
\eea
Inserting $\tilde{t}_{p,q}$ into the expressions of \refoth{\ref{cj:ab}}
for $\Psi_{p,q}^{\pm}(t)$ we obtain $\Psi_{p,q}^+(\tilde{t}_{p,q})=
(2q,-q)_{\frac{pq}{2}}^+$ and $\Psi_{p,q}^-(\tilde{t}_{p,q})=
(\frac{8}{p},2q)_{\frac{pq}{2}}^-$. We use equations \refoth{\ref{th:Gprod}}
to act with $G_0$ on the vectors $\Psi_{p,q}^{\pm}(\tilde{t}_{p,q})$,
which leads for both parities to the zero vector $(0,0)_{\frac{pq}{2}}^{\pm}$.
Hence for all possible cases $t=\tilde{t}_{p,q}$ for which
$\Psi_{p,q}^{\pm}(t)$ may be $G$-closed it turns out that it really is.
Furthermore, the ordering kernels for $G$-closed singular vectors, 
given  in table
\tab{\ref{tab:adkern2}}, 
are just one-dimensional and therefore $G$-closed singular vectors are
defined by one coefficient only.

\bth \label{th:Gsvec}
The singular vectors $\Psi_{p,q}^{\pm}(t)$ always become $G$-closed
for $t=\tilde{t}_{p,q}=-\frac{q}{p}$ and these are the only cases such that
$\Delta_{p,q}+\frac{pq}{2}=\frac{c}{24}$.
$G$-closed singular vectors $\Psi_l^{\pm}$ are uniquely defined by fixing the coefficient
of $L_{-1}^l$ or $L_{-1}^{l}G_{-0}$ depending on the parity.
\eth

Because $t<0$ and $t=-\frac{q}{p}$ these cases belong to the type $\cs_{e+}$.
One easily finds that the straight line $\cL$ has only one relevant solution
$(p,q)$. The shifted line for descendant solutions goes through the origin
and has $t<0$ and therefore does not have any descendant solutions at all. 
Hence, the embedding diagram contains only the singular vectors $\Psi_{p,q}^{\pm}$
unconnected as they are both $G$-closed.
 
The amazing finding is that whenever an embedded
singular vector $\Psi_l^{\pm}$ in $\vm_{\Delta}$ happens to have
the conformal weight $\Delta+l=\frac{c}{24}$ then it becomes $G$-closed.
Hence the module generated by this singular vector is not isomorphic
to the full Verma module $\vm_{\frac{c}{24}}$ but only to the smaller 
$G$-closed module $\vm_{\frac{c}{24}}^{G}$. 

As in the previous section it is evident that among the cases of $\Psi_{p_1,q_1}^{\pm}$
that become $G$-closed there are none of the degenerate cases. For the latter
we require $t=\pm\frac{q_1}{p_2}=\pm\frac{q_2}{p_1}$, whilst the former
need $t=-\frac{q_1}{p_1}$ and ultimately $(p_1,q_1)=(p_2,q_2)$.
Hence, the degenerate cases do
not correspond to $G$-closed singular vectors,
i.e. two-dimensional singular spaces are never $G$-closed, not even partly.


\section{Conclusions and prospects}
\label{sec:conclusions}

By introducing an adapted ordering on the weight spaces of Ramond
Verma modules we derived the ordering kernels of the $N=1$ Ramond
algebra. These ordering kernels contain in the most general cases two
elements and consequently Ramond singular vectors are uniquely defined
by just two particular coefficients. In addition, Ramond singular vector spaces
can be degenerate (2-dimensional).
We derived expressions for these two coefficients
for all primitive Ramond singular vectors. It turned out that these two
coefficients are closed under multiplication. The corresponding multiplication
formulae therefore yield the whole Ramond embedding diagrams.

We argued that the structure of the highest weight representations of the (extended)
Ramond algebra can easily be transformed into the corresponding representations
of the unextended Ramond algebra. The only case where this transformation breaks
down is the supersymmetric case $\Delta=\frac{c}{24}$ for which the unextended module is 
already the quotient module of the extended module divided by its level $0$ singular
vector. In some of these cases we find that the (extended) Verma 
module has subsingular
vectors but not the unextended Verma module. We argued further that
one should not expect any other subsingular vectors. This was related to the fact that
the product of singular vector operators has proven to be always non-trivial with
the only exception of the supersymmetric case $\Delta=\frac{c}{24}$. 
A rigorous proof of this claim will be given in Ref. \icite{progress}
through a detailed analysis of the multiplicities of the roots of the determinant and their
intersections.

The embedding diagram patterns proven in this paper reveal that the derivation
of Ramond embedding diagrams is significantly more complicated than in the 
Virasoro case and the conjectures given in the literature\cite{kiritsis2} 
for the Ramond unitary models can only be proven by using the multiplication rules
for singular vector operators derived via the ordering kernel coefficients. 
We showed that degenerate singular vector spaces do appear for the Ramond algebra
which arise from ordering kernels with two elements. The structure of the
embedding diagrams is very similar to the Virasoro embedding diagrams,
nevertheless the rules for embedding patterns to collapse to simpler patterns
are very different than in the Virasoro case. As a consequence, we have embedding diagrams
where the two lowest level singular vectors are at the same level but the
embedding diagram still does not simplify to the straight line embedding diagram
- as it would do in the Virsoro case - but simply stays as the typical crossed
embedding pattern of conformal rational models.

We have once more demonstrated that the adapted ordering method is a very simple
mechanism that uniquely defines all singular vectors and reveals much
information about the embedding structure of the Verma modules. 
One easily manages even complications through subsingular vectors
and degenerate singular vectors and finally obtains the whole structure
of the embedding
patterns. In principle, it should easily generalise to any other kind of 
Lie (super)-algebra.

It was shown that the $N=1$ Ramond algebra contains many of the interesting
features that had only been discovered for the $N=2$ superconformal
algebras so far. Surprisingly enough,
the results obtained for the $N=1$ Ramond algebra resemble
much more the corresponding results for the twisted $N=2$ algebra
than for the Ramond $N=2$ algebra. These facts are one more reason
to claim that Ramond field theories
are still much less understood than Neveu-Schwarz field theories.
Nevertheless, the structure finally proven has still much in common
with the Virasoro structure and the $N=1$ Neveu-Schwarz structure.


\appendix
\section{Appendix: Level $1$, level $2$, and level $3$ singular vectors.}
\label{app:a}
Explicit expressions for the singular vectors $\Psi_{1,q}^{\times}$,
which are the analogues of  $\Psi_{1,q}^{\pm}$, in $\vm_{\lambda}^{\times}$
for $\lambda^2=\Delta_{1,q}-\frac{c}{24}$,
were given by Watts\cite{gerard} using the concept of fusion
in conformal field theory. These vectors can easily be transformed to
the basis used in this paper via the basis transformation discussed earlier.
In this section of the appendix we shall give all singular vectors 
in $\vm_{\Delta_{p,q}}$ until
level $3$. With the exception of $\Psi_{3,2}^{\pm}(t)$, 
these vectors are of the type $\Psi_{1,q}^{\pm}$ (and its symmetric
partners)
given by Watts in the unextended basis.
As mentioned earlier, at level 3 we observe for the
first time degenerate singular vectors arising from the intersection of
$\Psi^{\pm}_{6,1}$ and $\Psi^{\pm}_{2,3}$ and also from the intersection of
$\Psi^{\pm}_{1,6}$ and $\Psi^{\pm}_{3,2}$. Let us note again that intersections
of singular vectors of different type $\epsilon$ 
(see theorem \refoth{\ref{th:prsvecs}}), 
such as $\Psi^{\pm}_{2,1}$ and 
$\Psi^{\pm}_{1,2}$ only lead to one-dimensional singular spaces. 
 By explicit calculations\cite{maple} one finds the following 
singular vectors for parity $+$.

\noi
Level $1$:
\bea
\Psi^+_{1,2}(t) &=& \Bigl\{ (2-t) L_{-1} - 2 G_{-1}G_0 \Bigr\} 
\ket{-\frac{3}{16} (1-\frac{2}{t})} \com \label{sv12+} \\
\Psi^+_{2,1}(t) &=& \Bigl\{ (1-2t) L_{-1} + 2t G_{-1}G_0 \Bigr\} 
\ket{-\frac{3}{16} (1-2t)} \pkt \label{sv21+} 
\eea 

We recall the symmetry $\Psi^+_{p,q}(t)=-\frac{1}{t}\Psi^+_{q,p}(\frac{1}{t})$ 
in our normalisation, which we can 
observe in \eqs{\ref{sv12+}} and \eqoth{\ref{sv21+}}. In the following
we shall therefore give
only the results for $q$ even. 
\noi
Level $2$:
\bea
\Psi^+_{1,4}(t) &=& \Bigl\{ (4-t) L_{-1}^2 -4 L_{-1} 
G_{-1}G_0 +(\frac{3}{2}-\frac{6}{t}) L_{-2}   \nn \\
&& +(1+\frac{6}{t}) G_{-2}G_0 \Bigr\}
\ket{-\frac{11}{16}+\frac{15}{8t}} \pkt \label{eq:psi14}
\eea 

\noi
Level $3$:
\bea
\Psi^+_{1,6}(t) &=& \Bigl\{  (6-t) L_{-1}^3 -6  L_{-1}^2 G_{-1}G_0 
+(3+\frac{24}{t}) G_{-2}L_{-1}G_0  +(\frac{13}{2}-\frac{39}{t}) L_{-2}L_{-1} \nn \\
&& +(3-\frac{28}{t}+\frac{60}{t^2}) L_{-3} 
+(3+\frac{18}{t}-\frac{60}{t^2}) G_{-3}G_0 
+(\frac{1}{4}-\frac{3}{2t}) G_{-2}G_{-1} \nn \\
&& +\frac{15}{t} L_{-2}G_{-1}G_0 \Bigr\} 
\ket{\frac{1}{16}(\frac{70}{t}-19)} \com \\
\Psi^+_{3,2}(t) &=& \Bigl\{  
 (2-3t) L_{-1}^3 
-2  L_{-1}^2 G_{-1}G_0 
+(1+8t) G_{-2}L_{-1}G_0  
+(6t^2-4t+\frac{3}{2}-\frac{1}{t}) L_{-2}L_{-1} \nn \\
&& +(-3t^3+11t^2-9t+2) L_{-3} 
+(-2t^2+6t-1) G_{-3}G_0 
+(-4t+\frac{1}{t}) L_{-2}G_{-1}G_0 
 \nn\\
&& +(-6t^2+4t+\frac{3}{4}-\frac{1}{2t}) G_{-2}G_{-1} 
\Bigr\} 
\ket{t-\frac{19}{16}+\frac{3}{8t}} \pkt
\label{sv32+}  \eea 

One easily verifies that these examples can be written using the
notation introduced in definition \refoth{\ref{def:ab}}, where we only 
need to consider the coefficients with respect to the ordering kernel.
\bea
\Psi^+_{1,2}(t) &=& (2-t,-2)^+_{1} \ket{-\frac{3}{16}(1-\frac{2}{t})}  \com \\
\Psi^+_{1,4}(t) &=& (4-t,-4)^+_{2} \ket{-\frac{11}{16}+\frac{15}{8t}}  \com \\
\Psi^+_{1,6}(t) &=& (6-t,-6)^+_{3} \ket{\frac{1}{16}(\frac{70}{t}-19)} \com \\
\Psi^+_{3,2}(t) &=& (2-3t,-2)^+_{3} \ket{t-\frac{19}{16}+\frac{3}{8t}}  \pkt
\eea
At the intersection point $t=\pm 2$ the vectors $\Psi^+_{1,6}(\pm 2)$ and
$\Psi^+_{3,2}(\pm 2)$ obviously define a two-dimensional singular space
in $\vm_1$ or $\vm_{-\frac{27}{8}}$ for $t=2$ or $t=-2$ respectively.
The same is of course true for $\Psi^+_{6,1}(\pm \frac{1}{2})$
and $\Psi^+_{2,3}(\pm \frac{1}{2})$ for $t=\pm\frac{1}{2}$.  However,
$\Psi^+_{1,6}(\pm 1)$ and $\Psi^+_{6,1}(\pm 1)$ are proportional as well
as  $\Psi^+_{1,6}(\pm 3)$ and $\Psi^+_{2,3}(\pm 3)$ and other combinations
of this type. 

Besides the vectors $\Psi_{p,q}$ we should - of course - have 
at least also found the descendant singular vectors $\theta_{p_1,q_1}
\Psi_{p_2,q_2}$ for $\Delta_{p_1,q_1}=\Delta_{p_2,q_2}+
\frac{p_2 q_2}{2}$ and descendants thereof. The latter equation
simplifies to $t=\frac{\pm q_1-q_2}{p_2\pm p_1}$. If the two vectors
are of different type, then obviously the values for $t$ in question
have $t=\frac{u}{v}$ with $u$ and $v$ both odd. Furthermore,
at level $1$, $2$, and $3$ all possible cases simplify to either
$\cs_{o\pm}^{0(u)}$ or $\cs_{o\pm}^{0(v)}$ and hence the corresponding
embedding patterns contain descendant singular vectors already
among the pattern for the (generically) primitive singular vectors.
In the case that the two singular vectors are of the same type, we 
obtain both numerator and denominator of $t$ even and hence
one of $u$ or $v$ may or may not be even. For the levels $1$,
$2$, $3$ these cases are $(p_1,q_1)=(1,4)$, $(p_2,q_2)=(1,2)$ with the
corresponding values for $t=1$ or $(p_1,q_1)=(1,2)$, $(p_2,q_2)=(1,4)$
with $t=-1$. Hence both cases again contain no additional 
descendant singular vectors at level $3$. Therefore there are
no additional descendant singular vectors until level $3$. The
first examples of descendant singular vectors not contained
in the pattern of (generically) primitive singular vectors
appear at level $4$, for example for  $(p_1,q_1)=(1,6)$,
$(p_2,q_2)=(1,2)$ with $t=2$ which is of type $\cs_{e-}$
and does not simplify to the straight line embedding pattern.

We conclude this section of the appendix with
all singular vectors of negative parity at level $1$, $2$, and $3$ .
The above remarks about degenerate singular vectors and descendant singular
vectors are equally valid for these cases. Also for the negative parity
vectors the condition $\Psi^-_{p,q}(t)=-\frac{1}{t}\Psi^-_{q,p}(\frac{1}{t})$ holds 
and we hence give the singular vectors only for $q$ even.

\noi
Level $1$:
\bea
\Psi^-_{1,2}(t) &=& \Bigl\{ -4t L_{-1} G_0 +(2-t) G_{-1} \Bigr\} 
\ket{-\frac{3}{16} (1-\frac{2}{t})} \com \label{sv12-} 
\eea 

\noi
Level $2$:
\bea
\Psi^-_{1,4}(t) &=& \Bigl\{ -2t L_{-1}^2G_0 +(4-t) L_{-1} 
G_{-1} +3 L_{-2} G_0  \nn \\
&& +(\frac{t}{4}+\frac{1}{2}-\frac{6}{t}) G_{-2} \Bigr\}
\ket{-\frac{11}{16}+\frac{15}{8t}} \pkt
\eea 

\noi
Level $3$:
\bea
\Psi^-_{1,6}(t) &=& \Bigl\{  -\frac{4}{3}t L_{-1}^3G_0 +(6-t)  L_{-1}^2 G_{-1} 
+(1+\frac{t}{2}-\frac{24}{t}) G_{-2}L_{-1}  \nn \\
&& +\frac{26}{3} L_{-2}L_{-1}G_0 
+(4-\frac{40}{3t}) L_{-3} G_0
+(-\frac{28}{t}+\frac{t}{2}+\frac{60}{t^2}) G_{-3} \nn\\
&& +\frac{1}{3} G_{-2}G_{-1}G_0 +
(\frac{5}{2} -\frac{15}{t}) L_{-2}G_{-1} \Bigr\} 
\ket{\frac{1}{16}(\frac{70}{t}-19)} \com \\
\Psi^-_{3,2}(t) &=& \Bigl\{  
 -4t L_{-1}^3 G_0
+(2-3t)  L_{-1}^2 G_{-1} 
+(12t^2-\frac{13}{2}t-1) G_{-2}L_{-1}  \nn \\
&& +(2+8t^2) L_{-2}L_{-1}G_0 
+(12t^2-4t-4t^3) L_{-3} G_0
+(1-\frac{15}{2}t+11t^2-3t^3) G_{-3} \nn\\
&& +(1-8t^2) G_{-2}G_{-1}G_0 
+(4t+\frac{3}{2}-6t^2-\frac{1}{t}) L_{-2}G_{-1} 
\Bigr\} 
\ket{t-\frac{19}{16}+\frac{3}{8t})} \pkt
\label{sv32-}  \eea 
\noi
Once again, we can use the notation of
definition \refoth{\ref{def:ab}} to write these singular
vectors with ordering kernel coefficients only.
\bea
\Psi^-_{1,2}(t) &=& (-4t,2-t)^-_{1} \ket{-\frac{3}{16}(1-\frac{2}{t})}  \com \\
\Psi^-_{1,4}(t) &=& (-2t,4-t)^-_{2} \ket{-\frac{11}{16}+\frac{15}{8t}}  \com \\
\Psi^-_{1,6}(t) &=& (-\frac{4t}{3},6-t)^-_{3} \ket{\frac{1}{16}(\frac{70}{t}-19)} \com \\
\Psi^-_{3,2}(t) &=& (-4t,2-3t)^-_{3} \ket{t-\frac{19}{16}+\frac{3}{8t}}  \pkt
\eea
For $t=\pm 2$ and $t=\pm\half$ we find two-dimensional singular spaces with
highest weight vectors $\ket{1}$ or $\ket{-\frac{27}{8}}$ for $+2$ (and $-\frac{1}{2}$)
or $-2$ (and $-\frac{1}{2}$) respectively. 


\section{Appendix: Examples of subsingular vectors in the supersymmetric case}
\label{app:b}

We have shown in section  \refoth{\ref{sec:GclosedVM}} that for the
supersymmetric case $\Delta=\frac{c}{24}$ the singular 
vectors $\Psi_{p,q}^{\pm}$ for $\Delta=\Delta_{p,q}$ always lie in
the embedded submodule built on $G_0\ket{\frac{c}{24}}$, which is isomorphic
to $\vm_{0}^{\times}$. They therefore vanish in the $G$-closed quotient 
module $\frac{\vm_{\frac{c}{24}}}{U(\ram)G_0\ket{\frac{c}{24}}}$. The singular
vectors, which we find in the quotient module are hence subsingular vectors in
$\vm_{\frac{c}{24}}$. Since the quotient module 
$\frac{\vm_{\frac{c}{24}}}{U(\ram)G_0\ket{\frac{c}{24}}}$ is also isomorphic to
$\vm_{0}^{\times}$ it has singular vectors at the same levels as the embedded
module built on $G_0\ket{\frac{c}{24}}$. We shall give here all examples at
levels $1$, $2$, and $3$. Let us recall again that $c(t)=c(\frac{1}{t})$ and therefore
the $t\leftrightarrow\frac{1}{t}$ symmetric solutions of $\Delta_{p,q}(t)=\frac{c}{24}$
lead to the same value of $\Delta$ and hence to the same vector. 

\noi
Level $1$ has subsingular vectors for $t=2$:
\bea
\Upsilon_{1,2}^{+}(2) &=& L_{-1}\ket{0} \com \\
\Upsilon_{1,2}^{-}(2) &=& G_{-1}\ket{0} \pkt
\eea

\noi
Level $2$ has subsingular vectors for $t=4$:
\bea
\Upsilon_{1,4}^{+}(4) &=&  \Bigl\{ 8L_{-1}^2-3L_{-2}  \Bigr\} 
\ket{-\frac{7}{32}} \com \\
\Upsilon_{1,4}^{-}(4) &=& \Bigl\{ 8L_{-1}G_{-1}-5 G_{-2} 
\Bigr\} 
\ket{-\frac{7}{32}} \pkt
\eea

\noi
Level $3$ has subsingular vectors for $t=6$ and for
$t=\frac{2}{3}$:
\bea
\Upsilon_{1,6}^{+}(6) &=& \Bigl\{ 
-24 L_{-1}^3 
+26 L_{-2}L_{-1} 
+\frac{16}{3} L_{-3} 
+G_{-2}G_{-1} \Bigr\}
\ket{-\frac{11}{24}} \com \\
\Upsilon_{1,6}^{-}(6) &=& \Bigl\{ 
-12 L_{-1}^2 G_{-1}
+5 L_{-2}G_{-1} 
+\frac{26}{3} G_{-3} 
+14 G_{-2}L_{-1} \Bigr\}
\ket{-\frac{11}{24}} \pkt
\eea
\bea
\Upsilon_{3,2}^{+}(\frac{2}{3}) &=& \Bigl\{
-36 L_{-1}^3 
+75 L_{-2}L_{-1} 
+20 L_{-3} 
-\frac{69}{2} G_{-2}G_{-1} \Bigr\}
\ket{\frac{1}{24}} \com \\
\Upsilon_{3,2}^{-}(\frac{2}{3}) &=& \Bigl\{ 
-12 L_{-1}^2 G_{-1}
-7 L_{-2}G_{-1} 
+\frac{38}{3} G_{-3} 
+38 G_{-2}L_{-1} \Bigr\}
\ket{\frac{1}{24}} \pkt
\eea

\noi
These vectors are easily obtained from their singular vector counterpart of $\vm_{p,q}$.
Evaluating for example the singular vector $\Psi_{1,4}^+$ of \eq{\ref{eq:psi14}}
for $t=4$ leads to 
\bea
\Psi_{1,4}^{+}(4) & = & \Bigl\{ -4 L_{-1}G_{-1}G_0+\frac{5}{2}G_{-2}G_0 \Bigr\} \ket{-\frac{7}{32}} \com
\eea
which can be rewritten as 
$\Psi_{1,4}^{+}(4)  = -\half \{ 8 L_{-1}G_{-1} - 5 G_{-2} \} G_0\ket{-\frac{7}{32}}$ and  therefore
$\Upsilon_{1,4}^-= \{8 L_{-1}G_{-1} - 5 G_{-2} \} \ket{-\frac{7}{32}}$. 


\acknowledgements
I am extremely grateful to Beatriz Gato-Rivera for a very fruitful collaboration
and for many discussions and 
comments on the entire work.
I am also very grateful to Adrian Kent for many 
discussions on the ordering method 
and to Victor Kac for his insightful explanations on superconformal algebras. I am
indebted to the Deutsche Forschungsgemeinschaft (DFG) for financial
support and to DAMTP for its kind hospitality. 
This work was also supported by PPARC.

\noi



\end{document}